\documentclass[reprint,amsmath,amssymb,aps,prb]{revtex4-2}
\usepackage{docmute}
\usepackage[caption=false]{subfig}
\usepackage{graphicx}
\usepackage[percent]{overpic}
\usepackage{tikz}
\usepackage{orcidlink}
\newcommand{\myciteauthor}[1]{{\protect\NoHyper\citeauthor{#1}\endNoHyper}}
\hypersetup{
  colorlinks=true,
  linkcolor=blue,
  citecolor=blue,
  urlcolor=blue,
  pdfborder={0 0 0}
}
\newcommand{\tabcitesup}[2]{\textsuperscript{\hyperlink{cite.#2}{#1}}}

\begin{document}

\preprint{APS/123-QED}
\title{Lattice dynamics and structural phase stability of group-IV elemental solids with the r\texorpdfstring{$\mathbf{^2}$}{2}SCAN functional}
\author{Adonis Haxhijaj\orcidlink{0009-0006-6113-9542}}
 \thanks{\hspace{-0.4em}Contact author: {\hypersetup{urlcolor=black}\href{mailto:adonis.haxhijaj@epfl.ch}{adonis.haxhijaj@epfl.ch}}}
\author{Stefan Riemelmoser\orcidlink{0000-0002-9202-2906}}%
\author{Alfredo Pasquarello\orcidlink{0000-0002-9142-2799}}%
\affiliation{%
Chaire de Simulation à l'Echelle Atomique (CSEA),
Ecole Polytechnique Fédérale de Lausanne (EPFL),
CH-1015 Lausanne, Switzerland
}
\date{\today}

\begin{abstract}
The strongly constrained and appropriately normed (SCAN) meta-generalized gradient approximation (meta-GGA) functional is a milestone achievement of electronic structure theory. Recently, a revised and restored form (r$^2$SCAN) has been suggested as a replacement for SCAN in high-throughput applications. Here, we assess the accuracy and reliability of the r$^2$SCAN meta-GGA functional for the group-IV elemental solids carbon (C), silicon (Si), germanium (Ge), and tin (Sn). We show that the r$^2$SCAN functional agrees closely with its parent functional SCAN for elastic constants, bulk moduli, and phonon dispersions, but the numerical stability of r$^2$SCAN is superior. Both meta-GGA functionals outperform standard GGA (Perdew-Burke-Ernzerhof) in terms of accuracy and approach the level of common hybrid functionals (Heyd-Scuseria-Ernzerhof). However, we find that r$^2$SCAN performs much worse than SCAN for the $\alpha\leftrightarrow \beta$ phase transition of both Ge and Sn, yielding larger phase energy differences and transition pressures.
\end{abstract}

\maketitle

\onecolumngrid
\vspace{-1cm}
\begin{widetext}
\noindent\small\textit{%
\begingroup
\spaceskip=0.8\fontdimen2\font plus 0.9\fontdimen3\font minus 0.9\fontdimen4=\spaceskip
This is a postprint of an article published by the American Physical Society in
\textbf{Phys.\ Rev.\ B 113, 104105 (2026)}. The final version of record is available at DOI: \href{https://doi.org/10.1103/znkf-3g37}{10.1103/znkf-3g37}.
This article may be downloaded for personal use only. Any other use requires prior
permission of the author and the American Physical Society. Copyright
\textcopyright\ 2026 American Physical Society.%
\endgroup}
\end{widetext}

\vspace{0.5em}

\twocolumngrid

\section{\label{sec:Introduction}INTRODUCTION}

The quest for an ever-improving exchange-correlation functional of density functional theory (DFT) has been a major challenge for generations of researchers in theoretical physics and chemistry~\cite{Burke2012,Jones2015,Medvedev2017}.
With the development of DFT in the 1960s~\cite{Hohenberg1964,Kohn1965}, Kohn and Sham also proposed the local density approximation (LDA)~\cite{Kohn1965}. In the 1990s, the generalized gradient approximation (GGA) revolutionized DFT applications. In particular, the Perdew-Burke-Ernzerhof functional (PBE)~\cite{Perdew1996} has become the default choice for many first-principles materials studies. 
PBE combines an elegant functional form, numerical stability, and is well balanced for different types of chemical bonding. However, it was quickly realized that PBE suffers from systematic underbinding~\cite{Favot1999}. That is, PBE overestimates bond lengths~\cite{Schimka2011}, and underestimates bulk moduli as well as elastic constants~\cite{Raasander2015}. Moreover, PBE underestimates phonon frequencies~\cite{Favot1999,Grabowski2007,Hummer2009} and formation energies of solids~\cite{Schimka2011,Tran2016}.
The wish to cure this underbinding has led to the development of other GGA functionals that are optimized for solid state applications~\cite{Armiento2005,Wu2006,Perdew2008}. However, these improvements come with some loss of accuracy for other properties. Broadly speaking, these ``GGA functionals for solids'' perform worse than PBE for atoms and molecules. More surprisingly, the GGAs for solids can also be less accurate than PBE for solid-solid phase transitions~\cite{Xiao2013,Mehl2021}. For a universal increase in accuracy, one has to step outside the GGA. \\
Jacob's Ladder of DFT~\cite{Perdew2001} characterizes the path toward chemical accuracy, with LDA presenting the first rung, and GGA presenting the second rung. The third rung consists of meta-GGA functionals, while hybrid functionals make up the fourth rung. The strongly constrained and appropriately normed (SCAN) meta-GGA functional of \citet{Sun2015,Sun2016} is a milestone achievement of theoretical development. SCAN largely corrects the PBE underbinding error and takes a large step toward chemical accuracy~\cite{Tran2016,Zhang2017,Zhang2018,Shahi2018}. 
However, SCAN has been plagued by numerical issues~\cite{Yao2017,Bartok2019,Furness2020,Furness2022} such as poor basis set convergence. This is particularly problematic for applications that require high numerical accuracy; for instance phonon calculations~\cite{Ning2022}. To fix SCAN's numerical problems, \myciteauthor{Bartok2019} proposed a revised version of SCAN (rSCAN), which however sacrifices some functional accuracy~\cite{Bartok2019}. Furness \textit{et al.} (partially) restored exact constraints that are violated by rSCAN, and proposed the r$^2$SCAN functional~\cite{Furness2020,Furness2022}. r$^2$SCAN promises to combine the accuracy of SCAN with the numerical stability of rSCAN, and has already been employed in several high-throughput studies~\cite{Kingsbury2022,Kothakonda2022}. Still, the r$^2$SCAN and SCAN functionals are not completely equivalent, but it is generally difficult to disentangle real physical differences due to the numerical noise inherent to SCAN calculations. One difference that has been established is that r$^2$SCAN, on average, yields slightly larger bond lengths than SCAN~\cite{Kingsbury2022,Kothakonda2022}.  However, it is not entirely clear yet if this is beneficial or not. \\
In this study, we assess the SCAN and r$^2$SCAN meta-GGA functionals for elastic properties of group-IV elemental solids (C, Si, Ge, and Sn). We put emphasis on high numerical accuracy, which is necessary to work out physical distinctions between the otherwise similar SCAN and r$^2$SCAN functionals. There are two reasons why the group-IV elemental solids are ideal for this purpose: (i) it is actually feasible to obtain converged SCAN results for these nonpolar materials and (ii) accurate references from experiments and higher-level theories are available, which makes it possible to assess which meta-GGA functional yields better results.\\
The rest of this study is structured as follows: In Sec.\@~\ref{sec:Theory}, we recall some relevant fundamentals of elastic theory.
Our computational setup and treatment of zero-point expansion (ZPE) is detailed in Sec.\@~\ref{sec:Methodology}. The results for elastic constants, phonon frequencies, and transition pressures are given in Secs.\@~\ref{sec:Results: Elastic properties},~\ref{sec:Results: phonons}, and~\ref{sec:Results: Transition-pressures}. In Sec.\@~\ref{sec:Discussion}, we discuss the results and present an outlook. Convergence tests are given in the Supplemental Material (SM)~\cite{SM}.

\section{\label{sec:Theory}THEORY}

\subsection{\label{sec:Th:Elastic constant}Elastic constants}

In what follows, we consider small elastic deformations of a homogeneous cubic crystal, described by the linearized strain tensor~\cite{Kittel} 
\begin{equation}
e_{ij} \;=\;\frac12\Bigl(\frac{\partial u_i}{\partial x_j}+\frac{\partial u_j}{\partial x_i}\Bigr)\,,
\end{equation}
where \(\mathbf{u}\) is the displacement field and \((x_1,x_2,x_3)\equiv(x,y,z)\) are Cartesian coordinates aligned with the crystal axes.  The elastic stiffness tensor \(C_{ijkl}\) relates the stress \(\sigma_{ij}\) to the strain via  
\begin{equation}
\sigma_{ij}=C_{ijkl}\,e_{kl}\,.
\end{equation}
By cubic symmetry, there are only three independent elastic constants: one diagonal constant, 
\begin{equation}
C_{11}=C_{1111}=C_{2222}=C_{3333},
\end{equation}
one off-diagonal constant,  
\begin{equation}
C_{12}=C_{1122}=C_{1133}=C_{2233},
\end{equation}
and the shear constant,
\begin{equation}
C_{44}=C_{2323}=C_{3131}=C_{1212}.
\end{equation}
We define the elastic energy density $U$ by
\begin{equation}
U=\tfrac12C_{ijkl}e_{ij}e_{kl},
\end{equation} 
and the (isothermal) bulk modulus \(B_0\) by  
\begin{equation}\label{eq: bulk modulus definition}
B_0\;\equiv\;-\left. V\,\frac{\mathrm{d}P}{\mathrm{d}V} \right|_{V=V_0}
\;=\;\left.\frac{\mathrm{d}^2 U}{\mathrm{d}(\Delta V/V)^2} \right|_{V=V_0}\,,
\end{equation}
where \(\Delta V/V=e_{xx}+e_{yy}+e_{zz}\). It can be shown that the elastic energy density of the cubic crystal can be expressed as~\cite{Kittel}
\begin{equation}
\begin{aligned}
U = \tfrac12\big[ & \, C_{11} \left(e_{xx}^2 + e_{yy}^2 + e_{zz}^2\right) \\
                    & + 2C_{12} \left(e_{xx}e_{yy} + e_{yy}e_{zz} + e_{zz}e_{xx}\right) \big] \,.
\end{aligned}
\end{equation}
\noindent For a uniform  dilation, the strain is the same in all directions: $ e_{x x}=e_{y y}=e_{z z} \equiv \delta $, and all shear strains are zero. Substituting these equal strains into the energy density gives
\begin{equation}\label{eq:U uniform dilation}
U=\tfrac12\left[3 C_{11} \delta^2+6 C_{12} \delta^2\right]=\tfrac32 \delta^2\left(C_{11}+2 C_{12}\right).
\end{equation}
For a small uniform dilation, the fractional change in volume is
\begin{equation}
\frac{\Delta V}{V}=e_{x x}+e_{y y}+e_{z z}=3 \delta.
\end{equation}
Combining this expression with Eqs.\@~\eqref{eq: bulk modulus definition} and~\eqref{eq:U uniform dilation}, we obtain the important identity
\begin{equation}\label{eq:bulk-with-C}
B_0\equiv B_0^{\text{elastic}}=\tfrac13\left(C_{11}+2 C_{12}\right) .
\end{equation}

\subsection{\label{sec:Th:Phonons}Phonon supercell method}

To compute the lattice dynamics within the harmonic approximation, we start with the interatomic force constants. That is, the Hessian matrix defined by~\cite{Kresse1995,Hummer2009}
\begin{equation}
\Phi_{\textbf{R}_{n i}\alpha,\, \textbf{R}_{m j}\beta}
\;=\;
-\frac{\partial F_{\textbf{R}_{m j}\beta}}{\partial u_{\textbf{R}_{n i}\alpha}}
\;=\;
\frac{\partial^2 E}{\partial u_{\textbf{R}_{n i}\alpha}\,\partial u_{\textbf{R}_{m j}\beta}} ,
\end{equation}
where $F_{\textbf{R}_{m j}\beta}$ is the Hellmann-Feynman force on atom $j$ in the cell located at $\textbf{R}_m$ due to a small displacement $u_{\textbf{R}_{n i}\alpha}$ of atom $i$ in the cell at $\textbf{R}_n$. Cartesian indices are denoted by $\alpha$ and $\beta$. Owing to translational invariance of the crystal, the force constants depend only on the difference $\textbf{R}_m-\textbf{R}_n$. Next, a Bloch wave ansatz is introduced for the atomic displacements:
\begin{equation}
\tilde{u}_{i\alpha}^{(\textbf{q})}=u_{i\alpha} e^{i \textbf{q} \cdot \textbf{R}_n} ,   
\end{equation}
where $\textbf{q}$ denotes the wave vector. This directly leads to the reciprocal-space representation of the force constant matrix (we drop Cartesian indices for simplicity)
\begin{equation}
\tilde{\Phi}_{i j}(\textbf{q})=\sum_{\textbf{R}_m-\textbf{R}_n} e^{i \textbf{q} \cdot\left(\textbf{R}_m-\textbf{R}_n\right)} \Phi_{i j}\!\left(\textbf{R}_m-\textbf{R}_n\right) .
\end{equation}
From here, we define the dynamical matrix $D$ by normalizing the force constants with the atomic masses $M_i$, 
\begin{equation}\label{eq:dynamical matrix}
D_{i j}(\textbf{q})=\frac{1}{\sqrt{M_i M_j}}\,\tilde{\Phi}_{i j}(\textbf{q}) .
\end{equation}
Finally, the lattice vibrational frequencies are determined by solving the eigenvalue problem 
\begin{equation}
D(\textbf{q}) \cdot \textbf{u}^{(\textbf{q})}=\omega^2(\textbf{q}) \textbf{u}^{(\textbf{q})} ,
\end{equation} 
so that the phonon frequencies are obtained as the square roots of the eigenvalues $\omega^2(\textbf{q})$. \\
In practical terms, using a supercell with finite displacements (“direct” approach), one calculates the real-space force constant matrix~\cite{Kresse1995,Hummer2009}. A Fourier transform yields the dynamical matrix, which is then diagonalized to find the phonon dispersion. The phonon frequencies are explicitly calculated at specific wave vectors $\textbf{q}$ that are commensurate with the supercell. Phonon frequencies at other wave vectors are interpolated, and convergence of the full phonon dispersion is achieved in the limit of large supercells. \\ 
Finally, the speeds of acoustic waves along high-symmetry directions are related to the elastic constants. For longitudinal acoustic (LA) and transverse acoustic (TA) waves propagating along the [100] direction, one can show that~\cite{Kittel}
\begin{equation}\label{eq:v_100}
v_{\text{LA},[100]}=\sqrt{\frac{C_{11}}{\rho}} \quad \text { and } \quad v_{\text{TA},[100]}=\sqrt{\frac{C_{44}}{\rho}},
\end{equation}
Furthermore, for LA waves along the [110] direction, the longitudinal speed is given by~\cite{Kittel}
\begin{equation}\label{eq:v_110}
v_{\text{LA},[110]}=\sqrt{\frac{C_{11}+C_{12}+2 C_{44}}{2 \rho}}.
\end{equation}
We will use this approach as an alternative method for computing the elastic constants. This will allow us to compare the results and evaluate whether it provides a more efficient or accurate means of determining the elastic properties without the need to apply external strain.

\begin{figure}[t]
\centering
\subfloat{%
  \begin{overpic}[width=0.45\linewidth,keepaspectratio]{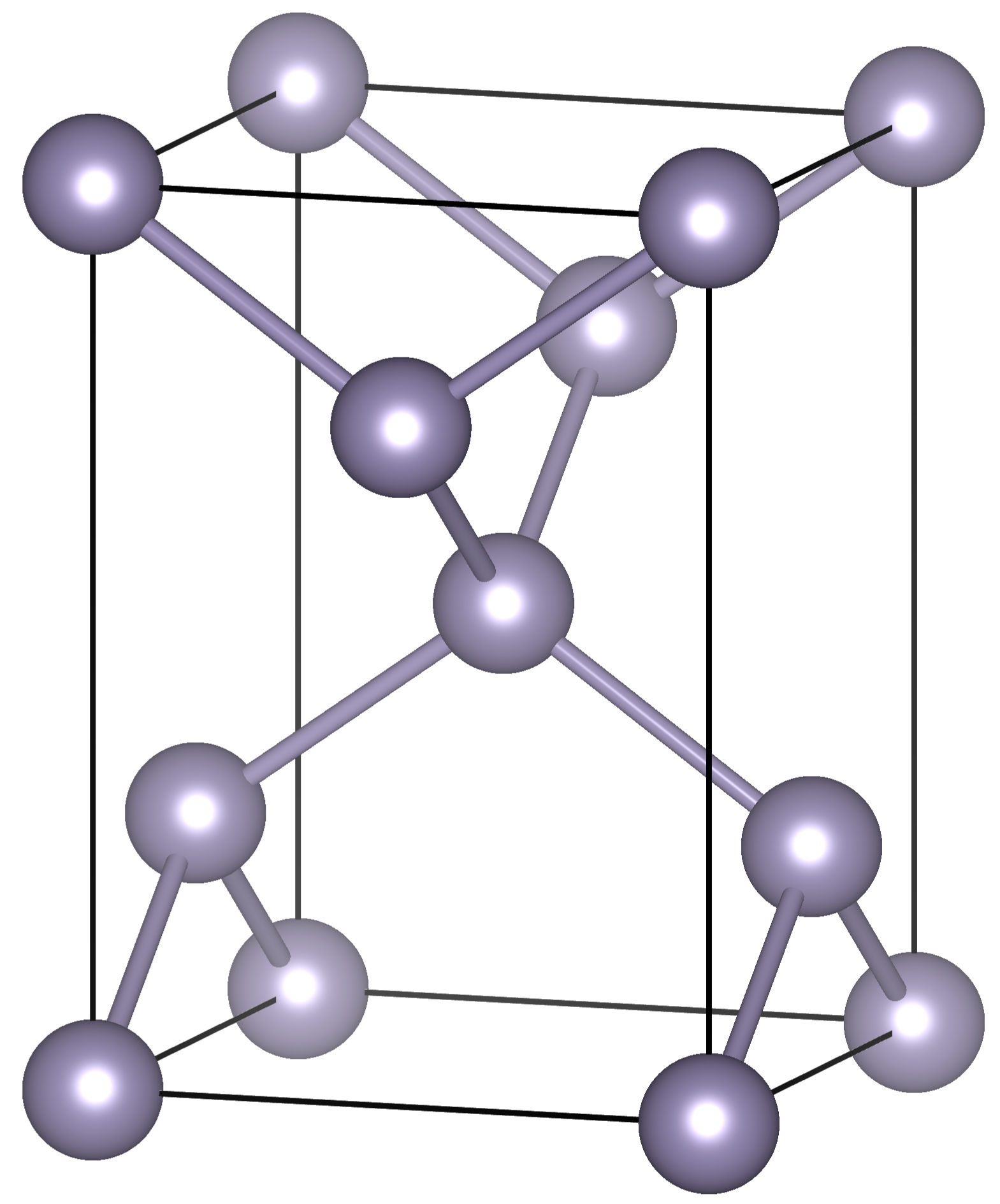}
    \put(30,0){\Large $a$}%
    \put(69,3){\Large $a$}%
    \put(-7,25){\rotatebox{90}{\Large$c=a\sqrt{2}$}}
  \end{overpic}%
  \label{alpha}%
}
\hspace{0.6cm}
\subfloat{%
  \begin{overpic}[width=0.41\linewidth,keepaspectratio]{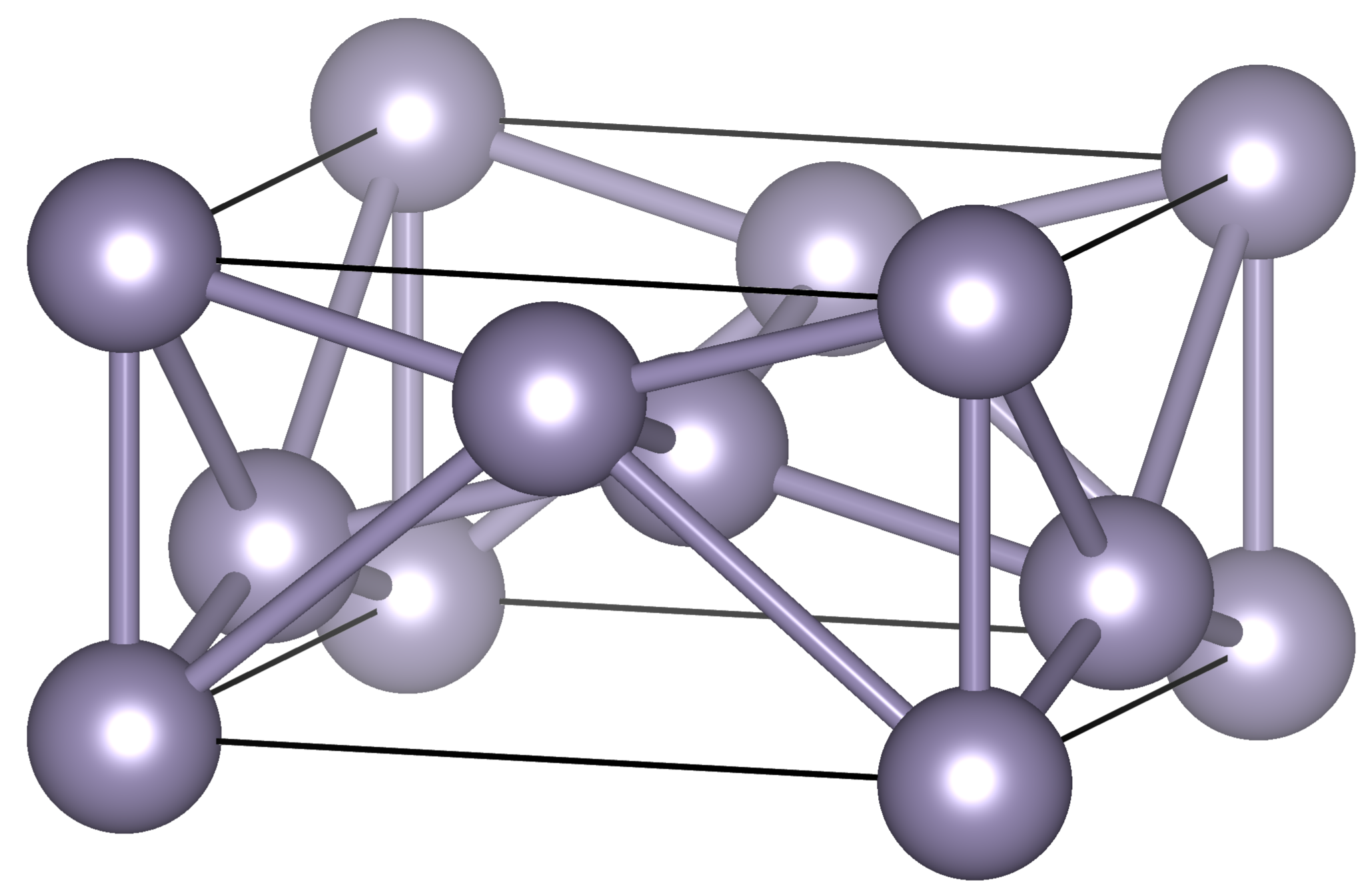}
    \put(37,0){\Large $a$}%
    \put(82,3){\Large $a$}%
    \put(-14,8){\rotatebox{90}{\Large$c<a\sqrt{2}$}}
  \end{overpic}%
  \label{beta}%
}
\caption{\label{fig:phases}Comparison of the (left)  $\alpha$-Sn  and (right) $\beta$-Sn crystal structures~\cite{Momma2011}. The $\beta$-Sn structure can be obtained from the $\alpha$-Sn structure  (diamond, shown here in a body-centered tetragonal unit cell) by applying a tetragonal compression along (001)~\cite{Yin1982a,Moll1995,Mehl2021}. The two phases coincide at $c/a = \sqrt{2} \approx 1.414$, while the ideal $\beta$ phase corresponds to sixfold coordination at $c/a = \sqrt{4/15} \approx 0.516$~\cite{Christensen1993}.}
\end{figure}
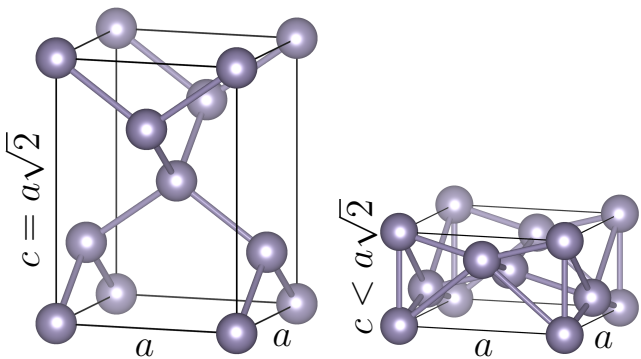

\subsection{\label{sec:Th:Transition-pressure}Transition pressures}

Elemental tin (Sn) is known to exhibit two allotropes, which are illustrated in
Fig.\@~\ref{fig:phases}. The thermodynamically stable phase at ambient conditions is a white, ductile metal ($\beta$-Sn, Strukturberich A5). At temperatures below $13\,^\circ\mathrm{C}$ (286~K), an entropy-driven phase transition occurs, and tin degrades to a brittle, gray semimetal ($\alpha$-Sn, Strukturberich A4, diamond structure). This phenomenon is commonly known as “tin pest”~\cite{Cornelius2017,Mehl2021}. The $\beta \rightarrow \alpha$ phase transition is anomalous, as the low temperature $\alpha$ phase is about 20\% larger in volume than the high-temperature $\beta$ phase (negative thermal expansion). At low temperatures, one can also induce the $\alpha \rightarrow \beta$  phase transition by applying pressure~\cite{Mujica2003}. Experimental transition pressures have not been measured directly, but the phase energy difference at $T = 0$~K has been estimated to be 15~meV~\cite{Ihm1981}. Elemental silicon (Si) and germanium (Ge) are semiconductors at ambient conditions and crystallize in the diamond structure. A pressure-driven phase transition to metallic $\beta$ phases is known to occur at 11.7~GPa for Si and 10.6~GPa for Ge, respectively~\cite{Mujica2003}. Since the pioneering studies of Cohen and coworkers~\cite{Yin1980,Ihm1981,Yin1982,Yin1982a}, the $\alpha\leftrightarrow\beta$ phase transitions of Si, Ge, and Sn have been the subject of numerous first-principles studies~\cite{Christensen1993,Moll1995,DalCorso1996,GaalNagy1999,Hennig2010,Xiao2013}. Early calculations and experimental data have been reviewed by \citet{Mujica2003}, while recent studies also investigated the SCAN functional~\cite{Sun2016,Yao2017,Sengupta2018,Shahi2018,Mehl2021}. 
A primary objective of this study is to assess whether the r$^2$SCAN meta‐GGA functional can accurately describe this structural phase transition. For the sake of completeness, we note that elemental carbon (C) does not exhibit a $\beta$ phase in the experimentally accessible pressure range~\cite{Mujica2003}. For the application of r$^2$SCAN to graphite, we refer to the recent study of \citet{Ning2022a}. \\
For the $\alpha$ phase, we perform a series of static total‐energy calculations over a range of volumes around the expected equilibrium. The resulting energy-volume data are fitted to the third‐order Birch-Murnaghan equation of state (EOS)~\cite{BM-potential}
\begin{equation}\label{eq:BM EOS}
\begin{aligned}
E_\alpha(V) & = E_{0,\alpha} + \frac{9V_{0,\alpha}B_{0,\alpha}}{16}
\Bigl\{\bigl[(V_{0,\alpha}/V)^{2/3}-1\bigr]^3 B_{0,\alpha}' \\
& + \bigl[(V_{0,\alpha}/V)^{2/3}-1\bigr]^2\bigl[6-4(V_{0,\alpha}/V)^{2/3}\bigr]\Bigr\}.
\end{aligned}
\end{equation}
\noindent The $\beta$ phase depends on two lattice parameters, $a$ and $c$, or equivalently $V$ and $c/a$.  For a set of volumes $V$, we optimize the $c/a$ ratio by relaxing the cell shape at fixed volume. Then, we fit these points $E(V) = \text{min}_{c/a} \; E(V,c/a)$ to the same third‐order Birch-Murnaghan equation of state [Eq.\@~\eqref{eq:BM EOS}] to extract $(E_{0,\beta},V_{0,\beta},B_{0,\beta},B_{0,\beta}')$. The transition pressure is obtained from the common‐tangent construction on the $E(V)$ curves. The pressure-volume curves $P(V)$ can be obtained analytically by taking the derivative of Eq.\@~\eqref{eq:BM EOS},
\begin{equation}\label{eq:pressure}
\begin{aligned}
P(V) &= -\frac{\partial E}{\partial V}
       = \frac{3B_0}{2}\Bigl[(V_0/V)^{7/3} - (V_0/V)^{5/3}\Bigr]\\
     &\quad\times \Bigl\{1 + \frac{3}{4}(B_0'-4)\bigl[(V_0/V)^{2/3}-1\bigr]\Bigr\},
\end{aligned}
\end{equation}
and the transition pressure $P_{\rm t}$ satisfies
\begin{equation}\label{eq:transition pressure}
\begin{aligned}
P_{\rm t} = P_\alpha\bigl(V_{\alpha}^t\bigr) &\;=\; P_\beta\bigl(V_{\beta}^t\bigr),\\
H_\alpha\bigl(V_{\alpha}^t\bigr) &\;=\; H_\beta\bigl(V_{\beta}^t\bigr).
\end{aligned}
\end{equation}
Here, $H=E+PV$ is the enthalpy per atom, and $V_{\alpha}^t$, $V_{\beta}^t$ are the volumes at which the two pressure curves coincide. 

\section{\label{sec:Methodology}METHODOLOGY}

All calculations are performed with the Vienna \emph{ab initio} Simulation Package (vasp)~\cite{vasp,KresseJoubert1999}.  We employ the projector augmented‐wave (PAW) method~\cite{Blochl1994} and consider three semilocal exchange-correlation functionals: the PBE generalized gradient approximation~\cite{Perdew1996}, the SCAN meta‐GGA~\cite{Sun2015}, and its regularized-restored form r$^2$SCAN~\cite{Furness2020}. For comparison, we also consider the Heyd-Scuseria-Ernzerhof (HSE) hybrid functional~\cite{Heyd2003,*Heyd2006,Paier2006,*Paier2006a,Krukau2006}.
Throughout, we use highly accurate PBE PAW pseudopotentials~\cite{KresseJoubert1999} as distributed with \texttt{vasp.6.4.2}.
For Si, Ge, and Sn, the semicore shell is fully treated as valence; further details are given in the SM~\cite{SM}.
We note that previous studies~\cite{Moll1995,Yao2017} have emphasized the importance of accurate pseudopotentials for transition pressures in particular.\\
Following~\citet{Hummer2009}, we employ 24$\times$24$\times$24 $\Gamma$-centered \textbf{k}-points for diamond structures in primitive two-atom cells. For $\beta$-Sn structures, we use 24 \textbf{k}-points along $x$ and $y$, with denser sampling along the $z$ direction, accounting for the smaller $c/a$ ratio.  For HSE calculations, we employ a fourfold downsampling of the Fock operator~\cite{Paier2006}. Thus, Fock exchange is sampled on a reduced \textbf{k}-point mesh (6$\times$6$\times$6 for the diamond structures).
Our tests indicate that the $\textbf{k}$-point error for elastic constants and phonon frequencies is below that of the basis set error (not shown).  Unless specified otherwise, we raise the default energy cutoff (ENCUT in vasp) by a factor 1.8 to ensure tight basis set convergence of our results. The energy differences between $\alpha$ and $\beta$ phases are converged to 1~meV/atom or better, see also Refs.\@~\cite{Shahi2018,Sengupta2018,Mehl2021}. \\
Our phonon calculations employ the direct method implemented in vasp, details of the implementations are given in Refs.\@~\cite{Kresse1995,Hummer2009,Engel2020}. For phonon calculations, we reduce the \textbf{k}-point mesh, commensurate with the supercell size (for instance, 12$\times$12$\times$12 \textbf{k}-points and 3$\times$3$\times$3 Fock \textbf{k}-points for a 16-atom supercell). 
Similar to \citet{Ning2022}, we consider two numerical setups for phonon calculations; details are discussed in the SM~\cite{SM}. In agreement with \citet{Ning2022}, we find that a single-grid strategy (PREC=Accurate in vasp) cannot yield reliable SCAN phonon dispersions at reasonable plane-wave cutoffs. Following Ref.\@~\cite{Ning2022}, we fix this by employing a double-grid technique (PREC=High in vasp), which uses a higher plane-wave cutoff for the PAW augmentation charges.
Furthermore, calculations of phonons and elastic constants use a fourth-order finite difference stencil with a step size of 0.060~\AA. In the SM~\cite{SM}, we show that this double-grid strategy achieves a numerical accuracy of 1~GPa for bulk moduli, while elastic constants have residual basis set errors of $\approx$~2 to 3~GPa due to residual Pulay stress~\cite{Dacosta1986,G_P_Francis_1990}. Phonon frequencies are converged within an accuracy of 0.1~THz.

\subsection{Zero-point expansion corrections}\label{app:ZPE correction}

Throughout this work, we account for the effect of ZPE. We do so by adding ZPE corrections to the experimental measurements. In principle, this approach assumes that the ZPE corrections do not depend on the density functional. This is a reasonable approximation given that the ZPE corrections considered here are consistently smaller than the DFT errors themselves. The ZPE corrections for the lattice constants and bulk moduli are taken from Ref.\@~\cite{Schimka2011}. The corrections in their study were obtained in the harmonic approximation, combining the PBE functional and phonon supercell calculations (16 atoms). \\
To correct the elastic constants $C_{ij}$ for zero-point expansion (Table~\ref{tab:elastic_constants_EXPT}), we adopt the following protocol. 
Using the r$^2$SCAN meta-GGA functional, we calculate the elastic constants twice. A first calculation is done at the equilibrium lattice parameter $a_0$ obtained from our Birch-Murnaghan EOS fit, and a second calculation is done at the slightly expanded cell $a_0+\Delta a_0^{\rm ZPE} $ (with $\Delta a_0^{\rm ZPE}$ taken from Ref.\@~\cite{Schimka2011}). The ZPE correction for each independent elastic constant is then calculated as the difference
\begin{equation}
\Delta C_{ij}^{\mathrm{ZPE}} = C_{ij}(a_0 + \Delta a_0^{\rm ZPE})-C_{ij}(a_0),
\end{equation}
and the corrected experimental elastic constants are simply
\begin{equation}\label{eq:ZPE elastic constants}
C_{ij}^{\mathrm{Expt.\!-\!ZPE}} = C_{ij}^{\mathrm{Expt.}} - \Delta C_{ij}^{\mathrm{ZPE}}.
\end{equation}
We can also use Eq.\@~\eqref{eq:bulk-with-C} and compare the quantity $(\Delta C_{11}^{\rm ZPE}+2\Delta C_{12}^{\rm ZPE})/3$, calculated here with r$^2$SCAN, with the results for $\Delta B_0^{\rm ZPE}$ obtained with PBE in Ref.\@~\cite{Schimka2011}. The differences are small: 2.6~GPa for C, 0.1~GPa for Si, 0.2~GPa for Ge, and 0.1~GPa for Sn. Considering our numerical errors as well as uncertainties in the experimental results, this confirms our assumption that the ZPE dependence on the DFT functional can be safely neglected.
In the same manner, we correct the experimental phonon frequencies $\nu$. For each point $\textbf{q}$ in the Brillouin zone and for each phonon branch $b$, we calculate the ZPE corrections
\begin{equation}
\Delta \nu_{\textbf{q},b} =\nu_{\textbf{q},b}\!\left(a_0+\Delta a_0^{\mathrm{ZPE}}\right) - \nu_{\textbf{q},b}(a_0) ,
\end{equation}
and subtract them from the experimental measurements, 
\begin{equation}\label{eq:phonon-zpe}
\nu^{\mathrm{Expt.-ZPE}}_{\textbf{q},b} \;=\; \nu^{\mathrm{Expt.}}_{\textbf{q},b} \;-\; \Delta \nu_{\textbf{q},b}.
\end{equation}
The corrections $\Delta \nu_{\textbf{q},b}$ are calculated by combining r$^2$SCAN and 16 atom supercells at the high-symmetry points $\Gamma$, X, and L. These points are commensurate with both the 16-atom and 128-atom supercells, and thus the finite size error is negligible here  (see also the SM~\cite{SM}).\\
Finally, ZPE corrections for the transition pressures of Si and Ge are taken from the work of \myciteauthor{Sengupta2018}~\cite{Sengupta2018}. Their study employed the quasi-harmonic approximation and also accounted for temperature effects at $T = 300$~K. Their phonon spectra were calculated with  density functional perturbation theory using the PBE functional, and the corresponding vibrational contributions to the enthalpy were added to their calculated equations of state for different DFT functionals. Here, we choose $\Delta P_{\rm t}^{\rm ZPE}$ calculated using the random-phase approximation (1.0~GPa for Si, 1.0~GPa for Ge), which predicts transition pressures in close agreement with experiment, but other functionals yield similar corrections. For instance, PBE yields 0.8~GPa for Si and 1.0~GPa for Ge, and SCAN yields 0.7~GPa for Si and 0.9~GPa for Ge~\cite{Sengupta2018}. Similarly, Ref.\@~\cite{Mehl2021} reported weak DFT dependence of the zero-point correction to the energy difference $\Delta E_0$ between $\alpha$-Sn and $\beta$-Sn. Here, we use the value of 7~meV/atom reported for SCAN (PBE also yields 7~meV/atom, see Ref.~\cite{Mehl2021}). 

\section{\label{sec:Results: Elastic properties}Elastic properties of the diamond structures}

\subsection{Lattice constants}
Table~\ref{tab:lattice_constants} lists the lattice constants of the diamond phases of C, Si, Ge, and Sn, obtained by Birch-Murnaghan fits [see Eq.\@~\eqref{eq:BM EOS}].  Relative to experiment~\cite{madelung2004}, PBE systematically overestimates the equilibrium lattice parameter $a_0$,  reflecting the well-known PBE underbinding. Both r$^2$SCAN and SCAN reduce this error, yielding smaller $a_0$, in better agreement with experimental lattice constants. The HSE functional also yields similar improvements. Inclusion of ZPE~\cite{Schimka2011} slightly increases the first-principles $a_0$ (corresponding to a negative correction to $a_0^{\rm expt}$). ZPE is most significant for the lightest element C and diminishes toward $\alpha$-Sn. 
We note that the lattice constants obtained here with PBE and HSE are in good agreement with earlier studies~\cite{Hummer2009,Schimka2011, Raasander2015,Shahi2018,Mehl2021} (differences  $\lesssim 0.005$~\AA). 
Larger deviations are found with respect to the SCAN and r$^2$SCAN lattice constants reported in previous studies.
In particular, our SCAN lattice constants for Si, Ge, and Sn are 0.01-0.02$~\text{\AA}$ larger than the corresponding results in the literature (5.428~\AA\@~\cite{Sengupta2018}, 5.428~\AA\@~\cite{Shahi2018}, and 5.426~\AA\@~\cite{Furness2020} for Si; 5.658~\AA\@~\cite{Sengupta2018}, 5.660~\AA\@~\cite{Shahi2018}, and 5.673~\AA\@~\cite{Furness2020} for Ge; 6.540~\AA\@ for Sn~\cite{Mehl2021}). Most of these differences can be traced back to the use of more accurate pseudopotentials in the present work. That is, by repeating the calculations with the pseudopotentials used in these previous studies (this involves freezing of semicore electrons with respect to our setup), we obtain smaller lattice constants in better agreement with the values in the literature (not shown). Given the accuracy of the present calculations, we confidently confirm that the r$^2$SCAN lattice constants are slightly larger than those of its parent functional SCAN~\cite{Furness2020,Kingsbury2022,Kothakonda2022}.Comparing with experimental lattice constants, Table~\ref{tab:lattice_constants} shows that SCAN is slightly more accurate than r$^2$SCAN. This is in line with the general trend found for a larger set of materials~\cite{Furness2020,Kingsbury2022,Kothakonda2022}. \\
Finally, we note that the PBE underbinding error is especially severe for Ge and Sn. This can be connected to qualitative failures of the electronic band structure~\cite{Hummer2009}. Namely, PBE falsely predicts the diamond phases of Ge and Sn to be metals, whereas in reality Ge is a semiconductor, and Sn is a semimetal. The false metallic ground state is accompanied by a spurious band inversion of the valence and conduction bands. Broadly speaking, electronic binding is weaker in metals than in semiconductors. Thus, the false metallic ground state leads to strongly overestimated lattice constants in PBE.
References~\cite{Zhang2025,Riemelmoser2025} have discussed that SCAN and r$^2$SCAN can open a band gap for challenging narrow-gap semiconductors such as Ge. In turn, the meta-GGA lattice constants of Ge and Sn are significantly more accurate. The same argument also applies to HSE.

\begin{table}[t]
\caption{\label{tab:lattice_constants}
Equilibrium lattice constants $a_{0,\alpha}$ (in angstrom) for group-IV elemental solids in the diamond phase. Calculated lattice constants are obtained by fitting the energy-volume curves to a third-order Birch-Murnaghan equation of state [see Eq.\@~\eqref{eq:BM EOS}].
Experimental references are taken from Ref.~\cite{madelung2004}, and corrected for zero-point expansion (ZPE; see Ref.~\cite{Schimka2011}). Uncorrected experimental lattice constants are given in the last column.}
\begin{ruledtabular}
\begin{tabular}{lcccccc}
\\[-2.5\medskipamount]
    & PBE & r$^2$SCAN & SCAN & HSE & Expt.-ZPE & Expt. \\
    \colrule
C   & 3.571  & 3.558  & 3.551  & 3.545  & 3.553  & 3.567  \\
Si  & 5.472  & 5.447  & 5.438  & 5.442  & 5.421  & 5.430  \\
Ge  & 5.765  & 5.689  & 5.678  & 5.686  & 5.644  & 5.652  \\
Sn  & 6.654  & 6.568  & 6.548  & 6.556  & 6.474  & 6.482  \\
\end{tabular}
\end{ruledtabular}
\end{table}

\subsection{Bulk moduli and elastic constants}

Table~\ref{tab:bulk moduli} lists the bulk moduli of the diamond phases of C, Si, Ge, and Sn. First, we discuss the estimated accuracy of our calculations before commenting on the performance of the various DFT functionals. On the one hand, we obtain bulk moduli as a fit parameter from the Birch-Murnaghan equation of state [$B_0^{\rm EOS}$; see Eq.\@~\eqref{eq:BM EOS}]. On the other hand, the bulk moduli can be calculated from the elastic constants [$B_0^{\rm elastic}$; see Eq.\@~\eqref{eq:bulk-with-C}]. 
Agreement between these two values is a powerful consistency check for our calculations~\cite{Shang2010}. That is, the difference $\Delta B_0$=$B_0^{\rm elastic}-B_0^{\rm EOS}$ should be exactly zero (in the limit of numerical convergence and assuming exact equation-of-state fits). It is satisfying that our calculations achieve close agreement ($\Delta B_0\approx 1$~GPa) for most functionals and materials. Residual differences are mostly due to basis set errors of the elastic constants, see the SM~\cite{SM}. We also find good agreement when comparing our PBE and HSE results for $B_0^{\rm EOS}$ with those of previous studies~\cite{Hummer2009,Schimka2011} (differences $\lesssim$~1~GPa). Further, the SCAN calculations of Ref.\@~\cite{Yao2017} yielded 99.3~GPa for Si, and 73.4~GPa for Ge, which differ by 2 to 3~GPa from our calculated $B_0^{\rm EOS}$. Their PBE bulk moduli are 88.0~GPa for Si and 58.7~GPa for Ge, in good agreement with our results (differences $\le$ 0.5~GPa). Finally, we note that zero-point corrections are significant only for C~\cite{Schimka2011}. \\
Turning now to the comparison of calculated bulk moduli and experimental data, we first note that the PBE bulk moduli are too small, especially for Ge and Sn. This reflects the underbinding problem of PBE, in line with the overestimated lattice constants. In agreement with previous studies, we find that larger bulk moduli are obtained by SCAN and HSE~\cite{Schimka2011,Raasander2015,Tran2016}. More generally, \myciteauthor{Tran2016} have established a broader inverse trend~\cite{Tran2016}. That is, DFT functionals that yield larger $a_0$ also yield smaller $B_0$, and vice versa.
Next, the bulk moduli of C, Si, and Ge obtained by r$^2$SCAN are slightly smaller than the corresponding SCAN bulk moduli, in line with the slightly larger lattice constants obtained by r$^2$SCAN. At variance with the general trend, the r$^2$SCAN bulk modulus of Sn is larger than the corresponding SCAN value, despite the fact that the lattice constant of r$^2$SCAN is larger for Sn.\\
Good agreement of calculated bulk moduli with experimental data is unambiguous for C, Si, and Ge. For Sn, however, the experimental data for $B_0^{\rm elastic}$ and $B_0^{\rm EOS}$ differ significantly. Table~\ref{tab:bulk moduli} shows that the relative discrepancy $\Delta B_0$/$B_0$ amounts to $\approx$ 25\%. The calculated bulk moduli are in better agreement with the experimental $B_0^{\rm elastic}$, while the experimental $B_0^{\rm EOS}$ is much larger. Such an observation was already made by \citet{Ihm1981}. At the time, the experimental $B_0^{\rm EOS}$ was even larger (111~GPa), while their LDA calculations predicted a bulk modulus of 45.6~GPa, closer to the experimental $B_0^{\rm elastic}$. Here, we reconduct their argument and recommend $B_0$ = 42.5~GPa as the experimental reference for the bulk modulus of $\alpha$-Sn~\cite{Ihm1981,Price1971}. Further measurements are required to completely resolve this discrepancy in the experimental data, but we find it noteworthy that the current value of $B_0^{\rm EOS}$ for $\alpha$-Sn is very close to that of $\beta$-Sn (55-58~GPa; see Ref.\@~\cite{Christensen1993}). Given the small energy difference between the $\alpha$ and $\beta$ phases (the experimental estimate is 15~meV~\cite{Ihm1981}), it seems plausible that the experimental measurements of $B_0^{\rm EOS}$ could be influenced by the coexistence of structural phases.\\
\begin{table}[tbh]
\caption{\label{tab:bulk moduli}
Bulk moduli $B_0$~(GPa) for group-IV elemental solids in the diamond phase. The first column reports bulk moduli obtained by equation-of-state fits [Eq.\@~\eqref{eq:BM EOS}].
The second column reports bulk moduli computed from the elastic constants [see Eq.\@~\eqref{eq:bulk-with-C}]. The third column reports the difference $\Delta B_0 = B_0^{\rm elastic}-B_0^{\rm EOS}$. Experimental references for $B_0^{\rm elastic}$ are taken from Refs.\@~\cite{McSkimin1972,McSkimin1964,McSkimin1963,Price1971}, while experimental data for $B_0^{\rm EOS}$ and zero-point expansion (ZPE) corrections are taken from Ref.~\cite{Schimka2011}.}
\begin{ruledtabular}
\begin{tabular}{lrrr}
\\[-2.5\medskipamount]
                & $B_0^{\rm EOS}$ & $B_0^{\rm elastic}$ & $\Delta B_0$ \\ 
\colrule
\multicolumn{4}{l}{C} \\
PBE             & 433.2 &  433.5 & +0.3 \\
r$^2$SCAN       &  451.2 &  450.8 & $-$0.4 \\
SCAN            &  459.7 & 459.0 & $-$0.7 \\
HSE             & 469.4      & 469.3 & -0.1 \\
Expt.-ZPE    &  454.7       & 455.3 & +0.6 \\
Expt.       &   443.0       & 443.6 & +0.6 \\[2pt]

\multicolumn{4}{l}{Si} \\
PBE             &  88.5 & 88.5 & $+0.0$\\
r$^2$SCAN        &  97.8 & 97.8 & $+$0.0 \\
SCAN            & 102.4 & 100.5 & $-$1.9 \\
HSE             &  97.6      & 100.1 & +2.5 \\
Expt.-ZPE~    &  100.8    & 100.8 & 0.0 \\
Expt.        & 99.2       & 99.2 & 0.0 \\[2pt]

\multicolumn{4}{l}{Ge} \\
PBE             & 58.5 &  59.0 & +0.5 \\
r$^2$SCAN       & 69.6 &  73.9 & +4.3 \\
SCAN            & 71.4 &  74.1 & +2.7 \\
HSE             &   71.2     & 73.3 & +2.1 \\
Expt.-ZPE      & 77.3   & 78.0 & +0.7 \\
Expt.         & 75.8   & 76.5 & +0.7 \\[2pt]

\multicolumn{4}{l}{Sn} \\
PBE             & 35.8  & 36.5 & +0.7 \\
r$^2$SCAN       & 42.9   & 44.3 & +1.4 \\
SCAN  &                               41.7  & 42.7 & $+1.0$ \\
HSE   & 43.6     & 44.8 & +1.2 \\
Expt.-ZPE    & 55.5     & 43.4 & $-$12.1 \\
Expt.       & 54.6     & 42.5 & $-$12.1 \\
\end{tabular}
\end{ruledtabular}
\end{table}
\begin{table}[tbh]
\caption{\label{tab:elastic_constants_EXPT}
Elastic constants $C_{11}$, $C_{12}$, and $C_{44}$ (in~GPa) of group-IV elemental solids in the diamond phase. Experimental elastic constants are taken from Refs.\@~\cite{McSkimin1972,McSkimin1964,McSkimin1963,Price1971}, and corrected for zero-point expansion [ZPE; see Eq.\@~\eqref{eq:ZPE elastic constants}]. See also Fig.\@~\ref{fig:combined_properties}(a) in the SM for data visualization~\cite{SM}.}
\begin{ruledtabular}
\begin{tabular}{lrrr}
\\[-2.5\medskipamount]
                & $C_{11}$ & $C_{12}$ & $C_{44}$ \\ 
                \colrule
\multicolumn{4}{l}{C} \\
PBE             & 1051   & 124   & 559   \\
r$^2$SCAN       & 1115   & 119   & 589   \\
SCAN            & 1130   & 124   & 598   \\
HSE             & 1137   & 135   & 609   \\
Expt.-ZPE     & 1104   & 135   & 595   \\
Expt.\tabcitesup{a}{McSkimin1972}       & 1081   & 125   & 579   \\[2pt]

\multicolumn{4}{l}{Si} \\
PBE             & 158    & 58    & 76    \\
r$^2$SCAN       & 170    & 62    & 82    \\
SCAN            & 174    & 64    & 85    \\
HSE             & 173    & 64    & 82    \\
Expt.-ZPE     & 169    & 67    & 81    \\
Expt.\tabcitesup{b}{McSkimin1964}       & 168    & 65    & 80    \\[2pt]

\multicolumn{4}{l}{Ge} \\
PBE             & 108    & 37    & 56    \\
r$^2$SCAN       & 132    & 45    & 69    \\
SCAN            & 130    & 46    & 67    \\
HSE             & 131    & 45    & 69    \\
Expt.-ZPE     & 133    & 50    & 69    \\
Expt.\tabcitesup{c}{McSkimin1963}       & 131    & 49    & 68    \\[2pt]

\multicolumn{4}{l}{Sn} \\
PBE             & 58     & 27    & 27    \\
r$^2$SCAN       & 70     & 32    & 33    \\
SCAN            & 66     & 31    & 29    \\
HSE             & 72     & 31    & 35    \\
Expt.-ZPE     & 70     & 30    & 36    \\
Expt.\tabcitesup{d}{Price1971}       & 69     & 29    & 36    \\
\end{tabular}
\end{ruledtabular} 
\noindent\begin{minipage}{\linewidth}
\raggedright
\vspace{0.05cm}
$^\text{a}$Reference\@~\cite{McSkimin1972}. \\
$^\text{b}$Reference\@~\cite{McSkimin1964}. \\
$^\text{c}$Reference\@~\cite{McSkimin1963}. \\
$^\text{d}$Reference\@~\cite{Price1971}.
\end{minipage}
\end{table}
We proceed by discussing the elastic constants in the diamond phase. Table~\ref{tab:elastic_constants_EXPT} lists the calculated elastic constants $C_{11}$, $C_{12}$, and $C_{44}$.
Given the relation between bulk moduli and elastic constants, it is not surprising that PBE underestimates the experimental elastic constants, especially for Ge and Sn, while SCAN and HSE yield larger elastic constants, in better agreement with experiment. Our PBE and HSE elastic constants are in good agreement  with the reference results from \citet{Raasander2015} (mean absolute deviation of 2~GPa). 
More recently, \myciteauthor{Xing2021} performed a high throughput study of elastic constants~\cite{Xing2021}. We find again good agreement for the PBE elastic constants  (mean absolute deviation of 2~GPa), but the agreement for the SCAN elastic constants is slightly worse (mean absolute deviation of 3~GPa). We note that both Refs.~\cite{Raasander2015} and~\cite{Xing2021} studied C, Si, and Ge, but not Sn.
Here, we also find that the r$^2$SCAN elastic constants are slightly smaller than the respective SCAN elastic constants for C and Si, very close for Ge, and slightly larger for Sn. This mirrors the results for the bulk moduli. 
In summary, the GGA functional PBE shows the largest deviations with respect to experimental data, with a mean absolute relative error (MARE) of 13\% and a root-mean-square relative error (RMSRE) of 15\%. The meta-GGA functionals SCAN  (MARE = 6\%, RMSRE = 7\%) and r$^2$SCAN (MARE = 4\%, RMSRE = 6\%) perform significantly better than PBE. HSE yields the lowest errors (MARE = 3\%, RMSRE = 4\%).\\
Finally, we note that the elastic constants can also be obtained from fitting the acoustic modes of the phonon dispersions [Eqs.\@~\eqref{eq:v_100} and \eqref{eq:v_110}]. In the SM, we discuss that very large supercells are needed to converge the elastic constants~\cite{SM}. The disadvantage due to the slow supercell convergence outweighs potential benefits in terms of basis set convergence. Therefore, the supercell approach is recommended only as an additional consistency check.

\section{\label{sec:Results: phonons}Lattice dynamics in the diamond phase}

\begin{figure*}[tbh]
\includegraphics[width=0.97\textwidth]{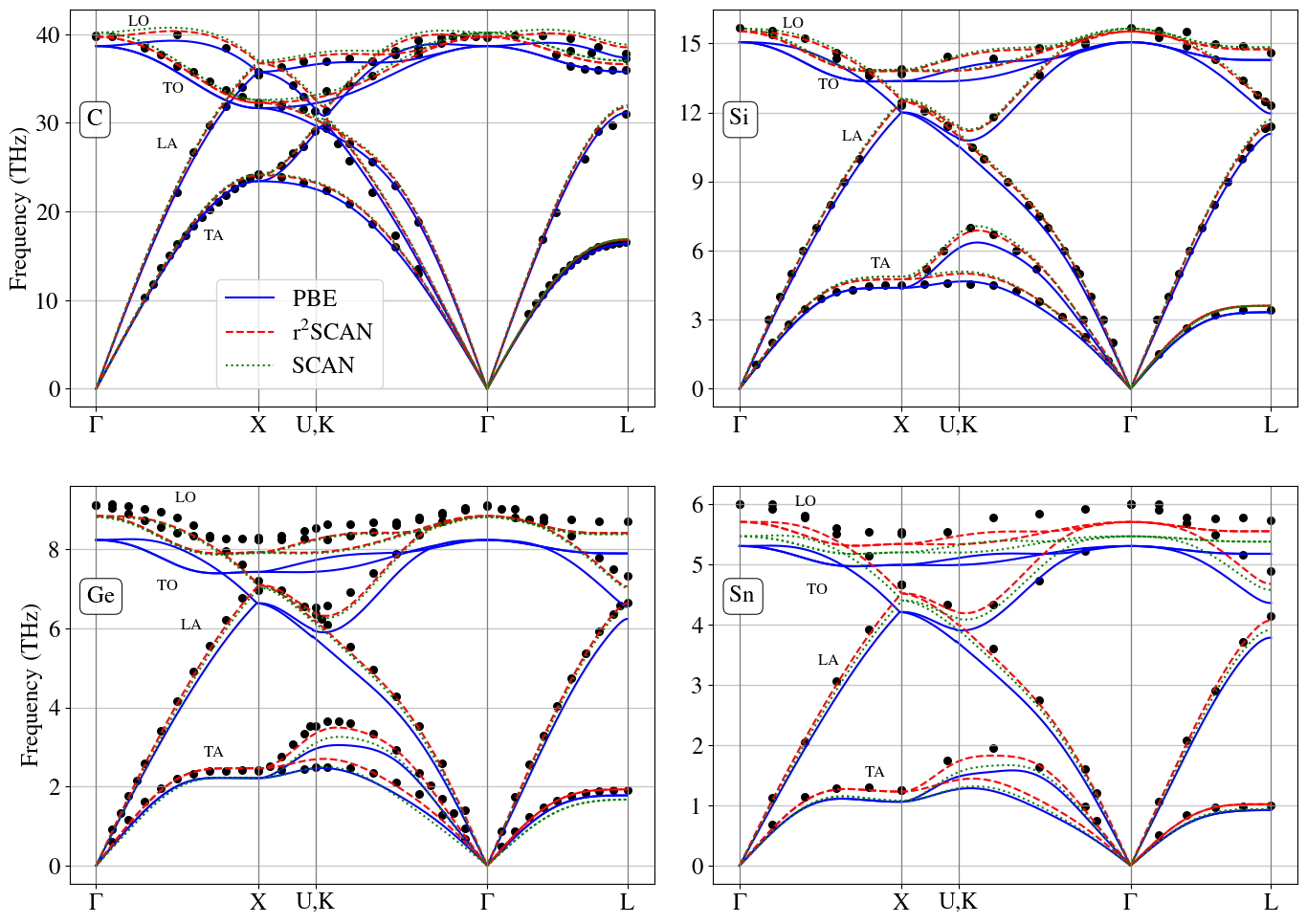}
        \begin{tikzpicture}[remember picture, overlay]
          \node[anchor=north west, inner sep=0pt] at (-0.926 \textwidth, 0.7\textwidth) {(a)};
        \end{tikzpicture}
        \begin{tikzpicture}[remember picture, overlay]
          \node[anchor=north west, inner sep=0pt] at (-0.931 \textwidth, 0.347\textwidth) {(c)};
        \end{tikzpicture}
        \begin{tikzpicture}[remember picture, overlay]
          \node[anchor=north west, inner sep=0pt] at (-0.461 \textwidth, 0.7\textwidth) {(b)};
        \end{tikzpicture}
        \begin{tikzpicture}[remember picture, overlay]
          \node[anchor=north west, inner sep=0pt] at (-0.467 \textwidth, 0.347\textwidth) {(d)};
        \end{tikzpicture}
\caption{\label{fig:phonons_vs_exp}Phonon dispersion relations for 128-atom supercells of group-IV elemental solids in the diamond phase: (a) C, (b) Si, (c) Ge, and (d) Sn. Phonon dispersions calculated with the PBE, r$^2$SCAN, and SCAN functionals are given as lines, whereas symbols indicate experimental data~\cite{Warren1967,Kulda2002,Kulda1994,Nilsson1971,Nilsson1972,Dolling1963,Price1971}. For Sn, experimental phonon data are not available for all branches along the $\Sigma$ direction (X-U,K-$\Gamma$)~\cite{Price1971}. }
\end{figure*}
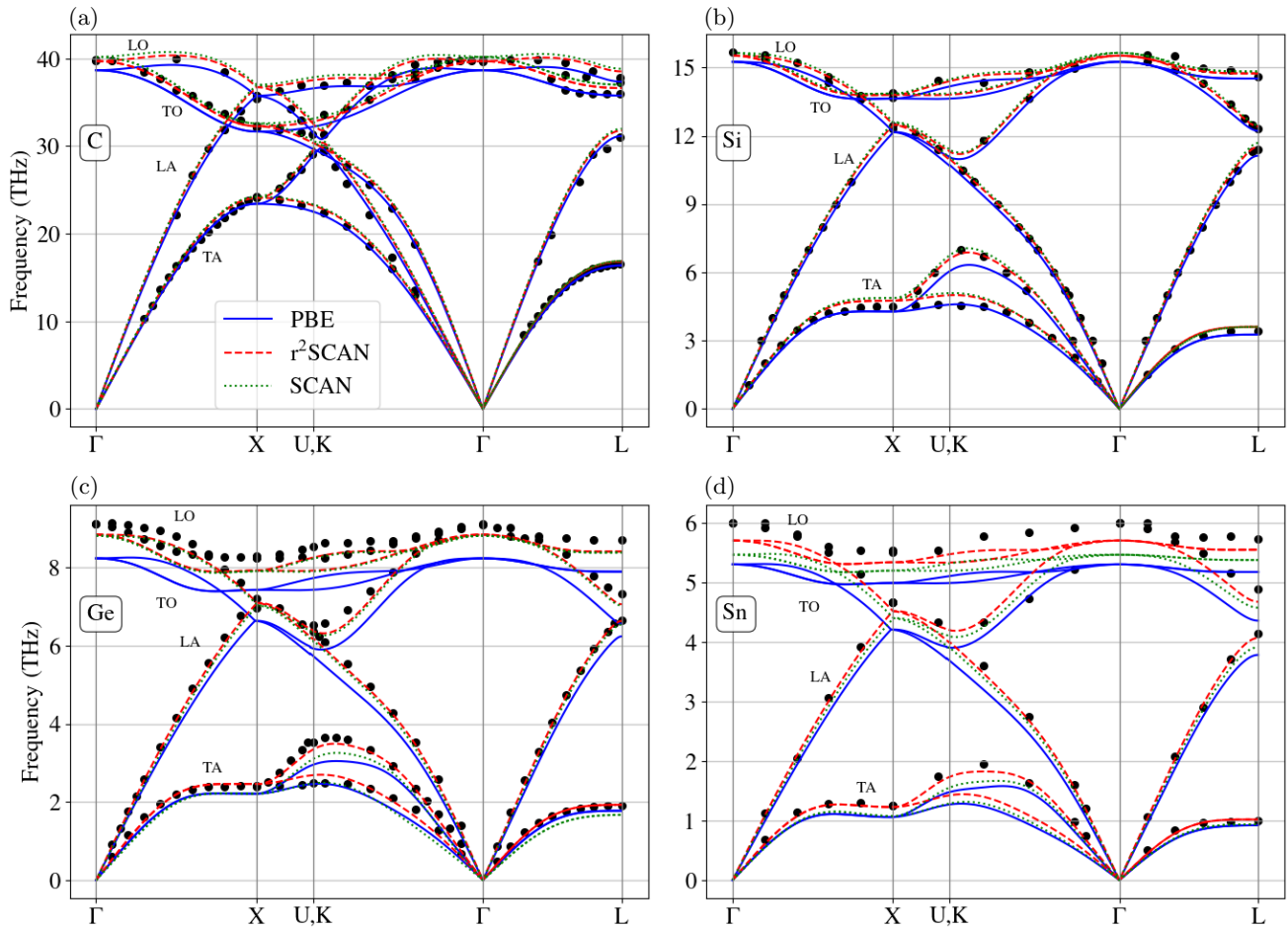

\begin{table*}[tbh]
\caption{\label{tab:vibrational_properties_combined}Phonon frequencies~(THz) at high-symmetry points calculated using PBE, r\(^2\)SCAN, SCAN, and HSE functionals. Experimental values are taken from Ref.\@~\cite{Hummer2009}, and corrected for zero-point expansion [ZPE; see Eq.\@~\eqref{eq:phonon-zpe}]. See also Fig.\@~\ref{fig:combined_properties}(b) in the SM for data visualization~\cite{SM}.}
\begin{ruledtabular}
\begin{tabular}{lcccccccc}
\\[-2.5\medskipamount]
 & \(\Gamma_{\text{LO/TO}}\) & \(X_{\text{TA}}\) & \(X_{\text{LA/LO}}\) & \(X_{\text{TO}}\) & \(L_{\text{TA}}\) & \(L_{\text{LA}}\) & \(L_{\text{LO}}\) & \(L_{\text{TO}}\) \\ 
\colrule
\multicolumn{9}{l}{C} \\
PBE          & 38.7 & 23.4 & 35.7 & 31.7 & 16.4 & 31.2 & 37.4 & 35.8 \\
r$^2$SCAN    & 39.7 & 24.1 & 36.7 & 32.3 & 16.8 & 31.8 & 38.5 & 36.6 \\
SCAN         & 40.2 & 24.2 & 37.0 & 32.6 & 16.9 & 32.0 & 38.8 & 37.0 \\
HSE          & 41.1 & 24.2 & 37.3 & 33.6 & 16.9 & 32.7 & 38.9 & 38.0 \\
Expt.-ZPE & 40.8 & 24.3 & 36.5 & 33.2 & 16.5 & 31.6 & 37.5 & 36.8 \\
Expt.        & 40.3 & 24.2 & 36.1 & 32.6 & 16.4 & 31.0 & 37.2 & 36.3 \\[2pt]
\multicolumn{9}{l}{Si} \\
PBE          & 15.1 & 4.4  & 12.0 & 13.4 & 3.3  & 11.1 & 12.0 & 14.3 \\
r$^2$SCAN    & 15.5 & 4.8  & 12.5 & 13.8 & 3.6  & 11.5 & 12.4 & 14.7 \\
SCAN         & 15.6 & 4.9  & 12.6 & 13.9 & 3.6  & 11.7 & 12.5 & 14.8 \\
HSE          & 15.8 & 4.7  & 12.6 & 14.1 & 3.5  & 11.6 & 12.6 & 15.0 \\
Expt.-ZPE & 15.6 & 4.5 & 12.4 & 14.0 & 3.4 & 11.4 & 12.7 & 14.8 \\
Expt.        & 15.5 & 4.5  & 12.3 & 13.9 & 3.4  & 11.4 & 12.6 & 14.7 \\[2pt]
\multicolumn{9}{l}{Ge} \\
PBE          & 8.2 & 2.2 & 6.6 & 7.4 & 1.8 & 6.2 & 6.6 & 7.9 \\
r$^2$SCAN    & 8.8 & 2.5 & 7.1 & 7.9 & 1.9 & 6.6 & 7.1 & 8.4 \\
SCAN         & 8.8 & 2.2 & 7.0 & 7.9 & 1.7 & 6.5 & 7.0 & 8.4 \\
HSE          & 9.0 & 2.5 & 7.2 & 8.1 & 1.9 & 6.7 & 7.2 & 8.6 \\
Expt.-ZPE & 9.1 & 2.4 & 7.2 & 8.4 & 1.9 & 6.7 & 7.4 & 8.7 \\
Expt.        & 9.1 & 2.4 & 7.2 & 8.3 & 1.9 & 6.7 & 7.3 & 8.7 \\[2pt]

\multicolumn{9}{l}{Sn} \\
PBE          & 5.3 & 1.1 & 4.2 & 5.0 & 0.9 & 3.8 & 4.4 & 5.2 \\
r$^2$SCAN    & 5.7 & 1.2 & 4.5 & 5.3 & 1.0 & 4.1 & 4.7 & 5.6 \\
SCAN         & 5.5 & 1.1 & 4.4 & 5.2 & 0.9 & 3.9 & 4.6 & 5.4 \\
HSE          & 5.8 & 1.3 & 4.6 & 5.4 & 1.1 & 4.1 & 4.8 & 5.7 \\
Expt.-ZPE  & 6.0 & 1.3 & 4.7 & 5.5 & 1.0 & 4.2 & 4.9 & 5.7 \\
Expt.        & 6.0  & 1.3  & 4.7  & 5.5  & 1.0  & 4.2  & 4.9  & 5.7 \\
\end{tabular}
\end{ruledtabular}
\end{table*}

To illustrate the phonon dispersion relations obtained from our calculations, we focus on the 128-atom supercell, with supercell convergence being discussed in the SM~\cite{SM}. 
Figure~\ref{fig:phonons_vs_exp} shows the calculated phonon dispersion curves compared to experimental data. 
Before commenting on particular results, we first discuss general trends. Overall, the phonon dispersions of the group-IV elemental solids are very similar and exhibit the same qualitative shapes. The frequencies show an inverse trend with atomic number [see Eq.\@~\eqref{eq:dynamical matrix}]. Specifically, the LO/TO frequency  for C is approximately 40~THz at the $\Gamma$ point, while for Si, Ge, and Sn, the respective frequencies are approximately 15, 9, and 6~THz. It is noteworthy that the phonon dispersion for C exhibits distinct maxima along \(\Gamma\)-X. Such maxima also appear along $\Gamma$-L and along $\Gamma$-K but in a less distinct fashion. Furthermore, the order of the optical branches at the X and L points is reversed, and there is a near degeneracy at the K point.
These anomalies for C have been discussed in detail in Refs.\@~\cite{Aouissi2006} and~\cite{Hummer2009}. \\
Now, we proceed with the comparison of different density functionals. PBE generally underestimates the experimental frequencies, a manifestation of the underbinding error. While PBE phonons are still fairly close to experiment for C and Si, the underestimation for Ge and Sn is more severe. \myciteauthor{Hummer2009} have explicitly connected this to the qualitative failures of the PBE electronic band structure~\cite{Hummer2009}. The SCAN and r$^2$SCAN functionals reduce the underbinding error, and yield systematically larger phonon frequencies. This generally improves the agreement with the experimental data, especially for Ge and Sn. Concerning the differences between the meta-GGA functionals, we note that SCAN yields slightly larger phonon frequencies for C, and somewhat smaller phonon frequencies for Sn. However, we note that the differences for Sn are only slightly larger than our numerical accuracy of 0.1~THz. For Si and Ge, SCAN and r$^2$SCAN yield essentially identical results (Ref.\@~\cite{Ning2022} reported this previously for Si).\\
For a more quantitative comparison, Table~\ref{tab:vibrational_properties_combined} provides frequencies at the high-symmetry points $\Gamma$, X, and L. These results can be compared with HSE results and experimental data. HSE phonon frequencies are calculated in a 16-atom supercell, which is commensurate with these high-symmetry points. We note that our HSE phonon frequencies are in excellent agreement with those reported by \citet{Hummer2009} (mean absolute deviation 0.1~THz). ZPE corrections increase the optical phonon frequencies, in particular for C. The ZPE corrections have the opposite sign for some of the acoustic modes of Si and Ge, corresponding to negative Grüneisen parameters for these modes~\cite{Favot1999}.
The trends of the relative errors are the same as those for the elastic constants. PBE exhibits the largest error, with a MARE of 7\% and a RMSRE of 8\%.
The meta-GGA functionals reduce these deviations. r\(^2\)SCAN yields a MARE of ~3\% and a RMSRE of ~4\%, very similar to the performance of  
SCAN (MARE~=~5\% and RMSRE~=~6\%).
HSE delivers the best overall agreement, with a MARE of only 2\%  and a RMSRE of 3\%.
For a meaningful comparison with experimental frequencies, it is important to note that the experimental uncertainties reported for Si, Ge, and Sn are $\lesssim$ 0.1~THz~\cite{Nilsson1971,Nilsson1972,Kulda1994,Dolling1963,Price1971}, and the LO frequencies for C are also very accurate ($< 0.1$~THz)~\cite{Warren1967,Kulda2002}. However, the rest of the C data has larger experimental uncertainties ($\sim2$\%-3\%)~\cite{Warren1967}. 

\section{\label{sec:Results: Transition-pressures} Structural phase transitions}

\begin{table}[tbh]
\caption{\label{tab:transition_pressures}
Structural $\alpha\leftrightarrow\beta$ phase transition for Si, Ge, and Sn. The first and the second columns show the equilibrium lattice parameters $a_{0,\beta}$ (in~\AA) and the tetragonal ratio $(c/a)_\beta$ of the $\beta$ phases (see Table~\ref{tab:lattice_constants} for the $\alpha$ phases), respectively. The third and the fourth columns show the volume differences $\Delta V_0 = V_{0,\beta}- V_{0,\alpha}$ (in~\AA$^3$), and the energy differences $\Delta E_0 = E_{0,\beta}-E_{0,\alpha}$ (in~eV/atom), respectively. The final column shows the transition pressure $P_{\rm t}$ (in~GPa), see Eq.\@~\eqref{eq:transition pressure}. Reference transition pressures for Si and Ge obtained using the random-phase approximation (RPA) are taken from Ref.\@~\cite{Sengupta2018}. The zero-point expansion (ZPE) corrections to the experimental transition pressures for Si and Ge also account for temperature effects at $T=$ 300~K~\cite{Sengupta2018}.}
\begin{ruledtabular}
\begin{tabular}{llllll}
\\[-2.5\medskipamount]
                            & $a_{0,\beta}$        & $(c/a)_\beta$       & $\Delta V_0$     & $\Delta E_0$      & $P_\text{t}$ \\
\colrule
\multicolumn{5}{l}{Si} \\
PBE                         & 4.839                & 0.551               & $-5.10$          & 0.288             & \hspace{4pt}9.7          \\
r$^2$SCAN                   & 4.802                & 0.549               & $-4.97$          & 0.423             & 14.9                     \\
SCAN                        & 4.776                & 0.554               & $-5.11$          & 0.402             & 13.7                     \\
HSE                         & 4.794                & 0.549               & $-$5.03          &  0.380            &   13.2                   \\
RPA                         &        &       & $-4.86\tabcitesup{a}{Sengupta2018}$        & $0.384\tabcitesup{a}{Sengupta2018}$         & $13.8\tabcitesup{a}{Sengupta2018}$                 \\
Expt.-ZPE                 &        &       &   &     & 12.7$\textsuperscript{\hyperlink{cite.Sengupta2018}{a},\hyperlink{cite.Mujica2003}{b}}$                 \\
Expt.                       &       & 0.552$\tabcitesup{c}{Moll1995}$           &   &     & 11.7$\tabcitesup{b}{Mujica2003}$             \\[2pt]
\multicolumn{5}{l}{Ge} \\
PBE                         & 5.192                & 0.552              & $-4.62$           & 0.228             & \hspace{4pt}8.7          \\
r$^2$SCAN                   & 5.133                & 0.553              &  $-4.26$          & 0.357             & 14.9         \\
SCAN                        & 5.149                & 0.545              & $-4.27$           & 0.266             & 10.8        \\
HSE                         &  5.132               & 0.550              & $-$4.39           & 0.329             &  13.2     \\
RPA                         &        &      & $-4.28\tabcitesup{a}{Sengupta2018}$         & $0.276\tabcitesup{a}{Sengupta2018}$         & $11.2\tabcitesup{a}{Sengupta2018}$ \\
Expt.-ZPE                 &        &      &    &     & 11.6\textsuperscript{\hyperlink{cite.Sengupta2018}{a},\hyperlink{cite.Mujica2003}{b}}            \\
Expt.                       &        & 0.551\tabcitesup{c}{Moll1995}          &    &     & 10.6\tabcitesup{b}{Mujica2003}           \\[2pt]
\multicolumn{5}{l}{Sn} \\
PBE                         & 5.939                & 0.542             & $-8.41$            & 0.042             & \hspace{4pt}0.8         \\
r$^2$SCAN                   & 5.879                & 0.543             & $-7.81$            & 0.125             & \hspace{4pt}2.7          \\
SCAN                        & 5.879                & 0.539             & $-7.69$            & 0.080             & \hspace{4pt}1.7          \\
HSE                         & 5.863                & 0.542             & $-$7.91            &  0.110            &  \hspace{4pt}2.3       \\
Expt.-ZPE                 &        &     &     & 0.022\tabcitesup{d}{Mehl2021}         &       \\
Expt.                       & 5.812\tabcitesup{e}{Rayne1960}            & 0.543\tabcitesup{e}{Rayne1960}        & $-$7.39\textsuperscript{\hyperlink{cite.Rayne1960}{e},\hyperlink{cite.madelung2004}{f}}            & 0.015\tabcitesup{g}{Ihm1981}         &      \\
\end{tabular}
\end{ruledtabular}
\noindent\begin{minipage}{\linewidth}
\raggedright
\vspace{0.05cm}
$^\text{a}$Reference\@~\cite{Sengupta2018}. \\
$^\text{b}$Reference\@~\cite{Mujica2003}. \\
$^\text{c}$Reference\@~\cite{Moll1995}. \\
$^\text{d}$Reference\@~\cite{Mehl2021}. \\
$^\text{e}$Reference\@~\cite{Rayne1960}. \\
$^\text{f}$Reference\@~\cite{madelung2004}. \\
$^\text{g}$Reference\@~\cite{Ihm1981}. 
\end{minipage}
\end{table}

\begin{figure*}[tbh]
  \centering
  \begin{tikzpicture}
    \node[anchor=south west, inner sep=0] (img) at (0,0)
      {\includegraphics[width=\textwidth]{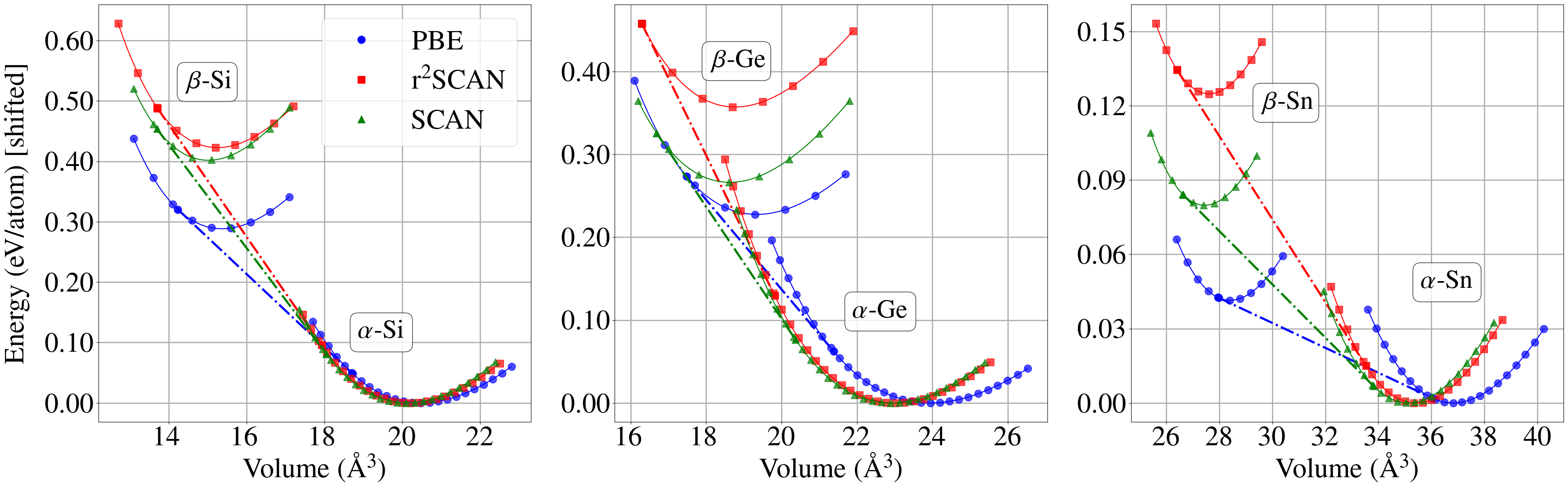}};
    \begin{scope}[x={(img.south east)}, y={(img.north west)}]
      \node[anchor=north west, inner sep=2pt]
        at (0.058, 1.068) {(a)}; 
    \end{scope}
    \begin{scope}[x={(img.south east)}, y={(img.north west)}]
      \node[anchor=north west, inner sep=2pt]
        at (0.386, 1.068) {(b)};  
    \end{scope}
    \begin{scope}[x={(img.south east)}, y={(img.north west)}]
      \node[anchor=north west, inner sep=2pt]
        at (0.717, 1.068) {(c)};  
    \end{scope}
  \end{tikzpicture}

  \caption{\label{fig:pressure}
    Energy-volume curves for the $\alpha$ and $\beta$ phases of (a) Si, (b) Ge, and (c) Sn, calculated with PBE, r$^2$SCAN, and SCAN. Energies are referenced to the minimum of the $\alpha$-phase curve for each element and functional. The solid lines are Birch-Murnaghan fits to the data. The dash-dotted lines show the common tangent to each pair of $E(V)$ curves; its negative slope equals the transition pressure $P_{\rm t}$ reported in Table~\ref{tab:transition_pressures}.
}
\end{figure*}
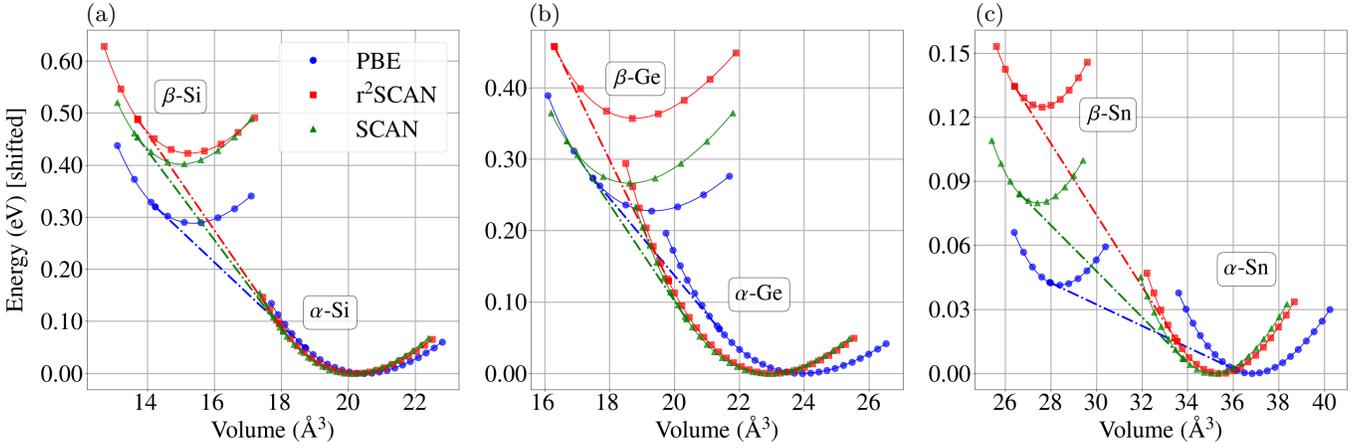

\begin{figure*}[tbh]
  \centering
  \begin{tikzpicture}
    \node[anchor=south west, inner sep=0] (img) at (0,0)
      {\includegraphics[width=\textwidth]{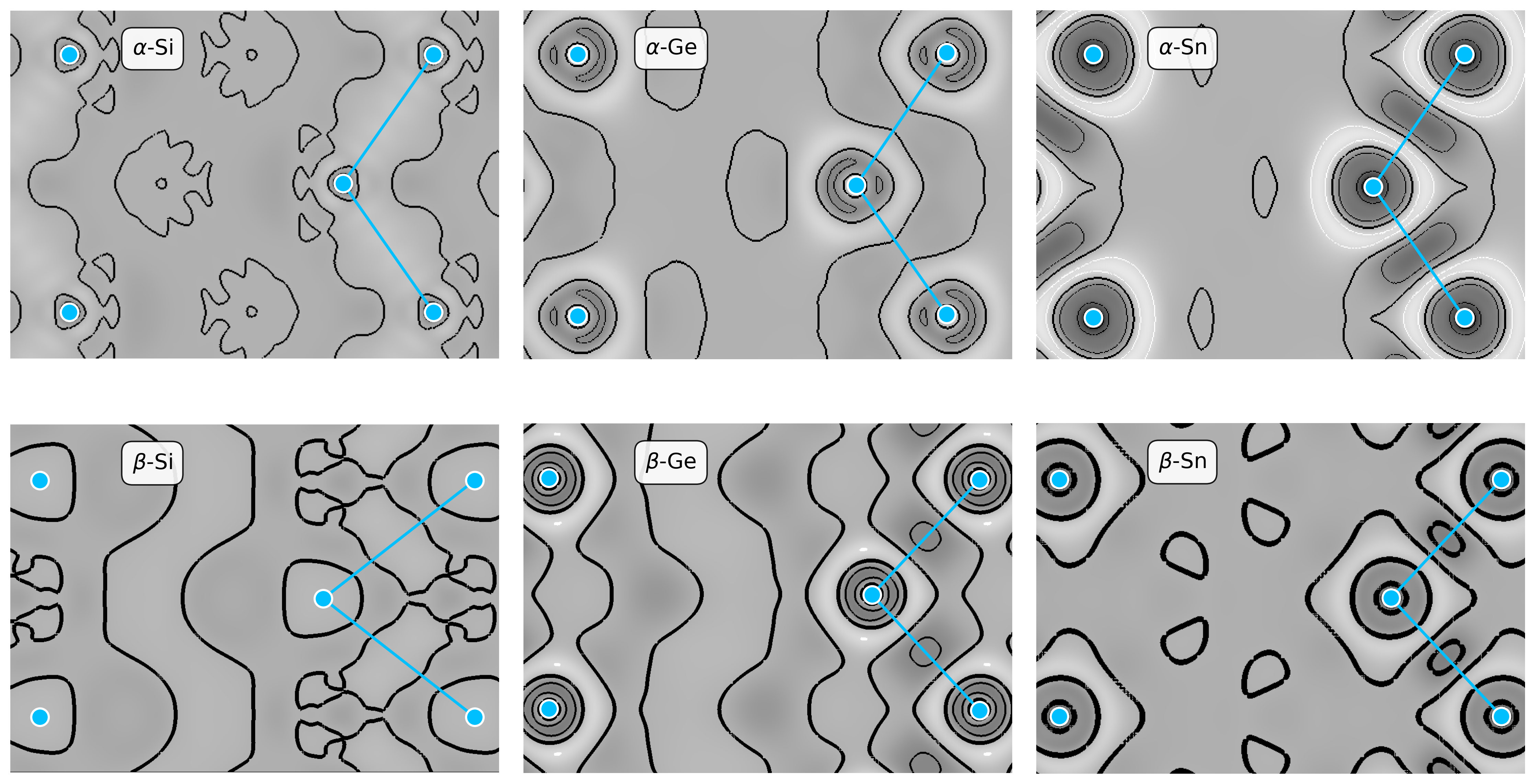}};
    \begin{scope}[x={(img.south east)}, y={(img.north west)}]
      \node[anchor=north west, inner sep=2pt]
        at (0.002, 1.035) {(a)};
    \end{scope}
    \begin{scope}[x={(img.south east)}, y={(img.north west)}]
      \node[anchor=north west, inner sep=2pt]
        at (0.335, 1.035) {(b)};
    \end{scope}
    \begin{scope}[x={(img.south east)}, y={(img.north west)}]
      \node[anchor=north west, inner sep=2pt]
        at (0.670, 1.035) {(c)};
    \end{scope}
    \begin{scope}[x={(img.south east)}, y={(img.north west)}]
      \node[anchor=north west, inner sep=2pt]
        at (0.002, 0.508) {(d)}; 
    \end{scope}
    \begin{scope}[x={(img.south east)}, y={(img.north west)}]
      \node[anchor=north west, inner sep=2pt]
        at (0.335, 0.508) {(e)}; 
    \end{scope}
    \begin{scope}[x={(img.south east)}, y={(img.north west)}]
      \node[anchor=north west, inner sep=2pt]
        at (0.670, 0.508) {(f)}; 
    \end{scope}
  \end{tikzpicture}
  \caption{\label{fig:charges}
    Charge density differences $\Delta n(\mathbf{r}) = n_{\mathrm{r^2SCAN}}(\mathbf{r}) - n_{\mathrm{SCAN}}(\mathbf{r})$
    for the $\alpha$ phase of (a) Si, (b) Ge, and (c) Sn; and for the $\beta$ phase of (d) Si, (e) Ge, and (f) Sn. Top panels show the $\alpha$ phases in the $(0\bar{1}1)$ plane, and bottom panels show the $\beta$ phases in the (110) plane. Experimental geometries are used where available (see Tables~\ref{tab:lattice_constants} and~\ref{tab:transition_pressures}), while SCAN geometries are used for $\beta$-Si and $\beta$-Ge. Blue circles indicate the positions of the atomic nuclei, and the nearest-neighbor bonds are indicated by blue lines. Dark areas indicate positive charge density differences ($n_{\rm r^2SCAN}>n_{\rm SCAN} $), and light areas indicate negative charge density differences ($n_{\rm r^2SCAN}<n_{\rm SCAN} $). Contour lines are drawn in steps of 0.5 electrons per cell volume~\cite{Momma2011}. Negative contours lines are indicated in white.}
\end{figure*}
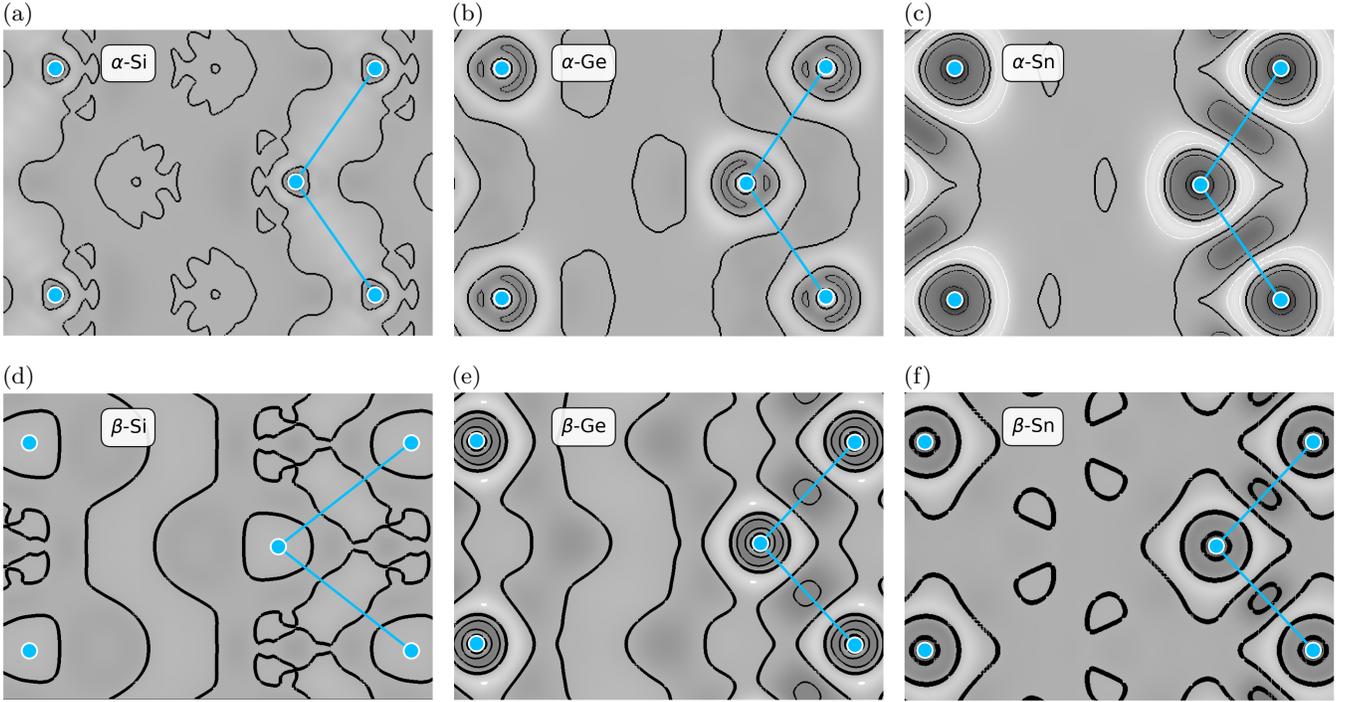

Figure~\ref{fig:pressure} presents the  $\alpha \leftrightarrow \beta $ phase transitions for Si, Ge, and Sn. Curves on the left represent the high-pressure $\beta$ phase, curves on the right the low-pressure $\alpha$ phase, and the negative transition pressure is indicated by the common tangent lines [see Eq.~\ref{eq:pressure}].  All energies have been offset so that the minimum of the $\alpha$-phase curve lies at zero, allowing the energy difference $\Delta E_0 = E_{0,\beta}-E_{0,\alpha}$ to be read directly; numerical values are presented in Table~\ref{tab:transition_pressures}. All functionals tested here correctly predict positive values for $\Delta E_0$, but we note that other functionals such as LDA give the reverse order for Sn~\cite{Mehl2021}.  Table~\ref{tab:transition_pressures} shows that PBE yields the smallest transition pressures, underestimating the experimental values for Si and Ge by $\approx$ 25\%. In agreement with previous studies~\cite{Shahi2018,Sengupta2018}, we find that SCAN yields larger transition pressures, improving the agreement with the experimental values. The relative errors for SCAN are +8\% for Si and $-$7\% for Ge.
Furthermore, SCAN transition pressures for Si and Ge are in close agreement with the RPA results from \citet{Sengupta2018} (RPA, a functional from the fifth and highest rung of Jacob's Ladder). No experimental transition pressure has been reported so far for Sn. However, SCAN strongly overestimates the experimental estimate for $\Delta E_0$ (relative error of 260\%; PBE yields 90\%), confirming a previous result from \citet{Mehl2021}. Their study also showed that this results in SCAN overestimating the $\alpha \leftrightarrow \beta$ transition temperature within the quasiharmonic approximation. Next, we find that r$^2$SCAN yields consistently larger transition pressures than SCAN. HSE also yields larger transition pressures than SCAN for Ge and Sn, but a slightly smaller value for Si. While differences are minor for Si, both r$^2$SCAN and HSE yield considerably worse transition pressures for Ge, and phase energy differences for Sn. \\
Concerning the accuracy of our calculations, we note that some of our PBE and SCAN transition pressures differ up to 1~GPa from previous calculations~\cite{Shahi2018,Sengupta2018}. In particular, Ref.\@ \onlinecite{Shahi2018} reported 7.9~GPa for PBE transition pressure for Ge, compared to 8.7~GPa in the present work. However, this seems to be a typographical error, as the equation-of-state parameters given in their SM are consistent with a transition pressure of 8.9~GPa, in good agreement with our result (8.7~GPa). Furthermore, Refs.\@~\cite{Shahi2018} and~\cite{Sengupta2018} reported 14.5~GPa for the SCAN transition pressure of Si, compared to 13.7~GPa in the present work. We have traced this difference back to our use of a more accurate Si pseudopotential. That is, by freezing the 2$s$ and 2$p$ semi-core electrons in the Si pseudopotential, we obtain a larger transition of 14.3~GPa, closer to the value of 14.5~GPa reported in Refs.\@~\cite{Shahi2018} and~\cite{Sengupta2018}. It is noteworthy that our present SCAN result (13.7~GPa) is in better agreement with the experimental transition pressure (12.7~GPa including ZPE corrections). We also note that \myciteauthor{Yao2017} commented on the sensitivity of $\Delta E_0$ to the Si pseudopotential~\cite{Yao2017}. However, their $\Delta E_0$, calculated with a SCAN pseudopotential, is 0.480~eV/atom, noticeably larger than our result (0.402~eV/atom) and also larger than the phase energy difference in Refs.\@~\cite{Shahi2018} and \cite{Sengupta2018} (both 0.417~eV/atom). This might relate to difficulties with the generation of SCAN pseudopotential, as discussed by \citet{Bartok2019}. Finally, our HSE transition pressure for Si (13.2~GPa) is in excellent agreement with the corresponding reference from \myciteauthor{Xiao2013} (13.3~GPa)~\cite{Xiao2013}.\\ 
Next, we return to discuss the poor performance of the r$^2$SCAN and HSE functionals for the transition pressures of Ge and Sn. 
We note that the reliability of HSE for transition pressures has not been established in a systematic fashion. Rather, \myciteauthor{Xiao2013} have shown that HSE also seriously overestimates the transition pressure for a metal-metal phase transition in zirconium~\cite{Xiao2013}. Furthermore, HSE includes a semiempirical parameter ($\mu = 0.208 {\; \text \AA}^{-1}$) that was adjusted to optimize performance for standard semiconductors~\cite{Heyd2003,*Heyd2006,Krukau2006}. It is perhaps not surprising that the accuracy of HSE declines when it is applied to materials that do not resemble those standard semiconductors (here the metallic $\beta$ phases of Ge and Sn).
The HSE transition pressure for Ge is clearly much larger than references from the RPA and experiment, see Table~\ref{tab:transition_pressures}. An experimental transition pressure for Sn is presently not available, but the overly large phase energy difference indicates that HSE performs poorly also for Sn. To the best of our knowledge, no precise theoretical estimate is yet available. High-level calculations using the RPA and/or quantum Monte Carlo are desirable to resolve this issue completely.\\
The fact that the r$^2$SCAN functional also yields consistently larger transition pressures than SCAN is rather unexpected. We remind the reader that r$^2$SCAN was proposed as a reliable estimate of SCAN~\cite{Furness2020}, and we have explicitly demonstrated that r$^2$SCAN closely approximates SCAN for the elastic properties of these materials. r$^2$SCAN yields slightly larger equilibrium volumes for both $\alpha$ and $\beta$-phases, and Table~\ref{tab:transition_pressures} shows that the differences $\Delta V_0=V_{0,\beta}-V_{0,\alpha}$ are very close for SCAN and r$^2$SCAN. Given that the transition pressure can be approximated as $P_{\rm t} \approx -\Delta E_0/\Delta V_0$ [see Eq.\@~\eqref{eq:pressure}], we conclude that r$^2$SCAN deteriorates $P_{\rm t}$ by overestimating $\Delta E_0$. This is also visually apparent in Fig.\@~\ref{fig:pressure}.
We further investigate the disagreement between these meta-GGA functionals by analyzing the charge density differences $\Delta n(\textbf{r}) = n_{\rm r^2SCAN}(\textbf{r})-n_{\rm SCAN}(\textbf{r})$. Figure~\ref{fig:charges} shows that the r$^2$SCAN density is generally enhanced along the atomic bonds (darker regions), generally indicating higher stability. On the contrary, the r$^2$SCAN density diminishes around the atomic nuclei (lighter regions), which generally indicates lower stability. By comparing the relative strengths of these opposing effects for the $\alpha$ and $\beta$ phases, we can get a rough, qualitative understanding of the phase energy differences $\Delta E_0$. Panels (c) and (f) show that the tighter electronic bond in $\alpha$-Sn seems to be the dominant effect for tin. Thus, r$^2$SCAN overestimates $\Delta E_0$ by stabilizing the $\alpha$ phase with respect to SCAN. Further support for the overstabilization of the $\alpha$-Sn phase by r$^2$SCAN is provided by the fact that r$^2$SCAN atypically yields larger phonon frequencies and a larger bulk modulus for this phase (see Fig.\@~\ref{fig:phonons_vs_exp} and Table~\ref{tab:bulk moduli}).  Next, panels (b) and (e) in Fig.\@~\ref{fig:charges} indicate that the dominant effect for germanium seems to be the contraction of electronic density around the nuclei in $\beta$-Ge. Therefore, r$^2$SCAN overestimates $\Delta E_0$ by destabilizing the $\beta$ phase with respect to SCAN. Finally, panels (a) and (d) demonstrate that the charge density differences are marginal for silicon. Thus, the small phase energy difference is due to r$^2$SCAN closely approximating SCAN for both the $\alpha$-Si and $\beta$-Si phases. 

\section{\label{sec:Discussion}Discussion and outlook}

We have benchmarked the lattice dynamics of the group-IV elemental solids C, Si, Ge, and Sn. To start, we have confirmed the underbinding error of PBE. Namely, PBE overestimates lattice constants, while it underestimates bulk moduli, elastic constants, and phonon frequencies.  This underbinding can be understood as a consequence of the self-interaction error. That is, semilocal exchange insufficiently counteracts the Hartree electronic repulsion~\cite{Zhang2017,Zhang2025}.
The underbinding error is particularly severe for Ge and Sn, as PBE falsely predicts Ge and Sn to be metals in the diamond phase~\cite{Hummer2009,Riemelmoser2025,Zhang2025}.
The electronic structure improvements lead to improved lattice dynamics for meta-GGA functionals. Improved band structures are also provided by hybrid functionals such as HSE, but at a much higher computational cost. Overall, our results demonstrate that r$^2$SCAN and SCAN yield very similar lattice dynamics for group-IV elemental solids, but r$^2$SCAN does not exhibit the severe numerical problems of SCAN (see the SM~\cite{SM}). We conclude that the lattice dynamics obtained by r$^2$SCAN offer a compelling level of accuracy ---nearly matching HSE--- while retaining the computational efficiency of a semilocal DFT functional. This conclusion is in line with recent findings in the literature~\cite{Ning2022,Wang2024,Zhang2025}. Putting these results in a broader context, we find a wider, emerging picture of r$^2$SCAN improving upon PBE in terms of linear response properties, which has been attributed to the reduction of self-interaction error~\cite{Ning2022,Riemelmoser2025,Zhang2025,Wang2024,Yates2025}. 

The fact that the PBE transition pressures are underestimated can also be related to underbinding and self-interaction errors. Namely, PBE underbinds the fourfold coordinated $\alpha$ phases more than the sixfold coordinated $\beta$ phases, and consequently, PBE underestimates $\Delta E_0$ and $P_{\rm t}$~\cite{Hennig2010,Xiao2013,Shahi2018,Sengupta2018}. Larger transition pressures are found with the meta-GGA functionals and HSE, due to a reduction of the self-interaction error. Surprisingly, we have found that r$^2$SCAN predicts overly large phase energy differences $\Delta E_0$, and drastically overestimates the transition pressures of Ge and Sn. As we have carefully converged our calculations, this difference between SCAN and r$^2$SCAN should be interpreted as a shortcoming of the r$^2$SCAN meta-GGA functional itself. While the functional forms of SCAN and r$^2$SCAN are very similar, they are obviously not identical. Importantly, SCAN fulfills the exact constraint for the exchange gradient expansion up to fourth order (GE4X), while r$^2$SCAN fulfills this constraint only up to second order (GE2X)~\cite{Furness2020,Furness2022}. PBE recovers the exchange gradient expansion only in leading order (GE0X). The relevance of GE4X has not yet been established for real systems~\cite{Furness2020,Furness2022}, but the  transition pressures of Ge and Sn seemingly present a rare situation where GE4X actually matters. At this point, it is not clear whether the overestimated transition pressures represent a general weakness of r$^2$SCAN, or rather represent singular shortcomings for the group-IV elemental solids studied here. Thus, it is important for future work to compare SCAN and r$^2$SCAN across a larger set of transition pressures, as done in Ref.\@~\cite{Shahi2018}. Given that the r$^2$SCAN functional is on its way of becoming the new DFT workhorse, assessing its the reliability for structural phase transitions is essential. We caution against relying on the HSE functional to assess the performance of the meta-GGA transition pressures, and advocate the use of experimental data or higher-level theories instead.

\section*{Acknowledgements}
The authors thank A. Bongiorno, M. Engel, and E. Kioupakis for fruitful discussions. The calculations were performed at the Swiss National Supercomputing Center (CSCS) under Project ID No.\@ lp60 and at SCITAS-EPFL.

\section*{Data availability}
Example input files and raw phonon data corresponding to Fig.\@~\ref{fig:phonons_vs_exp} are freely available on the Materials Cloud platform, see Ref.\@~\cite{MC}.

\bibliography{main}

\begin{thebibliography}{79}%
\makeatletter
\providecommand \@ifxundefined [1]{%
 \@ifx{#1\undefined}
}%
\providecommand \@ifnum [1]{%
 \ifnum #1\expandafter \@firstoftwo
 \else \expandafter \@secondoftwo
 \fi
}%
\providecommand \@ifx [1]{%
 \ifx #1\expandafter \@firstoftwo
 \else \expandafter \@secondoftwo
 \fi
}%
\providecommand \natexlab [1]{#1}%
\providecommand \enquote  [1]{``#1''}%
\providecommand \bibnamefont  [1]{#1}%
\providecommand \bibfnamefont [1]{#1}%
\providecommand \citenamefont [1]{#1}%
\providecommand \href@noop [0]{\@secondoftwo}%
\providecommand \href [0]{\begingroup \@sanitize@url \@href}%
\providecommand \@href[1]{\@@startlink{#1}\@@href}%
\providecommand \@@href[1]{\endgroup#1\@@endlink}%
\providecommand \@sanitize@url [0]{\catcode `\\12\catcode `\$12\catcode `\&12\catcode `\#12\catcode `\^12\catcode `\_12\catcode `\%12\relax}%
\providecommand \@@startlink[1]{}%
\providecommand \@@endlink[0]{}%
\providecommand \url  [0]{\begingroup\@sanitize@url \@url }%
\providecommand \@url [1]{\endgroup\@href {#1}{\urlprefix }}%
\providecommand \urlprefix  [0]{URL }%
\providecommand \Eprint [0]{\href }%
\providecommand \doibase [0]{https://doi.org/}%
\providecommand \selectlanguage [0]{\@gobble}%
\providecommand \bibinfo  [0]{\@secondoftwo}%
\providecommand \bibfield  [0]{\@secondoftwo}%
\providecommand \translation [1]{[#1]}%
\providecommand \BibitemOpen [0]{}%
\providecommand \bibitemStop [0]{}%
\providecommand \bibitemNoStop [0]{.\EOS\space}%
\providecommand \EOS [0]{\spacefactor3000\relax}%
\providecommand \BibitemShut  [1]{\csname bibitem#1\endcsname}%
\let\auto@bib@innerbib\@empty
\bibitem [{\citenamefont {Burke}(2012)}]{Burke2012}%
  \BibitemOpen
  \bibfield  {author} {\bibinfo {author} {\bibfnamefont {K.}~\bibnamefont {Burke}},\ }\bibfield  {title} {\bibinfo {title} {Perspective on density functional theory},\ }\href {https://doi.org/10.1063/1.4704546} {\bibfield  {journal} {\bibinfo  {journal} {J. Chem. Phys.}\ }\textbf {\bibinfo {volume} {136}},\ \bibinfo {pages} {150901} (\bibinfo {year} {2012})}\BibitemShut {NoStop}%
\bibitem [{\citenamefont {Jones}(2015)}]{Jones2015}%
  \BibitemOpen
  \bibfield  {author} {\bibinfo {author} {\bibfnamefont {R.~O.}\ \bibnamefont {Jones}},\ }\bibfield  {title} {\bibinfo {title} {Density functional theory: Its origins, rise to prominence, and future},\ }\href {https://doi.org/https://doi.org/10.1103/RevModPhys.87.897} {\bibfield  {journal} {\bibinfo  {journal} {Rev. Mod. Phys.}\ }\textbf {\bibinfo {volume} {87}},\ \bibinfo {pages} {897} (\bibinfo {year} {2015})}\BibitemShut {NoStop}%
\bibitem [{\citenamefont {Medvedev}\ \emph {et~al.}(2017)\citenamefont {Medvedev}, \citenamefont {Bushmarinov}, \citenamefont {Sun}, \citenamefont {Perdew},\ and\ \citenamefont {Lyssenko}}]{Medvedev2017}%
  \BibitemOpen
  \bibfield  {author} {\bibinfo {author} {\bibfnamefont {M.~G.}\ \bibnamefont {Medvedev}}, \bibinfo {author} {\bibfnamefont {I.~S.}\ \bibnamefont {Bushmarinov}}, \bibinfo {author} {\bibfnamefont {J.}~\bibnamefont {Sun}}, \bibinfo {author} {\bibfnamefont {J.~P.}\ \bibnamefont {Perdew}},\ and\ \bibinfo {author} {\bibfnamefont {K.~A.}\ \bibnamefont {Lyssenko}},\ }\bibfield  {title} {\bibinfo {title} {Density functional theory is straying from the path toward the exact functional},\ }\href {https://doi.org/10.1126/science.aah5975} {\bibfield  {journal} {\bibinfo  {journal} {Science}\ }\textbf {\bibinfo {volume} {355}},\ \bibinfo {pages} {49} (\bibinfo {year} {2017})}\BibitemShut {NoStop}%
\bibitem [{\citenamefont {Hohenberg}\ and\ \citenamefont {Kohn}(1964)}]{Hohenberg1964}%
  \BibitemOpen
  \bibfield  {author} {\bibinfo {author} {\bibfnamefont {P.}~\bibnamefont {Hohenberg}}\ and\ \bibinfo {author} {\bibfnamefont {W.}~\bibnamefont {Kohn}},\ }\bibfield  {title} {\bibinfo {title} {Inhomogeneous electron gas},\ }\href {https://doi.org/10.1103/PhysRev.136.B864} {\bibfield  {journal} {\bibinfo  {journal} {Phys. Rev.}\ }\textbf {\bibinfo {volume} {136}},\ \bibinfo {pages} {B864} (\bibinfo {year} {1964})}\BibitemShut {NoStop}%
\bibitem [{\citenamefont {Kohn}\ and\ \citenamefont {Sham}(1965)}]{Kohn1965}%
  \BibitemOpen
  \bibfield  {author} {\bibinfo {author} {\bibfnamefont {W.}~\bibnamefont {Kohn}}\ and\ \bibinfo {author} {\bibfnamefont {L.~J.}\ \bibnamefont {Sham}},\ }\bibfield  {title} {\bibinfo {title} {Self-consistent equations including exchange and correlation effects},\ }\href {https://doi.org/10.1103/PhysRev.140.A1133} {\bibfield  {journal} {\bibinfo  {journal} {Phys. Rev.}\ }\textbf {\bibinfo {volume} {140}},\ \bibinfo {pages} {A1133} (\bibinfo {year} {1965})}\BibitemShut {NoStop}%
\bibitem [{\citenamefont {Perdew}\ \emph {et~al.}(1996)\citenamefont {Perdew}, \citenamefont {Burke},\ and\ \citenamefont {Ernzerhof}}]{Perdew1996}%
  \BibitemOpen
  \bibfield  {author} {\bibinfo {author} {\bibfnamefont {J.~P.}\ \bibnamefont {Perdew}}, \bibinfo {author} {\bibfnamefont {K.}~\bibnamefont {Burke}},\ and\ \bibinfo {author} {\bibfnamefont {M.}~\bibnamefont {Ernzerhof}},\ }\bibfield  {title} {\bibinfo {title} {Generalized gradient approximation made simple},\ }\href {https://doi.org/10.1103/PhysRevLett.77.3865} {\bibfield  {journal} {\bibinfo  {journal} {Phys. Rev. Lett.}\ }\textbf {\bibinfo {volume} {77}},\ \bibinfo {pages} {3865} (\bibinfo {year} {1996})}\BibitemShut {NoStop}%
\bibitem [{\citenamefont {Favot}\ and\ \citenamefont {Dal~Corso}(1999)}]{Favot1999}%
  \BibitemOpen
  \bibfield  {author} {\bibinfo {author} {\bibfnamefont {F.}~\bibnamefont {Favot}}\ and\ \bibinfo {author} {\bibfnamefont {A.}~\bibnamefont {Dal~Corso}},\ }\bibfield  {title} {\bibinfo {title} {Phonon dispersions: {Performance} of the generalized gradient approximation},\ }\href {https://doi.org/10.1103/physrevb.60.11427} {\bibfield  {journal} {\bibinfo  {journal} {Phys. Rev. B}\ }\textbf {\bibinfo {volume} {60}},\ \bibinfo {pages} {11427} (\bibinfo {year} {1999})}\BibitemShut {NoStop}%
\bibitem [{\citenamefont {Schimka}\ \emph {et~al.}(2011)\citenamefont {Schimka}, \citenamefont {Harl},\ and\ \citenamefont {Kresse}}]{Schimka2011}%
  \BibitemOpen
  \bibfield  {author} {\bibinfo {author} {\bibfnamefont {L.}~\bibnamefont {Schimka}}, \bibinfo {author} {\bibfnamefont {J.}~\bibnamefont {Harl}},\ and\ \bibinfo {author} {\bibfnamefont {G.}~\bibnamefont {Kresse}},\ }\bibfield  {title} {\bibinfo {title} {Improved hybrid functional for solids: The {HSE}sol functional},\ }\href {https://doi.org/10.1063/1.3524336} {\bibfield  {journal} {\bibinfo  {journal} {J. Chem. Phys.}\ }\textbf {\bibinfo {volume} {134}},\ \bibinfo {pages} {024116} (\bibinfo {year} {2011})}\BibitemShut {NoStop}%
\bibitem [{\citenamefont {Råsander}\ and\ \citenamefont {Moram}(2015)}]{Raasander2015}%
  \BibitemOpen
  \bibfield  {author} {\bibinfo {author} {\bibfnamefont {M.}~\bibnamefont {Råsander}}\ and\ \bibinfo {author} {\bibfnamefont {M.~A.}\ \bibnamefont {Moram}},\ }\bibfield  {title} {\bibinfo {title} {On the accuracy of commonly used density functional approximations in determining the elastic constants of insulators and semiconductors},\ }\href {https://doi.org/10.1063/1.4932334} {\bibfield  {journal} {\bibinfo  {journal} {J. Chem. Phys.}\ }\textbf {\bibinfo {volume} {143}},\ \bibinfo {pages} {144104} (\bibinfo {year} {2015})}\BibitemShut {NoStop}%
\bibitem [{\citenamefont {Grabowski}\ \emph {et~al.}(2007)\citenamefont {Grabowski}, \citenamefont {Hickel},\ and\ \citenamefont {Neugebauer}}]{Grabowski2007}%
  \BibitemOpen
  \bibfield  {author} {\bibinfo {author} {\bibfnamefont {B.}~\bibnamefont {Grabowski}}, \bibinfo {author} {\bibfnamefont {T.}~\bibnamefont {Hickel}},\ and\ \bibinfo {author} {\bibfnamefont {J.}~\bibnamefont {Neugebauer}},\ }\bibfield  {title} {\bibinfo {title} {{$Ab$ $initio$} study of the thermodynamic properties of nonmagnetic elementary fcc metals: Exchange-correlation-related error bars and chemical trends},\ }\href {https://doi.org/10.1103/PhysRevB.76.024309} {\bibfield  {journal} {\bibinfo  {journal} {Phys. Rev. B}\ }\textbf {\bibinfo {volume} {76}},\ \bibinfo {pages} {024309} (\bibinfo {year} {2007})}\BibitemShut {NoStop}%
\bibitem [{\citenamefont {Hummer}\ \emph {et~al.}(2009)\citenamefont {Hummer}, \citenamefont {Harl},\ and\ \citenamefont {Kresse}}]{Hummer2009}%
  \BibitemOpen
  \bibfield  {author} {\bibinfo {author} {\bibfnamefont {K.}~\bibnamefont {Hummer}}, \bibinfo {author} {\bibfnamefont {J.}~\bibnamefont {Harl}},\ and\ \bibinfo {author} {\bibfnamefont {G.}~\bibnamefont {Kresse}},\ }\bibfield  {title} {\bibinfo {title} {{Heyd-Scuseria-Ernzerhof} hybrid functional for calculating the lattice dynamics of semiconductors},\ }\href {https://doi.org/10.1103/physrevb.80.115205} {\bibfield  {journal} {\bibinfo  {journal} {Phys. Rev. B}\ }\textbf {\bibinfo {volume} {80}},\ \bibinfo {pages} {115205} (\bibinfo {year} {2009})}\BibitemShut {NoStop}%
\bibitem [{\citenamefont {Tran}\ \emph {et~al.}(2016)\citenamefont {Tran}, \citenamefont {Stelzl},\ and\ \citenamefont {Blaha}}]{Tran2016}%
  \BibitemOpen
  \bibfield  {author} {\bibinfo {author} {\bibfnamefont {F.}~\bibnamefont {Tran}}, \bibinfo {author} {\bibfnamefont {J.}~\bibnamefont {Stelzl}},\ and\ \bibinfo {author} {\bibfnamefont {P.}~\bibnamefont {Blaha}},\ }\bibfield  {title} {\bibinfo {title} {Rungs 1 to 4 of {DFT Jacob’s ladder: Extensive} test on the lattice constant, bulk modulus, and cohesive energy of solids},\ }\href {https://doi.org/10.1063/1.4948636} {\bibfield  {journal} {\bibinfo  {journal} {J. Chem. Phys.}\ }\textbf {\bibinfo {volume} {144}},\ \bibinfo {pages} {204120} (\bibinfo {year} {2016})}\BibitemShut {NoStop}%
\bibitem [{\citenamefont {Armiento}\ and\ \citenamefont {Mattsson}(2005)}]{Armiento2005}%
  \BibitemOpen
  \bibfield  {author} {\bibinfo {author} {\bibfnamefont {R.}~\bibnamefont {Armiento}}\ and\ \bibinfo {author} {\bibfnamefont {A.~E.}\ \bibnamefont {Mattsson}},\ }\bibfield  {title} {\bibinfo {title} {Functional designed to include surface effects in self-consistent density functional theory},\ }\href {https://doi.org/10.1103/physrevb.72.085108} {\bibfield  {journal} {\bibinfo  {journal} {Phys. Rev. B}\ }\textbf {\bibinfo {volume} {72}},\ \bibinfo {pages} {085108} (\bibinfo {year} {2005})}\BibitemShut {NoStop}%
\bibitem [{\citenamefont {Wu}\ and\ \citenamefont {Cohen}(2006)}]{Wu2006}%
  \BibitemOpen
  \bibfield  {author} {\bibinfo {author} {\bibfnamefont {Z.}~\bibnamefont {Wu}}\ and\ \bibinfo {author} {\bibfnamefont {R.~E.}\ \bibnamefont {Cohen}},\ }\bibfield  {title} {\bibinfo {title} {More accurate generalized gradient approximation for solids},\ }\href {https://doi.org/10.1103/physrevb.73.235116} {\bibfield  {journal} {\bibinfo  {journal} {Phys. Rev. B}\ }\textbf {\bibinfo {volume} {73}},\ \bibinfo {pages} {235116} (\bibinfo {year} {2006})}\BibitemShut {NoStop}%
\bibitem [{\citenamefont {Perdew}\ \emph {et~al.}(2008)\citenamefont {Perdew}, \citenamefont {Ruzsinszky}, \citenamefont {Csonka}, \citenamefont {Vydrov}, \citenamefont {Scuseria}, \citenamefont {Constantin}, \citenamefont {Zhou},\ and\ \citenamefont {Burke}}]{Perdew2008}%
  \BibitemOpen
  \bibfield  {author} {\bibinfo {author} {\bibfnamefont {J.~P.}\ \bibnamefont {Perdew}}, \bibinfo {author} {\bibfnamefont {A.}~\bibnamefont {Ruzsinszky}}, \bibinfo {author} {\bibfnamefont {G.~I.}\ \bibnamefont {Csonka}}, \bibinfo {author} {\bibfnamefont {O.~A.}\ \bibnamefont {Vydrov}}, \bibinfo {author} {\bibfnamefont {G.~E.}\ \bibnamefont {Scuseria}}, \bibinfo {author} {\bibfnamefont {L.~A.}\ \bibnamefont {Constantin}}, \bibinfo {author} {\bibfnamefont {X.}~\bibnamefont {Zhou}},\ and\ \bibinfo {author} {\bibfnamefont {K.}~\bibnamefont {Burke}},\ }\bibfield  {title} {\bibinfo {title} {Restoring the density-gradient expansion for exchange in solids and surfaces},\ }\href {https://doi.org/10.1103/physrevlett.100.136406} {\bibfield  {journal} {\bibinfo  {journal} {Phys. Rev. Lett.}\ }\textbf {\bibinfo {volume} {100}},\ \bibinfo {pages} {136406} (\bibinfo {year} {2008})}\BibitemShut {NoStop}%
\bibitem [{\citenamefont {Xiao}\ \emph {et~al.}(2013)\citenamefont {Xiao}, \citenamefont {Sun}, \citenamefont {Ruzsinszky}, \citenamefont {Feng}, \citenamefont {Haunschild}, \citenamefont {Scuseria},\ and\ \citenamefont {Perdew}}]{Xiao2013}%
  \BibitemOpen
  \bibfield  {author} {\bibinfo {author} {\bibfnamefont {B.}~\bibnamefont {Xiao}}, \bibinfo {author} {\bibfnamefont {J.}~\bibnamefont {Sun}}, \bibinfo {author} {\bibfnamefont {A.}~\bibnamefont {Ruzsinszky}}, \bibinfo {author} {\bibfnamefont {J.}~\bibnamefont {Feng}}, \bibinfo {author} {\bibfnamefont {R.}~\bibnamefont {Haunschild}}, \bibinfo {author} {\bibfnamefont {G.~E.}\ \bibnamefont {Scuseria}},\ and\ \bibinfo {author} {\bibfnamefont {J.~P.}\ \bibnamefont {Perdew}},\ }\bibfield  {title} {\bibinfo {title} {Testing density functionals for structural phase transitions of solids under pressure: {Si, SiO$_2$}, and {Zr}},\ }\href {https://doi.org/10.1103/physrevb.88.184103} {\bibfield  {journal} {\bibinfo  {journal} {Phys. Rev. B}\ }\textbf {\bibinfo {volume} {88}},\ \bibinfo {pages} {184103} (\bibinfo {year} {2013})}\BibitemShut {NoStop}%
\bibitem [{\citenamefont {Mehl}\ \emph {et~al.}(2021)\citenamefont {Mehl}, \citenamefont {Ronquillo}, \citenamefont {Hicks}, \citenamefont {Esters}, \citenamefont {Oses}, \citenamefont {Friedrich}, \citenamefont {Smolyanyuk}, \citenamefont {Gossett}, \citenamefont {Finkenstadt},\ and\ \citenamefont {Curtarolo}}]{Mehl2021}%
  \BibitemOpen
  \bibfield  {author} {\bibinfo {author} {\bibfnamefont {M.~J.}\ \bibnamefont {Mehl}}, \bibinfo {author} {\bibfnamefont {M.}~\bibnamefont {Ronquillo}}, \bibinfo {author} {\bibfnamefont {D.}~\bibnamefont {Hicks}}, \bibinfo {author} {\bibfnamefont {M.}~\bibnamefont {Esters}}, \bibinfo {author} {\bibfnamefont {C.}~\bibnamefont {Oses}}, \bibinfo {author} {\bibfnamefont {R.}~\bibnamefont {Friedrich}}, \bibinfo {author} {\bibfnamefont {A.}~\bibnamefont {Smolyanyuk}}, \bibinfo {author} {\bibfnamefont {E.}~\bibnamefont {Gossett}}, \bibinfo {author} {\bibfnamefont {D.}~\bibnamefont {Finkenstadt}},\ and\ \bibinfo {author} {\bibfnamefont {S.}~\bibnamefont {Curtarolo}},\ }\bibfield  {title} {\bibinfo {title} {Tin-pest problem as a test of density functionals using high-throughput calculations},\ }\href {https://doi.org/10.1103/physrevmaterials.5.083608} {\bibfield  {journal} {\bibinfo  {journal} {Phys. Rev. Mater.}\ }\textbf {\bibinfo {volume} {5}},\ \bibinfo {pages} {083608} (\bibinfo {year} {2021})}\BibitemShut
  {NoStop}%
\bibitem [{\citenamefont {Perdew}\ and\ \citenamefont {Schmidt}(2001)}]{Perdew2001}%
  \BibitemOpen
  \bibfield  {author} {\bibinfo {author} {\bibfnamefont {J.~P.}\ \bibnamefont {Perdew}}\ and\ \bibinfo {author} {\bibfnamefont {K.}~\bibnamefont {Schmidt}},\ }\bibfield  {title} {\bibinfo {title} {Jacob's ladder of density functional approximations for the exchange-correlation energy},\ }\href {https://doi.org/10.1063/1.1390175} {\bibfield  {journal} {\bibinfo  {journal} {AIP Conf. Proc.}\ }\textbf {\bibinfo {volume} {577}},\ \bibinfo {pages} {1} (\bibinfo {year} {2001})}\BibitemShut {NoStop}%
\bibitem [{\citenamefont {Sun}\ \emph {et~al.}(2015)\citenamefont {Sun}, \citenamefont {Ruzsinszky},\ and\ \citenamefont {Perdew}}]{Sun2015}%
  \BibitemOpen
  \bibfield  {author} {\bibinfo {author} {\bibfnamefont {J.}~\bibnamefont {Sun}}, \bibinfo {author} {\bibfnamefont {A.}~\bibnamefont {Ruzsinszky}},\ and\ \bibinfo {author} {\bibfnamefont {J.~P.}\ \bibnamefont {Perdew}},\ }\bibfield  {title} {\bibinfo {title} {Strongly constrained and appropriately normed semilocal density functional},\ }\href {https://doi.org/10.1103/PhysRevLett.115.036402} {\bibfield  {journal} {\bibinfo  {journal} {Phys. Rev. Lett.}\ }\textbf {\bibinfo {volume} {115}},\ \bibinfo {pages} {036402} (\bibinfo {year} {2015})}\BibitemShut {NoStop}%
\bibitem [{\citenamefont {Sun}\ \emph {et~al.}(2016)\citenamefont {Sun}, \citenamefont {Remsing}, \citenamefont {Zhang}, \citenamefont {Sun}, \citenamefont {Ruzsinszky}, \citenamefont {Peng}, \citenamefont {Yang}, \citenamefont {Paul}, \citenamefont {Waghmare}, \citenamefont {Wu}, \citenamefont {Klein},\ and\ \citenamefont {Perdew}}]{Sun2016}%
  \BibitemOpen
  \bibfield  {author} {\bibinfo {author} {\bibfnamefont {J.}~\bibnamefont {Sun}}, \bibinfo {author} {\bibfnamefont {R.~C.}\ \bibnamefont {Remsing}}, \bibinfo {author} {\bibfnamefont {Y.}~\bibnamefont {Zhang}}, \bibinfo {author} {\bibfnamefont {Z.}~\bibnamefont {Sun}}, \bibinfo {author} {\bibfnamefont {A.}~\bibnamefont {Ruzsinszky}}, \bibinfo {author} {\bibfnamefont {H.}~\bibnamefont {Peng}}, \bibinfo {author} {\bibfnamefont {Z.}~\bibnamefont {Yang}}, \bibinfo {author} {\bibfnamefont {A.}~\bibnamefont {Paul}}, \bibinfo {author} {\bibfnamefont {U.}~\bibnamefont {Waghmare}}, \bibinfo {author} {\bibfnamefont {X.}~\bibnamefont {Wu}}, \bibinfo {author} {\bibfnamefont {M.~L.}\ \bibnamefont {Klein}},\ and\ \bibinfo {author} {\bibfnamefont {J.~P.}\ \bibnamefont {Perdew}},\ }\bibfield  {title} {\bibinfo {title} {Accurate first-principles structures and energies of diversely bonded systems from an efficient density functional},\ }\href {https://doi.org/10.1038/nchem.2535} {\bibfield  {journal} {\bibinfo  {journal} {Nat.
  Chem.}\ }\textbf {\bibinfo {volume} {8}},\ \bibinfo {pages} {831} (\bibinfo {year} {2016})}\BibitemShut {NoStop}%
\bibitem [{\citenamefont {Zhang}\ \emph {et~al.}(2017)\citenamefont {Zhang}, \citenamefont {Sun}, \citenamefont {Perdew},\ and\ \citenamefont {Wu}}]{Zhang2017}%
  \BibitemOpen
  \bibfield  {author} {\bibinfo {author} {\bibfnamefont {Y.}~\bibnamefont {Zhang}}, \bibinfo {author} {\bibfnamefont {J.}~\bibnamefont {Sun}}, \bibinfo {author} {\bibfnamefont {J.~P.}\ \bibnamefont {Perdew}},\ and\ \bibinfo {author} {\bibfnamefont {X.}~\bibnamefont {Wu}},\ }\bibfield  {title} {\bibinfo {title} {Comparative first-principles studies of prototypical ferroelectric materials by {LDA}, {GGA}, and {SCAN} {meta-GGA}},\ }\href {https://doi.org/10.1103/physrevb.96.035143} {\bibfield  {journal} {\bibinfo  {journal} {Phys. Rev. B}\ }\textbf {\bibinfo {volume} {96}},\ \bibinfo {pages} {035143} (\bibinfo {year} {2017})}\BibitemShut {NoStop}%
\bibitem [{\citenamefont {Zhang}\ \emph {et~al.}(2018)\citenamefont {Zhang}, \citenamefont {Kitchaev}, \citenamefont {Yang}, \citenamefont {Chen}, \citenamefont {Dacek}, \citenamefont {Sarmiento-Pérez}, \citenamefont {Marques}, \citenamefont {Peng}, \citenamefont {Ceder}, \citenamefont {Perdew},\ and\ \citenamefont {Sun}}]{Zhang2018}%
  \BibitemOpen
  \bibfield  {author} {\bibinfo {author} {\bibfnamefont {Y.}~\bibnamefont {Zhang}}, \bibinfo {author} {\bibfnamefont {D.~A.}\ \bibnamefont {Kitchaev}}, \bibinfo {author} {\bibfnamefont {J.}~\bibnamefont {Yang}}, \bibinfo {author} {\bibfnamefont {T.}~\bibnamefont {Chen}}, \bibinfo {author} {\bibfnamefont {S.~T.}\ \bibnamefont {Dacek}}, \bibinfo {author} {\bibfnamefont {R.~A.}\ \bibnamefont {Sarmiento-Pérez}}, \bibinfo {author} {\bibfnamefont {M.~A.~L.}\ \bibnamefont {Marques}}, \bibinfo {author} {\bibfnamefont {H.}~\bibnamefont {Peng}}, \bibinfo {author} {\bibfnamefont {G.}~\bibnamefont {Ceder}}, \bibinfo {author} {\bibfnamefont {J.~P.}\ \bibnamefont {Perdew}},\ and\ \bibinfo {author} {\bibfnamefont {J.}~\bibnamefont {Sun}},\ }\bibfield  {title} {\bibinfo {title} {Efficient first-principles prediction of solid stability: {Towards} chemical accuracy},\ }\href {https://doi.org/10.1038/s41524-018-0065-z} {\bibfield  {journal} {\bibinfo  {journal} {npj Comput. Mater.}\ }\textbf {\bibinfo {volume} {4}},\ \bibinfo
  {pages} {9} (\bibinfo {year} {2018})}\BibitemShut {NoStop}%
\bibitem [{\citenamefont {Shahi}\ \emph {et~al.}(2018)\citenamefont {Shahi}, \citenamefont {Sun},\ and\ \citenamefont {Perdew}}]{Shahi2018}%
  \BibitemOpen
  \bibfield  {author} {\bibinfo {author} {\bibfnamefont {C.}~\bibnamefont {Shahi}}, \bibinfo {author} {\bibfnamefont {J.}~\bibnamefont {Sun}},\ and\ \bibinfo {author} {\bibfnamefont {J.~P.}\ \bibnamefont {Perdew}},\ }\bibfield  {title} {\bibinfo {title} {Accurate critical pressures for structural phase transitions of group {IV, III-V, and II-VI} compounds from the {SCAN} density functional},\ }\href {https://doi.org/10.1103/physrevb.97.094111} {\bibfield  {journal} {\bibinfo  {journal} {Phys. Rev. B}\ }\textbf {\bibinfo {volume} {97}},\ \bibinfo {pages} {094111} (\bibinfo {year} {2018})}\BibitemShut {NoStop}%
\bibitem [{\citenamefont {Yao}\ and\ \citenamefont {Kanai}(2017)}]{Yao2017}%
  \BibitemOpen
  \bibfield  {author} {\bibinfo {author} {\bibfnamefont {Y.}~\bibnamefont {Yao}}\ and\ \bibinfo {author} {\bibfnamefont {Y.}~\bibnamefont {Kanai}},\ }\bibfield  {title} {\bibinfo {title} {Plane-wave pseudopotential implementation and performance of {SCAN} {meta-GGA} exchange-correlation functional for extended systems},\ }\href {https://doi.org/10.1063/1.4984939} {\bibfield  {journal} {\bibinfo  {journal} {J. Chem. Phys.}\ }\textbf {\bibinfo {volume} {146}},\ \bibinfo {pages} {224105} (\bibinfo {year} {2017})}\BibitemShut {NoStop}%
\bibitem [{\citenamefont {Bartók}\ and\ \citenamefont {Yates}(2019)}]{Bartok2019}%
  \BibitemOpen
  \bibfield  {author} {\bibinfo {author} {\bibfnamefont {A.~P.}\ \bibnamefont {Bartók}}\ and\ \bibinfo {author} {\bibfnamefont {J.~R.}\ \bibnamefont {Yates}},\ }\bibfield  {title} {\bibinfo {title} {Regularized {SCAN} functional},\ }\href {https://doi.org/10.1063/1.5094646} {\bibfield  {journal} {\bibinfo  {journal} {J. Chem. Phys.}\ }\textbf {\bibinfo {volume} {150}},\ \bibinfo {pages} {161101} (\bibinfo {year} {2019})}\BibitemShut {NoStop}%
\bibitem [{\citenamefont {Furness}\ \emph {et~al.}(2020)\citenamefont {Furness}, \citenamefont {Kaplan}, \citenamefont {Ning}, \citenamefont {Perdew},\ and\ \citenamefont {Sun}}]{Furness2020}%
  \BibitemOpen
  \bibfield  {author} {\bibinfo {author} {\bibfnamefont {J.~W.}\ \bibnamefont {Furness}}, \bibinfo {author} {\bibfnamefont {A.~D.}\ \bibnamefont {Kaplan}}, \bibinfo {author} {\bibfnamefont {J.}~\bibnamefont {Ning}}, \bibinfo {author} {\bibfnamefont {J.~P.}\ \bibnamefont {Perdew}},\ and\ \bibinfo {author} {\bibfnamefont {J.}~\bibnamefont {Sun}},\ }\bibfield  {title} {\bibinfo {title} {Accurate and numerically efficient {r$^2$SCAN} meta-generalized gradient approximation},\ }\href {https://doi.org/10.1021/acs.jpclett.0c02405} {\bibfield  {journal} {\bibinfo  {journal} {J. Phys. Chem. Lett.}\ }\textbf {\bibinfo {volume} {11}},\ \bibinfo {pages} {8208} (\bibinfo {year} {2020})}\BibitemShut {NoStop}%
\bibitem [{\citenamefont {Furness}\ \emph {et~al.}(2022)\citenamefont {Furness}, \citenamefont {Kaplan}, \citenamefont {Ning}, \citenamefont {Perdew},\ and\ \citenamefont {Sun}}]{Furness2022}%
  \BibitemOpen
  \bibfield  {author} {\bibinfo {author} {\bibfnamefont {J.~W.}\ \bibnamefont {Furness}}, \bibinfo {author} {\bibfnamefont {A.~D.}\ \bibnamefont {Kaplan}}, \bibinfo {author} {\bibfnamefont {J.}~\bibnamefont {Ning}}, \bibinfo {author} {\bibfnamefont {J.~P.}\ \bibnamefont {Perdew}},\ and\ \bibinfo {author} {\bibfnamefont {J.}~\bibnamefont {Sun}},\ }\bibfield  {title} {\bibinfo {title} {Construction of {meta-GGA} functionals through restoration of exact constraint adherence to regularized {SCAN} functionals},\ }\href {https://doi.org/10.1063/5.0073623} {\bibfield  {journal} {\bibinfo  {journal} {J. Chem. Phys.}\ }\textbf {\bibinfo {volume} {156}},\ \bibinfo {pages} {034109} (\bibinfo {year} {2022})}\BibitemShut {NoStop}%
\bibitem [{\citenamefont {Ning}\ \emph {et~al.}(2022{\natexlab{a}})\citenamefont {Ning}, \citenamefont {Furness},\ and\ \citenamefont {Sun}}]{Ning2022}%
  \BibitemOpen
  \bibfield  {author} {\bibinfo {author} {\bibfnamefont {J.}~\bibnamefont {Ning}}, \bibinfo {author} {\bibfnamefont {J.~W.}\ \bibnamefont {Furness}},\ and\ \bibinfo {author} {\bibfnamefont {J.}~\bibnamefont {Sun}},\ }\bibfield  {title} {\bibinfo {title} {Reliable lattice dynamics from an efficient density functional approximation},\ }\href {https://doi.org/10.1021/acs.chemmater.1c03222} {\bibfield  {journal} {\bibinfo  {journal} {Chem. Mater.}\ }\textbf {\bibinfo {volume} {34}},\ \bibinfo {pages} {2562} (\bibinfo {year} {2022}{\natexlab{a}})}\BibitemShut {NoStop}%
\bibitem [{\citenamefont {Kingsbury}\ \emph {et~al.}(2022)\citenamefont {Kingsbury}, \citenamefont {Gupta}, \citenamefont {Bartel}, \citenamefont {Munro}, \citenamefont {Dwaraknath}, \citenamefont {Horton},\ and\ \citenamefont {Persson}}]{Kingsbury2022}%
  \BibitemOpen
  \bibfield  {author} {\bibinfo {author} {\bibfnamefont {R.}~\bibnamefont {Kingsbury}}, \bibinfo {author} {\bibfnamefont {A.~S.}\ \bibnamefont {Gupta}}, \bibinfo {author} {\bibfnamefont {C.~J.}\ \bibnamefont {Bartel}}, \bibinfo {author} {\bibfnamefont {J.~M.}\ \bibnamefont {Munro}}, \bibinfo {author} {\bibfnamefont {S.}~\bibnamefont {Dwaraknath}}, \bibinfo {author} {\bibfnamefont {M.}~\bibnamefont {Horton}},\ and\ \bibinfo {author} {\bibfnamefont {K.~A.}\ \bibnamefont {Persson}},\ }\bibfield  {title} {\bibinfo {title} {Performance comparison of {r$^2$SCAN} and {SCAN} {metaGGA} density functionals for solid materials via an automated, high-throughput computational workflow},\ }\href {https://doi.org/10.1103/physrevmaterials.6.013801} {\bibfield  {journal} {\bibinfo  {journal} {Phys. Rev. Mater.}\ }\textbf {\bibinfo {volume} {6}},\ \bibinfo {pages} {013801} (\bibinfo {year} {2022})}\BibitemShut {NoStop}%
\bibitem [{\citenamefont {Kothakonda}\ \emph {et~al.}(2022)\citenamefont {Kothakonda}, \citenamefont {Kaplan}, \citenamefont {Isaacs}, \citenamefont {Bartel}, \citenamefont {Furness}, \citenamefont {Ning}, \citenamefont {Wolverton}, \citenamefont {Perdew},\ and\ \citenamefont {Sun}}]{Kothakonda2022}%
  \BibitemOpen
  \bibfield  {author} {\bibinfo {author} {\bibfnamefont {M.}~\bibnamefont {Kothakonda}}, \bibinfo {author} {\bibfnamefont {A.~D.}\ \bibnamefont {Kaplan}}, \bibinfo {author} {\bibfnamefont {E.~B.}\ \bibnamefont {Isaacs}}, \bibinfo {author} {\bibfnamefont {C.~J.}\ \bibnamefont {Bartel}}, \bibinfo {author} {\bibfnamefont {J.~W.}\ \bibnamefont {Furness}}, \bibinfo {author} {\bibfnamefont {J.}~\bibnamefont {Ning}}, \bibinfo {author} {\bibfnamefont {C.}~\bibnamefont {Wolverton}}, \bibinfo {author} {\bibfnamefont {J.~P.}\ \bibnamefont {Perdew}},\ and\ \bibinfo {author} {\bibfnamefont {J.}~\bibnamefont {Sun}},\ }\bibfield  {title} {\bibinfo {title} {Testing the {r$^2$SCAN} density functional for the thermodynamic stability of solids with and without a van der {Waals} correction},\ }\href {https://doi.org/10.1021/acsmaterialsau.2c00059} {\bibfield  {journal} {\bibinfo  {journal} {ACS Mater. Au}\ }\textbf {\bibinfo {volume} {3}},\ \bibinfo {pages} {102} (\bibinfo {year} {2022})}\BibitemShut {NoStop}%
\bibitem [{SM()}]{SM}%
  \BibitemOpen
  \href@noop {} {}\bibinfo {note} {See Supplemental Material at [URL will be inserted by publisher] for detailed numerical setup, convergence tests, and additional data visualization.}\BibitemShut {Stop}%
\bibitem [{\citenamefont {Kittel}(2005)}]{Kittel}%
  \BibitemOpen
  \bibfield  {author} {\bibinfo {author} {\bibfnamefont {C.}~\bibnamefont {Kittel}},\ }\href@noop {} {\emph {\bibinfo {title} {Introduction to Solid State Physics}}},\ \bibinfo {edition} {8th}\ ed.\ (\bibinfo  {publisher} {Wiley},\ \bibinfo {address} {New York},\ \bibinfo {year} {2005})\BibitemShut {NoStop}%
\bibitem [{\citenamefont {Kresse}\ \emph {et~al.}(1995)\citenamefont {Kresse}, \citenamefont {Furthmüller},\ and\ \citenamefont {Hafner}}]{Kresse1995}%
  \BibitemOpen
  \bibfield  {author} {\bibinfo {author} {\bibfnamefont {G.}~\bibnamefont {Kresse}}, \bibinfo {author} {\bibfnamefont {J.}~\bibnamefont {Furthmüller}},\ and\ \bibinfo {author} {\bibfnamefont {J.}~\bibnamefont {Hafner}},\ }\bibfield  {title} {\bibinfo {title} {Ab initio force constant approach to phonon dispersion relations of diamond and graphite},\ }\href {https://doi.org/10.1209/0295-5075/32/9/005} {\bibfield  {journal} {\bibinfo  {journal} {Europhys Lett}\ }\textbf {\bibinfo {volume} {32}},\ \bibinfo {pages} {729–734} (\bibinfo {year} {1995})}\BibitemShut {NoStop}%
\bibitem [{\citenamefont {Momma}\ and\ \citenamefont {Izumi}(2011)}]{Momma2011}%
  \BibitemOpen
  \bibfield  {author} {\bibinfo {author} {\bibfnamefont {K.}~\bibnamefont {Momma}}\ and\ \bibinfo {author} {\bibfnamefont {F.}~\bibnamefont {Izumi}},\ }\bibfield  {title} {\bibinfo {title} {{VESTA 3} for three-dimensional visualization of crystal, volumetric and morphology data},\ }\href {https://doi.org/10.1107/S0021889811038970} {\bibfield  {journal} {\bibinfo  {journal} {J. Appl. Crystallogr.}\ }\textbf {\bibinfo {volume} {44}},\ \bibinfo {pages} {1272} (\bibinfo {year} {2011})}\BibitemShut {NoStop}%
\bibitem [{\citenamefont {Yin}\ and\ \citenamefont {Cohen}(1982{\natexlab{a}})}]{Yin1982a}%
  \BibitemOpen
  \bibfield  {author} {\bibinfo {author} {\bibfnamefont {M.~T.}\ \bibnamefont {Yin}}\ and\ \bibinfo {author} {\bibfnamefont {M.~L.}\ \bibnamefont {Cohen}},\ }\bibfield  {title} {\bibinfo {title} {Theory of static structural properties, crystal stability, and phase transformations: {Application} to {Si} and {Ge}},\ }\href {https://doi.org/10.1103/physrevb.26.5668} {\bibfield  {journal} {\bibinfo  {journal} {Phys. Rev. B}\ }\textbf {\bibinfo {volume} {26}},\ \bibinfo {pages} {5668} (\bibinfo {year} {1982}{\natexlab{a}})}\BibitemShut {NoStop}%
\bibitem [{\citenamefont {Moll}\ \emph {et~al.}(1995)\citenamefont {Moll}, \citenamefont {Bockstedte}, \citenamefont {Fuchs}, \citenamefont {Pehlke},\ and\ \citenamefont {Scheffler}}]{Moll1995}%
  \BibitemOpen
  \bibfield  {author} {\bibinfo {author} {\bibfnamefont {N.}~\bibnamefont {Moll}}, \bibinfo {author} {\bibfnamefont {M.}~\bibnamefont {Bockstedte}}, \bibinfo {author} {\bibfnamefont {M.}~\bibnamefont {Fuchs}}, \bibinfo {author} {\bibfnamefont {E.}~\bibnamefont {Pehlke}},\ and\ \bibinfo {author} {\bibfnamefont {M.}~\bibnamefont {Scheffler}},\ }\bibfield  {title} {\bibinfo {title} {Application of generalized gradient approximations: {The} diamond–{$\beta$}-tin phase transition in {Si} and {Ge}},\ }\href {https://doi.org/10.1103/physrevb.52.2550} {\bibfield  {journal} {\bibinfo  {journal} {Phys. Rev. B}\ }\textbf {\bibinfo {volume} {52}},\ \bibinfo {pages} {2550} (\bibinfo {year} {1995})}\BibitemShut {NoStop}%
\bibitem [{\citenamefont {Christensen}\ and\ \citenamefont {Methfessel}(1993)}]{Christensen1993}%
  \BibitemOpen
  \bibfield  {author} {\bibinfo {author} {\bibfnamefont {N.~E.}\ \bibnamefont {Christensen}}\ and\ \bibinfo {author} {\bibfnamefont {M.}~\bibnamefont {Methfessel}},\ }\bibfield  {title} {\bibinfo {title} {Density-functional calculations of the structural properties of tin under pressure},\ }\href {https://doi.org/10.1103/physrevb.48.5797} {\bibfield  {journal} {\bibinfo  {journal} {Phys. Rev. B}\ }\textbf {\bibinfo {volume} {48}},\ \bibinfo {pages} {5797} (\bibinfo {year} {1993})}\BibitemShut {NoStop}%
\bibitem [{\citenamefont {Cornelius}\ \emph {et~al.}(2017)\citenamefont {Cornelius}, \citenamefont {Treivish}, \citenamefont {Rosenthal},\ and\ \citenamefont {Pecht}}]{Cornelius2017}%
  \BibitemOpen
  \bibfield  {author} {\bibinfo {author} {\bibfnamefont {B.}~\bibnamefont {Cornelius}}, \bibinfo {author} {\bibfnamefont {S.}~\bibnamefont {Treivish}}, \bibinfo {author} {\bibfnamefont {Y.}~\bibnamefont {Rosenthal}},\ and\ \bibinfo {author} {\bibfnamefont {M.}~\bibnamefont {Pecht}},\ }\bibfield  {title} {\bibinfo {title} {The phenomenon of tin pest: {A} review},\ }\href {https://doi.org/10.1016/j.microrel.2017.10.030} {\bibfield  {journal} {\bibinfo  {journal} {Microelectron. Reliab.}\ }\textbf {\bibinfo {volume} {79}},\ \bibinfo {pages} {175} (\bibinfo {year} {2017})}\BibitemShut {NoStop}%
\bibitem [{\citenamefont {Mujica}\ \emph {et~al.}(2003)\citenamefont {Mujica}, \citenamefont {Rubio}, \citenamefont {Muñoz},\ and\ \citenamefont {Needs}}]{Mujica2003}%
  \BibitemOpen
  \bibfield  {author} {\bibinfo {author} {\bibfnamefont {A.}~\bibnamefont {Mujica}}, \bibinfo {author} {\bibfnamefont {A.}~\bibnamefont {Rubio}}, \bibinfo {author} {\bibfnamefont {A.}~\bibnamefont {Muñoz}},\ and\ \bibinfo {author} {\bibfnamefont {R.~J.}\ \bibnamefont {Needs}},\ }\bibfield  {title} {\bibinfo {title} {High-pressure phases of group-{IV}, {III–V}, and {II–VI} compounds},\ }\href {https://doi.org/10.1103/revmodphys.75.863} {\bibfield  {journal} {\bibinfo  {journal} {Rev. Mod. Phys.}\ }\textbf {\bibinfo {volume} {75}},\ \bibinfo {pages} {863} (\bibinfo {year} {2003})}\BibitemShut {NoStop}%
\bibitem [{\citenamefont {Ihm}\ and\ \citenamefont {Cohen}(1981)}]{Ihm1981}%
  \BibitemOpen
  \bibfield  {author} {\bibinfo {author} {\bibfnamefont {J.}~\bibnamefont {Ihm}}\ and\ \bibinfo {author} {\bibfnamefont {M.~L.}\ \bibnamefont {Cohen}},\ }\bibfield  {title} {\bibinfo {title} {Equilibrium properties and the phase transition of grey and white tin},\ }\href {https://doi.org/10.1103/physrevb.23.1576} {\bibfield  {journal} {\bibinfo  {journal} {Phys. Rev. B}\ }\textbf {\bibinfo {volume} {23}},\ \bibinfo {pages} {1576} (\bibinfo {year} {1981})}\BibitemShut {NoStop}%
\bibitem [{\citenamefont {Yin}\ and\ \citenamefont {Cohen}(1980)}]{Yin1980}%
  \BibitemOpen
  \bibfield  {author} {\bibinfo {author} {\bibfnamefont {M.~T.}\ \bibnamefont {Yin}}\ and\ \bibinfo {author} {\bibfnamefont {M.~L.}\ \bibnamefont {Cohen}},\ }\bibfield  {title} {\bibinfo {title} {Microscopic theory of the phase transformation and lattice dynamics of {Si}},\ }\href {https://doi.org/10.1103/physrevlett.45.1004} {\bibfield  {journal} {\bibinfo  {journal} {Phys. Rev. Lett.}\ }\textbf {\bibinfo {volume} {45}},\ \bibinfo {pages} {1004} (\bibinfo {year} {1980})}\BibitemShut {NoStop}%
\bibitem [{\citenamefont {Yin}\ and\ \citenamefont {Cohen}(1982{\natexlab{b}})}]{Yin1982}%
  \BibitemOpen
  \bibfield  {author} {\bibinfo {author} {\bibfnamefont {M.~T.}\ \bibnamefont {Yin}}\ and\ \bibinfo {author} {\bibfnamefont {M.~L.}\ \bibnamefont {Cohen}},\ }\bibfield  {title} {\bibinfo {title} {Theory of lattice-dynamical properties of solids: {Application} to {Si} and {Ge}},\ }\href {https://doi.org/10.1103/physrevb.26.3259} {\bibfield  {journal} {\bibinfo  {journal} {Phys. Rev. B}\ }\textbf {\bibinfo {volume} {26}},\ \bibinfo {pages} {3259} (\bibinfo {year} {1982}{\natexlab{b}})}\BibitemShut {NoStop}%
\bibitem [{\citenamefont {Dal~Corso}\ \emph {et~al.}(1996)\citenamefont {Dal~Corso}, \citenamefont {Pasquarello}, \citenamefont {Baldereschi},\ and\ \citenamefont {Car}}]{DalCorso1996}%
  \BibitemOpen
  \bibfield  {author} {\bibinfo {author} {\bibfnamefont {A.}~\bibnamefont {Dal~Corso}}, \bibinfo {author} {\bibfnamefont {A.}~\bibnamefont {Pasquarello}}, \bibinfo {author} {\bibfnamefont {A.}~\bibnamefont {Baldereschi}},\ and\ \bibinfo {author} {\bibfnamefont {R.}~\bibnamefont {Car}},\ }\bibfield  {title} {\bibinfo {title} {Generalized-gradient approximations to density-functional theory: {A} comparative study for atoms and solids},\ }\href {https://doi.org/10.1103/physrevb.53.1180} {\bibfield  {journal} {\bibinfo  {journal} {Phys. Rev. B}\ }\textbf {\bibinfo {volume} {53}},\ \bibinfo {pages} {1180} (\bibinfo {year} {1996})}\BibitemShut {NoStop}%
\bibitem [{\citenamefont {Gaál-Nagy}\ \emph {et~al.}(1999)\citenamefont {Gaál-Nagy}, \citenamefont {Bauer}, \citenamefont {Schmitt}, \citenamefont {Karch}, \citenamefont {Pavone},\ and\ \citenamefont {Strauch}}]{GaalNagy1999}%
  \BibitemOpen
  \bibfield  {author} {\bibinfo {author} {\bibfnamefont {K.}~\bibnamefont {Gaál-Nagy}}, \bibinfo {author} {\bibfnamefont {A.}~\bibnamefont {Bauer}}, \bibinfo {author} {\bibfnamefont {M.}~\bibnamefont {Schmitt}}, \bibinfo {author} {\bibfnamefont {K.}~\bibnamefont {Karch}}, \bibinfo {author} {\bibfnamefont {P.}~\bibnamefont {Pavone}},\ and\ \bibinfo {author} {\bibfnamefont {D.}~\bibnamefont {Strauch}},\ }\bibfield  {title} {\bibinfo {title} {Temperature and dynamical effects on the high-pressure cubic-diamond {$\leftrightarrow \beta$}-tin phase transition in {Si and Ge}},\ }\href {https://doi.org/10.1002/(sici)1521-3951(199901)211:1<275::aid-pssb275>3.3.co;2-f} {\bibfield  {journal} {\bibinfo  {journal} {Phys. Status Solidi B}\ }\textbf {\bibinfo {volume} {211}},\ \bibinfo {pages} {275} (\bibinfo {year} {1999})}\BibitemShut {NoStop}%
\bibitem [{\citenamefont {Hennig}\ \emph {et~al.}(2010)\citenamefont {Hennig}, \citenamefont {Wadehra}, \citenamefont {Driver}, \citenamefont {Parker}, \citenamefont {Umrigar},\ and\ \citenamefont {Wilkins}}]{Hennig2010}%
  \BibitemOpen
  \bibfield  {author} {\bibinfo {author} {\bibfnamefont {R.~G.}\ \bibnamefont {Hennig}}, \bibinfo {author} {\bibfnamefont {A.}~\bibnamefont {Wadehra}}, \bibinfo {author} {\bibfnamefont {K.~P.}\ \bibnamefont {Driver}}, \bibinfo {author} {\bibfnamefont {W.~D.}\ \bibnamefont {Parker}}, \bibinfo {author} {\bibfnamefont {C.~J.}\ \bibnamefont {Umrigar}},\ and\ \bibinfo {author} {\bibfnamefont {J.~W.}\ \bibnamefont {Wilkins}},\ }\bibfield  {title} {\bibinfo {title} {Phase transformation in {Si} from semiconducting diamond to metallic $\ensuremath{\beta}\text{-Sn}$ phase in {QMC} and {DFT} under hydrostatic and anisotropic stress},\ }\href {https://doi.org/10.1103/PhysRevB.82.014101} {\bibfield  {journal} {\bibinfo  {journal} {Phys. Rev. B}\ }\textbf {\bibinfo {volume} {82}},\ \bibinfo {pages} {014101} (\bibinfo {year} {2010})}\BibitemShut {NoStop}%
\bibitem [{\citenamefont {Sengupta}\ \emph {et~al.}(2018)\citenamefont {Sengupta}, \citenamefont {Bates},\ and\ \citenamefont {Ruzsinszky}}]{Sengupta2018}%
  \BibitemOpen
  \bibfield  {author} {\bibinfo {author} {\bibfnamefont {N.}~\bibnamefont {Sengupta}}, \bibinfo {author} {\bibfnamefont {J.~E.}\ \bibnamefont {Bates}},\ and\ \bibinfo {author} {\bibfnamefont {A.}~\bibnamefont {Ruzsinszky}},\ }\bibfield  {title} {\bibinfo {title} {From semilocal density functionals to random phase approximation renormalized perturbation theory: {A} methodological assessment of structural phase transitions},\ }\href {https://doi.org/10.1103/physrevb.97.235136} {\bibfield  {journal} {\bibinfo  {journal} {Phys. Rev. B}\ }\textbf {\bibinfo {volume} {97}},\ \bibinfo {pages} {235136} (\bibinfo {year} {2018})}\BibitemShut {NoStop}%
\bibitem [{\citenamefont {Ning}\ \emph {et~al.}(2022{\natexlab{b}})\citenamefont {Ning}, \citenamefont {Kothakonda}, \citenamefont {Furness}, \citenamefont {Kaplan}, \citenamefont {Ehlert}, \citenamefont {Brandenburg}, \citenamefont {Perdew},\ and\ \citenamefont {Sun}}]{Ning2022a}%
  \BibitemOpen
  \bibfield  {author} {\bibinfo {author} {\bibfnamefont {J.}~\bibnamefont {Ning}}, \bibinfo {author} {\bibfnamefont {M.}~\bibnamefont {Kothakonda}}, \bibinfo {author} {\bibfnamefont {J.~W.}\ \bibnamefont {Furness}}, \bibinfo {author} {\bibfnamefont {A.~D.}\ \bibnamefont {Kaplan}}, \bibinfo {author} {\bibfnamefont {S.}~\bibnamefont {Ehlert}}, \bibinfo {author} {\bibfnamefont {J.~G.}\ \bibnamefont {Brandenburg}}, \bibinfo {author} {\bibfnamefont {J.~P.}\ \bibnamefont {Perdew}},\ and\ \bibinfo {author} {\bibfnamefont {J.}~\bibnamefont {Sun}},\ }\bibfield  {title} {\bibinfo {title} {Workhorse minimally empirical dispersion-corrected density functional with tests for weakly bound systems: {r$^2$SCAN+rVV10}},\ }\href {https://doi.org/10.1103/physrevb.106.075422} {\bibfield  {journal} {\bibinfo  {journal} {Phys. Rev. B}\ }\textbf {\bibinfo {volume} {106}},\ \bibinfo {pages} {075422} (\bibinfo {year} {2022}{\natexlab{b}})}\BibitemShut {NoStop}%
\bibitem [{\citenamefont {Poirier}(2000)}]{BM-potential}%
  \BibitemOpen
  \bibfield  {author} {\bibinfo {author} {\bibfnamefont {J.-P.}\ \bibnamefont {Poirier}},\ }\bibfield  {title} {\bibinfo {title} {Equations of state},\ }in\ \href@noop {} {\emph {\bibinfo {booktitle} {Introduction to the Physics of the Earth’s Interior}}}\ (\bibinfo  {publisher} {Cambridge University Press},\ \bibinfo {address} {Cambridge},\ \bibinfo {year} {2000})\ Chap.~\bibinfo {chapter} {4}, p.\ \bibinfo {pages} {63–109}\BibitemShut {NoStop}%
\bibitem [{\citenamefont {Kresse}\ and\ \citenamefont {Furthmüller}(1996)}]{vasp}%
  \BibitemOpen
  \bibfield  {author} {\bibinfo {author} {\bibfnamefont {G.}~\bibnamefont {Kresse}}\ and\ \bibinfo {author} {\bibfnamefont {J.}~\bibnamefont {Furthmüller}},\ }\bibfield  {title} {\bibinfo {title} {Efficient iterative schemes for {$ab$ $initio$} total‐energy calculations using a plane‐wave basis set},\ }\href {https://doi.org/10.1103/PhysRevB.54.11169} {\bibfield  {journal} {\bibinfo  {journal} {Phys. Rev. B}\ }\textbf {\bibinfo {volume} {54}},\ \bibinfo {pages} {11169} (\bibinfo {year} {1996})}\BibitemShut {NoStop}%
\bibitem [{\citenamefont {Kresse}\ and\ \citenamefont {Joubert}(1999)}]{KresseJoubert1999}%
  \BibitemOpen
  \bibfield  {author} {\bibinfo {author} {\bibfnamefont {G.}~\bibnamefont {Kresse}}\ and\ \bibinfo {author} {\bibfnamefont {D.}~\bibnamefont {Joubert}},\ }\bibfield  {title} {\bibinfo {title} {From ultrasoft pseudopotentials to the projector augmented‐wave method},\ }\href {https://doi.org/10.1103/PhysRevB.59.1758} {\bibfield  {journal} {\bibinfo  {journal} {Phys. Rev. B}\ }\textbf {\bibinfo {volume} {59}},\ \bibinfo {pages} {1758} (\bibinfo {year} {1999})}\BibitemShut {NoStop}%
\bibitem [{\citenamefont {Bl{\"o}chl}(1994)}]{Blochl1994}%
  \BibitemOpen
  \bibfield  {author} {\bibinfo {author} {\bibfnamefont {P.~E.}\ \bibnamefont {Bl{\"o}chl}},\ }\bibfield  {title} {\bibinfo {title} {Projector augmented‐wave method},\ }\href {https://doi.org/10.1103/PhysRevB.50.17953} {\bibfield  {journal} {\bibinfo  {journal} {Phys. Rev. B}\ }\textbf {\bibinfo {volume} {50}},\ \bibinfo {pages} {17953} (\bibinfo {year} {1994})}\BibitemShut {NoStop}%
\bibitem [{\citenamefont {Heyd}\ \emph {et~al.}(2003)\citenamefont {Heyd}, \citenamefont {Scuseria},\ and\ \citenamefont {Ernzerhof}}]{Heyd2003}%
  \BibitemOpen
  \bibfield  {author} {\bibinfo {author} {\bibfnamefont {J.}~\bibnamefont {Heyd}}, \bibinfo {author} {\bibfnamefont {G.~E.}\ \bibnamefont {Scuseria}},\ and\ \bibinfo {author} {\bibfnamefont {M.}~\bibnamefont {Ernzerhof}},\ }\bibfield  {title} {\bibinfo {title} {Hybrid functionals based on a screened {Coulomb} potential},\ }\href {https://doi.org/10.1063/1.1564060} {\bibfield  {journal} {\bibinfo  {journal} {J. Chem. Phys.}\ }\textbf {\bibinfo {volume} {118}},\ \bibinfo {pages} {8207} (\bibinfo {year} {2003})}\BibitemShut {NoStop}%
\bibitem [{\citenamefont {Heyd}\ \emph {et~al.}(2006)\citenamefont {Heyd}, \citenamefont {Scuseria},\ and\ \citenamefont {Ernzerhof}}]{Heyd2006}%
  \BibitemOpen
  \bibfield  {author} {\bibinfo {author} {\bibfnamefont {J.}~\bibnamefont {Heyd}}, \bibinfo {author} {\bibfnamefont {G.~E.}\ \bibnamefont {Scuseria}},\ and\ \bibinfo {author} {\bibfnamefont {M.}~\bibnamefont {Ernzerhof}},\ }\bibfield  {title} {\bibinfo {title} {Erratum: “{Hybrid} functionals based on a screened coulomb potential” [{J. Chem. Phys.} 118, 8207 (2003)]},\ }\href {https://doi.org/10.1063/1.2204597} {\bibfield  {journal} {\bibinfo  {journal} {J. Chem. Phys.}\ }\textbf {\bibinfo {volume} {124}},\ \bibinfo {pages} {219906} (\bibinfo {year} {2006})}\BibitemShut {NoStop}%
\bibitem [{\citenamefont {Paier}\ \emph {et~al.}(2006{\natexlab{a}})\citenamefont {Paier}, \citenamefont {Marsman}, \citenamefont {Hummer}, \citenamefont {Kresse}, \citenamefont {Gerber},\ and\ \citenamefont {{\'{A}}ngy{\'{a}}n}}]{Paier2006}%
  \BibitemOpen
  \bibfield  {author} {\bibinfo {author} {\bibfnamefont {J.}~\bibnamefont {Paier}}, \bibinfo {author} {\bibfnamefont {M.}~\bibnamefont {Marsman}}, \bibinfo {author} {\bibfnamefont {K.}~\bibnamefont {Hummer}}, \bibinfo {author} {\bibfnamefont {G.}~\bibnamefont {Kresse}}, \bibinfo {author} {\bibfnamefont {I.~C.}\ \bibnamefont {Gerber}},\ and\ \bibinfo {author} {\bibfnamefont {J.~G.}\ \bibnamefont {{\'{A}}ngy{\'{a}}n}},\ }\bibfield  {title} {\bibinfo {title} {Screened hybrid density functionals applied to solids},\ }\href {https://doi.org/10.1063/1.2187006} {\bibfield  {journal} {\bibinfo  {journal} {J. Chem. Phys.}\ }\textbf {\bibinfo {volume} {124}},\ \bibinfo {pages} {154709} (\bibinfo {year} {2006}{\natexlab{a}})}\BibitemShut {NoStop}%
\bibitem [{\citenamefont {Paier}\ \emph {et~al.}(2006{\natexlab{b}})\citenamefont {Paier}, \citenamefont {Marsman}, \citenamefont {Hummer}, \citenamefont {Kresse}, \citenamefont {Gerber},\ and\ \citenamefont {{\'{A}}ngy{\'{a}}n}}]{Paier2006a}%
  \BibitemOpen
  \bibfield  {author} {\bibinfo {author} {\bibfnamefont {J.}~\bibnamefont {Paier}}, \bibinfo {author} {\bibfnamefont {M.}~\bibnamefont {Marsman}}, \bibinfo {author} {\bibfnamefont {K.}~\bibnamefont {Hummer}}, \bibinfo {author} {\bibfnamefont {G.}~\bibnamefont {Kresse}}, \bibinfo {author} {\bibfnamefont {I.~C.}\ \bibnamefont {Gerber}},\ and\ \bibinfo {author} {\bibfnamefont {J.~G.}\ \bibnamefont {{\'{A}}ngy{\'{a}}n}},\ }\bibfield  {title} {\bibinfo {title} {Erratum: {“Screened} hybrid density functionals applied to solids” [{J. Chem. Phys.} 124, 154709 (2006)]},\ }\href {https://doi.org/10.1063/1.2403866} {\bibfield  {journal} {\bibinfo  {journal} {J. Chem. Phys.}\ }\textbf {\bibinfo {volume} {125}},\ \bibinfo {pages} {249901} (\bibinfo {year} {2006}{\natexlab{b}})}\BibitemShut {NoStop}%
\bibitem [{\citenamefont {Krukau}\ \emph {et~al.}(2006)\citenamefont {Krukau}, \citenamefont {Vydrov}, \citenamefont {Izmaylov},\ and\ \citenamefont {Scuseria}}]{Krukau2006}%
  \BibitemOpen
  \bibfield  {author} {\bibinfo {author} {\bibfnamefont {A.~V.}\ \bibnamefont {Krukau}}, \bibinfo {author} {\bibfnamefont {O.~A.}\ \bibnamefont {Vydrov}}, \bibinfo {author} {\bibfnamefont {A.~F.}\ \bibnamefont {Izmaylov}},\ and\ \bibinfo {author} {\bibfnamefont {G.~E.}\ \bibnamefont {Scuseria}},\ }\bibfield  {title} {\bibinfo {title} {Influence of the exchange screening parameter on the performance of screened hybrid functionals},\ }\href {https://doi.org/10.1063/1.2404663} {\bibfield  {journal} {\bibinfo  {journal} {J. Chem. Phys.}\ }\textbf {\bibinfo {volume} {125}},\ \bibinfo {pages} {224106} (\bibinfo {year} {2006})}\BibitemShut {NoStop}%
\bibitem [{\citenamefont {Engel}\ \emph {et~al.}(2020)\citenamefont {Engel}, \citenamefont {Marsman}, \citenamefont {Franchini},\ and\ \citenamefont {Kresse}}]{Engel2020}%
  \BibitemOpen
  \bibfield  {author} {\bibinfo {author} {\bibfnamefont {M.}~\bibnamefont {Engel}}, \bibinfo {author} {\bibfnamefont {M.}~\bibnamefont {Marsman}}, \bibinfo {author} {\bibfnamefont {C.}~\bibnamefont {Franchini}},\ and\ \bibinfo {author} {\bibfnamefont {G.}~\bibnamefont {Kresse}},\ }\bibfield  {title} {\bibinfo {title} {Electron-phonon interactions using the projector augmented-wave method and {Wannier} functions},\ }\href {https://doi.org/10.1103/physrevb.101.184302} {\bibfield  {journal} {\bibinfo  {journal} {Phys. Rev. B}\ }\textbf {\bibinfo {volume} {101}},\ \bibinfo {pages} {184302} (\bibinfo {year} {2020})}\BibitemShut {NoStop}%
\bibitem [{\citenamefont {Dacosta}\ \emph {et~al.}(1986)\citenamefont {Dacosta}, \citenamefont {Nielsen},\ and\ \citenamefont {Kunc}}]{Dacosta1986}%
  \BibitemOpen
  \bibfield  {author} {\bibinfo {author} {\bibfnamefont {P.~G.}\ \bibnamefont {Dacosta}}, \bibinfo {author} {\bibfnamefont {O.~H.}\ \bibnamefont {Nielsen}},\ and\ \bibinfo {author} {\bibfnamefont {K.}~\bibnamefont {Kunc}},\ }\bibfield  {title} {\bibinfo {title} {Stress theorem in the determination of static equilibrium by the density functional method},\ }\href {https://doi.org/10.1088/0022-3719/19/17/012} {\bibfield  {journal} {\bibinfo  {journal} {J. Phys. C: Solid State Phys.}\ }\textbf {\bibinfo {volume} {19}},\ \bibinfo {pages} {3163} (\bibinfo {year} {1986})}\BibitemShut {NoStop}%
\bibitem [{\citenamefont {Francis}\ and\ \citenamefont {Payne}(1990)}]{G_P_Francis_1990}%
  \BibitemOpen
  \bibfield  {author} {\bibinfo {author} {\bibfnamefont {G.~P.}\ \bibnamefont {Francis}}\ and\ \bibinfo {author} {\bibfnamefont {M.~C.}\ \bibnamefont {Payne}},\ }\bibfield  {title} {\bibinfo {title} {Finite basis set corrections to total energy pseudopotential calculations},\ }\href {https://doi.org/10.1088/0953-8984/2/19/007} {\bibfield  {journal} {\bibinfo  {journal} {J. Phys.: Condens. Matter}\ }\textbf {\bibinfo {volume} {2}},\ \bibinfo {pages} {4395} (\bibinfo {year} {1990})}\BibitemShut {NoStop}%
\bibitem [{\citenamefont {Madelung}(2004)}]{madelung2004}%
  \BibitemOpen
  \bibfield  {author} {\bibinfo {author} {\bibfnamefont {O.}~\bibnamefont {Madelung}},\ }\href {https://doi.org/10.1007/978-3-642-18865-7} {\emph {\bibinfo {title} {Semiconductors: Data Handbook}}},\ \bibinfo {edition} {3rd}\ ed.\ (\bibinfo  {publisher} {Springer-Verlag},\ \bibinfo {address} {Berlin Heidelberg},\ \bibinfo {year} {2004})\BibitemShut {NoStop}%
\bibitem [{\citenamefont {Zhang}\ \emph {et~al.}(2025)\citenamefont {Zhang}, \citenamefont {Ramasamy}, \citenamefont {Pokharel}, \citenamefont {Kothakonda}, \citenamefont {Xiao}, \citenamefont {Furness}, \citenamefont {Ning}, \citenamefont {Zhang},\ and\ \citenamefont {Sun}}]{Zhang2025}%
  \BibitemOpen
  \bibfield  {author} {\bibinfo {author} {\bibfnamefont {Y.}~\bibnamefont {Zhang}}, \bibinfo {author} {\bibfnamefont {A.}~\bibnamefont {Ramasamy}}, \bibinfo {author} {\bibfnamefont {K.}~\bibnamefont {Pokharel}}, \bibinfo {author} {\bibfnamefont {M.}~\bibnamefont {Kothakonda}}, \bibinfo {author} {\bibfnamefont {B.}~\bibnamefont {Xiao}}, \bibinfo {author} {\bibfnamefont {J.~W.}\ \bibnamefont {Furness}}, \bibinfo {author} {\bibfnamefont {J.}~\bibnamefont {Ning}}, \bibinfo {author} {\bibfnamefont {R.}~\bibnamefont {Zhang}},\ and\ \bibinfo {author} {\bibfnamefont {J.}~\bibnamefont {Sun}},\ }\bibfield  {title} {\bibinfo {title} {Advances and challenges of {SCAN} and {r$^2$SCAN} density functionals in transition‐metal compounds},\ }\href {https://doi.org/10.1002/wcms.70007} {\bibfield  {journal} {\bibinfo  {journal} {WIREs Comput. Mol. Sci.}\ }\textbf {\bibinfo {volume} {15}},\ \bibinfo {pages} {e70007} (\bibinfo {year} {2025})}\BibitemShut {NoStop}%
\bibitem [{\citenamefont {Riemelmoser}\ \emph {et~al.}()\citenamefont {Riemelmoser}, \citenamefont {Xu},\ and\ \citenamefont {Pasquarello}}]{Riemelmoser2025}%
  \BibitemOpen
  \bibfield  {author} {\bibinfo {author} {\bibfnamefont {S.}~\bibnamefont {Riemelmoser}}, \bibinfo {author} {\bibfnamefont {X.}~\bibnamefont {Xu}},\ and\ \bibinfo {author} {\bibfnamefont {A.}~\bibnamefont {Pasquarello}},\ }\bibfield  {title} {\bibinfo {title} {Climbing {Jacob's Ladder} for the {C}oulomb hole: Dielectric-dependent hybrid functional based on {meta-GGA}},\ }\href@noop {} {\bibinfo  {journal} {(unpublished)}\ }\BibitemShut {NoStop}%
\bibitem [{\citenamefont {Shang}\ \emph {et~al.}(2010)\citenamefont {Shang}, \citenamefont {Saengdeejing}, \citenamefont {Mei}, \citenamefont {Kim}, \citenamefont {Zhang}, \citenamefont {Ganeshan}, \citenamefont {Wang},\ and\ \citenamefont {Liu}}]{Shang2010}%
  \BibitemOpen
\bibfield  {journal} {  }\bibfield  {author} {\bibinfo {author} {\bibfnamefont {S.}~\bibnamefont {Shang}}, \bibinfo {author} {\bibfnamefont {A.}~\bibnamefont {Saengdeejing}}, \bibinfo {author} {\bibfnamefont {Z.}~\bibnamefont {Mei}}, \bibinfo {author} {\bibfnamefont {D.}~\bibnamefont {Kim}}, \bibinfo {author} {\bibfnamefont {H.}~\bibnamefont {Zhang}}, \bibinfo {author} {\bibfnamefont {S.}~\bibnamefont {Ganeshan}}, \bibinfo {author} {\bibfnamefont {Y.}~\bibnamefont {Wang}},\ and\ \bibinfo {author} {\bibfnamefont {Z.}~\bibnamefont {Liu}},\ }\bibfield  {title} {\bibinfo {title} {First-principles calculations of pure elements: {Equations} of state and elastic stiffness constants},\ }\href {https://doi.org/10.1016/j.commatsci.2010.03.041} {\bibfield  {journal} {\bibinfo  {journal} {Comput. Mater. Sci.}\ }\textbf {\bibinfo {volume} {48}},\ \bibinfo {pages} {813} (\bibinfo {year} {2010})}\BibitemShut {NoStop}%
\bibitem [{\citenamefont {Price}\ \emph {et~al.}(1971)\citenamefont {Price}, \citenamefont {Rowe},\ and\ \citenamefont {Nicklow}}]{Price1971}%
  \BibitemOpen
  \bibfield  {author} {\bibinfo {author} {\bibfnamefont {D.~L.}\ \bibnamefont {Price}}, \bibinfo {author} {\bibfnamefont {J.~M.}\ \bibnamefont {Rowe}},\ and\ \bibinfo {author} {\bibfnamefont {R.~M.}\ \bibnamefont {Nicklow}},\ }\bibfield  {title} {\bibinfo {title} {Lattice dynamics of grey tin and indium antimonide},\ }\href {https://doi.org/10.1103/physrevb.3.1268} {\bibfield  {journal} {\bibinfo  {journal} {Phys. Rev. B}\ }\textbf {\bibinfo {volume} {3}},\ \bibinfo {pages} {1268} (\bibinfo {year} {1971})}\BibitemShut {NoStop}%
\bibitem [{\citenamefont {McSkimin}\ and\ \citenamefont {Andreatch}(1972)}]{McSkimin1972}%
  \BibitemOpen
  \bibfield  {author} {\bibinfo {author} {\bibfnamefont {H.~J.}\ \bibnamefont {McSkimin}}\ and\ \bibinfo {author} {\bibfnamefont {P.}~\bibnamefont {Andreatch}},\ }\bibfield  {title} {\bibinfo {title} {Elastic moduli of diamond as a function of pressure and temperature},\ }\href {https://doi.org/10.1063/1.1661636} {\bibfield  {journal} {\bibinfo  {journal} {J. Appl. Phys.}\ }\textbf {\bibinfo {volume} {43}},\ \bibinfo {pages} {2944} (\bibinfo {year} {1972})}\BibitemShut {NoStop}%
\bibitem [{\citenamefont {McSkimin}\ and\ \citenamefont {Andreatch}(1964)}]{McSkimin1964}%
  \BibitemOpen
  \bibfield  {author} {\bibinfo {author} {\bibfnamefont {H.~J.}\ \bibnamefont {McSkimin}}\ and\ \bibinfo {author} {\bibfnamefont {P.}~\bibnamefont {Andreatch}},\ }\bibfield  {title} {\bibinfo {title} {Elastic moduli of silicon vs hydrostatic pressure at 25.0°{C} and {$-$}195.8°{C}},\ }\href {https://doi.org/10.1063/1.1702809} {\bibfield  {journal} {\bibinfo  {journal} {J. Appl. Phys.}\ }\textbf {\bibinfo {volume} {35}},\ \bibinfo {pages} {2161} (\bibinfo {year} {1964})}\BibitemShut {NoStop}%
\bibitem [{\citenamefont {McSkimin}\ and\ \citenamefont {Andreatch}(1963)}]{McSkimin1963}%
  \BibitemOpen
  \bibfield  {author} {\bibinfo {author} {\bibfnamefont {H.~J.}\ \bibnamefont {McSkimin}}\ and\ \bibinfo {author} {\bibfnamefont {P.}~\bibnamefont {Andreatch}},\ }\bibfield  {title} {\bibinfo {title} {Elastic moduli of germanium versus hydrostatic pressure at 25.0°{C} and {$-$}195.8°{C}},\ }\href {https://doi.org/10.1063/1.1729323} {\bibfield  {journal} {\bibinfo  {journal} {J. Appl. Phys.}\ }\textbf {\bibinfo {volume} {34}},\ \bibinfo {pages} {651} (\bibinfo {year} {1963})}\BibitemShut {NoStop}%
\bibitem [{\citenamefont {Xing}\ \emph {et~al.}(2021)\citenamefont {Xing}, \citenamefont {Meng}, \citenamefont {Ning}, \citenamefont {Sun},\ and\ \citenamefont {Yu}}]{Xing2021}%
  \BibitemOpen
  \bibfield  {author} {\bibinfo {author} {\bibfnamefont {W.}~\bibnamefont {Xing}}, \bibinfo {author} {\bibfnamefont {F.}~\bibnamefont {Meng}}, \bibinfo {author} {\bibfnamefont {J.}~\bibnamefont {Ning}}, \bibinfo {author} {\bibfnamefont {J.}~\bibnamefont {Sun}},\ and\ \bibinfo {author} {\bibfnamefont {R.}~\bibnamefont {Yu}},\ }\bibfield  {title} {\bibinfo {title} {Comparative first-principles study of elastic constants of covalent and ionic materials with {LDA, GGA, and meta-GGA} functionals and the prediction of mechanical hardness},\ }\href {https://doi.org/10.1007/s11431-021-1825-x} {\bibfield  {journal} {\bibinfo  {journal} {Sci. China Technol. Sci.}\ }\textbf {\bibinfo {volume} {64}},\ \bibinfo {pages} {2755} (\bibinfo {year} {2021})}\BibitemShut {NoStop}%
\bibitem [{\citenamefont {Warren}\ \emph {et~al.}(1967)\citenamefont {Warren}, \citenamefont {Yarnell}, \citenamefont {Dolling},\ and\ \citenamefont {Cowley}}]{Warren1967}%
  \BibitemOpen
  \bibfield  {author} {\bibinfo {author} {\bibfnamefont {J.~L.}\ \bibnamefont {Warren}}, \bibinfo {author} {\bibfnamefont {J.~L.}\ \bibnamefont {Yarnell}}, \bibinfo {author} {\bibfnamefont {G.}~\bibnamefont {Dolling}},\ and\ \bibinfo {author} {\bibfnamefont {R.~A.}\ \bibnamefont {Cowley}},\ }\bibfield  {title} {\bibinfo {title} {Lattice dynamics of diamond},\ }\href {https://doi.org/10.1103/physrev.158.805} {\bibfield  {journal} {\bibinfo  {journal} {Phys. Rev.}\ }\textbf {\bibinfo {volume} {158}},\ \bibinfo {pages} {805} (\bibinfo {year} {1967})}\BibitemShut {NoStop}%
\bibitem [{\citenamefont {Kulda}\ \emph {et~al.}(2002)\citenamefont {Kulda}, \citenamefont {Kainzmaier}, \citenamefont {Strauch}, \citenamefont {Dorner}, \citenamefont {Lorenzen},\ and\ \citenamefont {Krisch}}]{Kulda2002}%
  \BibitemOpen
  \bibfield  {author} {\bibinfo {author} {\bibfnamefont {J.}~\bibnamefont {Kulda}}, \bibinfo {author} {\bibfnamefont {H.}~\bibnamefont {Kainzmaier}}, \bibinfo {author} {\bibfnamefont {D.}~\bibnamefont {Strauch}}, \bibinfo {author} {\bibfnamefont {B.}~\bibnamefont {Dorner}}, \bibinfo {author} {\bibfnamefont {M.}~\bibnamefont {Lorenzen}},\ and\ \bibinfo {author} {\bibfnamefont {M.}~\bibnamefont {Krisch}},\ }\bibfield  {title} {\bibinfo {title} {Overbending of the longitudinal optical phonon branch in diamond as evidenced by inelastic neutron and x-ray scattering},\ }\href {https://doi.org/10.1103/physrevb.66.241202} {\bibfield  {journal} {\bibinfo  {journal} {Phys. Rev. B}\ }\textbf {\bibinfo {volume} {66}},\ \bibinfo {pages} {241202} (\bibinfo {year} {2002})}\BibitemShut {NoStop}%
\bibitem [{\citenamefont {Kulda}\ \emph {et~al.}(1994)\citenamefont {Kulda}, \citenamefont {Strauch}, \citenamefont {Pavone},\ and\ \citenamefont {Ishii}}]{Kulda1994}%
  \BibitemOpen
  \bibfield  {author} {\bibinfo {author} {\bibfnamefont {J.}~\bibnamefont {Kulda}}, \bibinfo {author} {\bibfnamefont {D.}~\bibnamefont {Strauch}}, \bibinfo {author} {\bibfnamefont {P.}~\bibnamefont {Pavone}},\ and\ \bibinfo {author} {\bibfnamefont {Y.}~\bibnamefont {Ishii}},\ }\bibfield  {title} {\bibinfo {title} {Inelastic-neutron-scattering study of phonon eigenvectors and frequencies in {Si}},\ }\href {https://doi.org/10.1103/PhysRevB.50.13347} {\bibfield  {journal} {\bibinfo  {journal} {Phys. Rev. B}\ }\textbf {\bibinfo {volume} {50}},\ \bibinfo {pages} {13347} (\bibinfo {year} {1994})}\BibitemShut {NoStop}%
\bibitem [{\citenamefont {Nilsson}\ and\ \citenamefont {Nelin}(1971)}]{Nilsson1971}%
  \BibitemOpen
  \bibfield  {author} {\bibinfo {author} {\bibfnamefont {G.}~\bibnamefont {Nilsson}}\ and\ \bibinfo {author} {\bibfnamefont {G.}~\bibnamefont {Nelin}},\ }\bibfield  {title} {\bibinfo {title} {Phonon dispersion relations in {Ge} at 80 °{K}},\ }\href {https://doi.org/10.1103/physrevb.3.364} {\bibfield  {journal} {\bibinfo  {journal} {Phys. Rev. B}\ }\textbf {\bibinfo {volume} {3}},\ \bibinfo {pages} {364} (\bibinfo {year} {1971})}\BibitemShut {NoStop}%
\bibitem [{\citenamefont {Nilsson}\ and\ \citenamefont {Nelin}(1972)}]{Nilsson1972}%
  \BibitemOpen
  \bibfield  {author} {\bibinfo {author} {\bibfnamefont {G.}~\bibnamefont {Nilsson}}\ and\ \bibinfo {author} {\bibfnamefont {G.}~\bibnamefont {Nelin}},\ }\bibfield  {title} {\bibinfo {title} {Study of the homology between silicon and germanium by thermal-neutron spectrometry},\ }\href {https://doi.org/10.1103/PhysRevB.6.3777} {\bibfield  {journal} {\bibinfo  {journal} {Phys. Rev. B}\ }\textbf {\bibinfo {volume} {6}},\ \bibinfo {pages} {3777} (\bibinfo {year} {1972})}\BibitemShut {NoStop}%
\bibitem [{\citenamefont {Dolling}(1963)}]{Dolling1963}%
  \BibitemOpen
  \bibfield  {author} {\bibinfo {author} {\bibfnamefont {G.}~\bibnamefont {Dolling}},\ }\bibfield  {title} {\bibinfo {title} {Lattice vibrations in crystals with the diamond structure},\ }in\ \href@noop {} {\emph {\bibinfo {booktitle} {Inelastic scattering of neutrons in solids and liquids}}},\ Vol.~\bibinfo {volume} {2}\ (\bibinfo  {publisher} {IAEA},\ \bibinfo {address} {Vienna},\ \bibinfo {year} {1963})\ pp.\ \bibinfo {pages} {37--48}\BibitemShut {NoStop}%
\bibitem [{\citenamefont {Aouissi}\ \emph {et~al.}(2006)\citenamefont {Aouissi}, \citenamefont {Hamdi}, \citenamefont {Meskini},\ and\ \citenamefont {Qteish}}]{Aouissi2006}%
  \BibitemOpen
  \bibfield  {author} {\bibinfo {author} {\bibfnamefont {M.}~\bibnamefont {Aouissi}}, \bibinfo {author} {\bibfnamefont {I.}~\bibnamefont {Hamdi}}, \bibinfo {author} {\bibfnamefont {N.}~\bibnamefont {Meskini}},\ and\ \bibinfo {author} {\bibfnamefont {A.}~\bibnamefont {Qteish}},\ }\bibfield  {title} {\bibinfo {title} {Phonon spectra of diamond, {Si}, {Ge}, and {$\alpha$-Sn}: Calculations with real-space interatomic force constants},\ }\href {https://doi.org/10.1103/physrevb.74.054302} {\bibfield  {journal} {\bibinfo  {journal} {Phys. Rev. B}\ }\textbf {\bibinfo {volume} {74}},\ \bibinfo {pages} {054302} (\bibinfo {year} {2006})}\BibitemShut {NoStop}%
\bibitem [{\citenamefont {Rayne}\ and\ \citenamefont {Chandrasekhar}(1960)}]{Rayne1960}%
  \BibitemOpen
  \bibfield  {author} {\bibinfo {author} {\bibfnamefont {J.~A.}\ \bibnamefont {Rayne}}\ and\ \bibinfo {author} {\bibfnamefont {B.~S.}\ \bibnamefont {Chandrasekhar}},\ }\bibfield  {title} {\bibinfo {title} {Elastic constants of $\ensuremath{\beta}$ tin from 4.2\ifmmode^\circ\else\textdegree\fi{}{K} to 300\ifmmode^\circ\else\textdegree\fi{}{K}},\ }\href {https://doi.org/10.1103/PhysRev.120.1658} {\bibfield  {journal} {\bibinfo  {journal} {Phys. Rev.}\ }\textbf {\bibinfo {volume} {120}},\ \bibinfo {pages} {1658} (\bibinfo {year} {1960})}\BibitemShut {NoStop}%
\bibitem [{\citenamefont {Wang}\ \emph {et~al.}(2026)\citenamefont {Wang}, \citenamefont {Engel}, \citenamefont {Lane}, \citenamefont {Zhang}, \citenamefont {Miranda}, \citenamefont {Hou}, \citenamefont {Barbiellini}, \citenamefont {Markiewicz}, \citenamefont {Zhu}, \citenamefont {Kresse}, \citenamefont {Bansil}, \citenamefont {Sun},\ and\ \citenamefont {Zhang}}]{Wang2024}%
  \BibitemOpen
  \bibfield  {author} {\bibinfo {author} {\bibfnamefont {Y.}~\bibnamefont {Wang}}, \bibinfo {author} {\bibfnamefont {M.}~\bibnamefont {Engel}}, \bibinfo {author} {\bibfnamefont {C.}~\bibnamefont {Lane}}, \bibinfo {author} {\bibfnamefont {Y.}~\bibnamefont {Zhang}}, \bibinfo {author} {\bibfnamefont {H.}~\bibnamefont {Miranda}}, \bibinfo {author} {\bibfnamefont {L.}~\bibnamefont {Hou}}, \bibinfo {author} {\bibfnamefont {B.}~\bibnamefont {Barbiellini}}, \bibinfo {author} {\bibfnamefont {R.~S.}\ \bibnamefont {Markiewicz}}, \bibinfo {author} {\bibfnamefont {J.-X.}\ \bibnamefont {Zhu}}, \bibinfo {author} {\bibfnamefont {G.}~\bibnamefont {Kresse}}, \bibinfo {author} {\bibfnamefont {A.}~\bibnamefont {Bansil}}, \bibinfo {author} {\bibfnamefont {J.}~\bibnamefont {Sun}},\ and\ \bibinfo {author} {\bibfnamefont {R.}~\bibnamefont {Zhang}},\ }\bibfield  {title} {\bibinfo {title} {Accurate electron-phonon interactions from advanced density functional theory},\ }\href {https://doi.org/10.1103/w56r-f5yy} {\bibfield  {journal}
  {\bibinfo  {journal} {PRX Energy}\ }\textbf {\bibinfo {volume} {5}},\ \bibinfo {pages} {013002} (\bibinfo {year} {2026})}\BibitemShut {NoStop}%
\bibitem [{\citenamefont {Yates}\ and\ \citenamefont {Bartók}(2025)}]{Yates2025}%
  \BibitemOpen
  \bibfield  {author} {\bibinfo {author} {\bibfnamefont {J.~R.}\ \bibnamefont {Yates}}\ and\ \bibinfo {author} {\bibfnamefont {A.~P.}\ \bibnamefont {Bartók}},\ }\bibfield  {title} {\bibinfo {title} {Accurate predictions of chemical shifts with the {rSCAN} and {r$^2$SCAN} {mGGA} exchange–correlation functionals},\ }\href {https://doi.org/10.1039/d4fd00142g} {\bibfield  {journal} {\bibinfo  {journal} {Faraday Discuss.}\ }\textbf {\bibinfo {volume} {255}},\ \bibinfo {pages} {192} (\bibinfo {year} {2025})}\BibitemShut {NoStop}%
\bibitem [{MC()}]{MC}%
  \BibitemOpen
  \href@noop {} {}\bibinfo {note} {See Supplementary Data at \url{https://doi.org/10.24435/materialscloud:dp-1f} for raw phonon data and VASP input files.}\BibitemShut {Stop}%
\end{thebibliography}%

\onecolumngrid
\clearpage
\onecolumngrid
\setcounter{page}{1} 
\renewcommand{\thefigure}{S\arabic{figure}}
\renewcommand{\thetable}{S\arabic{table}}
\renewcommand{\theequation}{S\arabic{equation}}
\setcounter{figure}{0}
\setcounter{table}{0}
\setcounter{equation}{0}

\cleardoublepage
\onecolumngrid
\setcounter{section}{0}
\setcounter{equation}{0}
\setcounter{figure}{0}
\setcounter{table}{0}
\setcounter{page}{1}
\renewcommand{\theequation}{S\arabic{equation}}
\renewcommand{\thefigure}{S\arabic{figure}}
\renewcommand{\theHfigure}{S\arabic{figure}}
\renewcommand{\thetable}{S\arabic{table}}
\renewcommand{\theHtable}{S\arabic{table}}
\begin{center}
{{\Large \bf Supplemental Material for}\\
{\Large \bf Lattice dynamics and structural phase stability of group-IV elemental solids with the
r$^2$SCAN functional }\\[2.\medskipamount]
{\large Adonis Haxhijaj,$^{*}$ Stefan Riemelmoser, and Alfredo Pasquarello}}\\
{\it Chaire de Simulation à l'Echelle Atomique (CSEA),\\ Ecole Polytechnique Fédérale de Lausanne (EPFL), CH-1015 Lausanne,
Switzerland}\\
(Dated: \today)\\[2.\medskipamount]
$^{\rm *}$Electronic mail: adonis.haxhijaj@epfl.ch
\end{center}    

\section{Numerical setup and basis-set convergence}\label{app:Appendix A}

Table~\ref{tab:pseudopotentials} lists the PBE PAW pseudopotentials employed in this work. By construction, these pseudopotentials ensure broad transferability, which is particularly important when PBE pseudopotentials are used in combination with meta-GGA functionals. The shorthands \_GW and \_h indicate accurate treatment of high-energy states. The shorthand \_sv indicates that the semicore shell is explicitly treated as valence. \\
As discussed in Sec.\@~\ref{sec:Methodology} in the main text, we compare two different numerical setups. Setup A in Table~\ref{tab:numerical setup}
represents a straightforward way to reach basis-set convergence in vasp. Specifically, setup A uses a single-grid strategy (PREC=Accurate), where the plane-wave basis-set cutoff is determined by a single plane-wave cutoff (ENCUT). Further, calculations of phonons and elastic constants use the default step size for the final displacements (POTIM = 0.015~\AA). In the following, we demonstrate that setup A is disadvantageous, in particular, for the meta-GGA functional SCAN. In contrast, satisfactory numerical accuracy can be achieved with setup B given in Table~\ref{tab:numerical setup}.  
Following \citet{Ning2022}, we employ a double-grid strategy (PREC=High in vasp), which uses a higher plane-wave cutoff for the PAW augmentation charges (ENAUG in vasp). Further, the finite displacement step (POTIM) is increased to 0.060~\AA. Finally, for both setups, a stringent energy criterion of $10^{-8}$~eV is used as a break condition for the electronic self-consistency cycle (EDIFF in vasp). This ensures high numerical accuracy for the forces in finite difference calculations.  \\

\begin{table}[htb]
\caption{Specification of pseudopotentials used. We employ PBE PAW pseudopotentials~\cite{KresseJoubert1999} as distributed with vasp.6.4.2. The first column lists the pseudopotentials, the second column the valence electron configuration. The default plane-wave basis-set cutoff (ENMAX in vasp) is given in the third column, while the fourth column lists the plane-wave cutoff used for the calculations in the main text (ENMAX in vasp). All cutoff energies are given in~eV. }\label{tab:pseudopotentials}
\begin{tabular}{l@{\qquad} l@{\qquad}r@{\qquad}r}
\hline\hline
\\[-2.5\medskipamount]
       PAW potential & valence & ENMAX & ENCUT \\ 
       \colrule
        C\_h\_GW      & 2$s^2$2$p^2$ &  741.7 & 1335.1 \\
        Si\_sv\_GW & 2$s^2$2$p^6$3$s^2$3$p^2$ & 547.6 & 985.7  \\ 
        Ge\_sv\_GW & 3$s^2$3$p^6$3$d^{10}$4$s^2$4$p^2$ & 410.4 & 738.7 \\
    Sn\_sv\_GW & 4$s^2$4$p^6$4$d^{10}$5$s^2$5$p^2$ & 368.8 & 663.8 \\
    \hline\hline
\end{tabular}
\end{table}

\begin{table}[htb]
\caption{Numerical setups for the phonon calculations in vasp. Setup A uses a single-grid (PREC=Accurate), with energy cutoff ENCUT chosen to be 1.8$\times$ the default cutoff (ENMAX). Setup B uses a double-grid strategy (PREC=High), which employs a different energy cutoff for the augmentation charges (ENAUG). The step size for the finite difference is increased in setup B, in both cases we use a fourth-order finite difference stencil (NFREE=4). Both setups use a stringent energy criterion as the break condition for electronic self-consistency (EDIFF).}\label{tab:numerical setup}
\begin{tabular}{l@{\qquad} c@{\qquad}c@{\qquad}l}
\hline\hline
\\[-2.5\medskipamount]
    & setup A & setup B \\
    \colrule
PREC & Accurate & High \\
ENCUT & 1.8 $\times $ ENMAX & 1.8 $\times$ ENMAX \\
ENAUG &     &   4000~eV       \\
POTIM & 0.015~\AA & 0.060~\AA \\
NFREE & 4          &   4       \\
EDIFF & $10^{-8}$~eV & $10^{-8}$~eV \\
\hline\hline
\end{tabular}
\end{table}
\begin{figure*}[thb]
  \centering
    \subfloat{
        \includegraphics[width=0.97\textwidth]{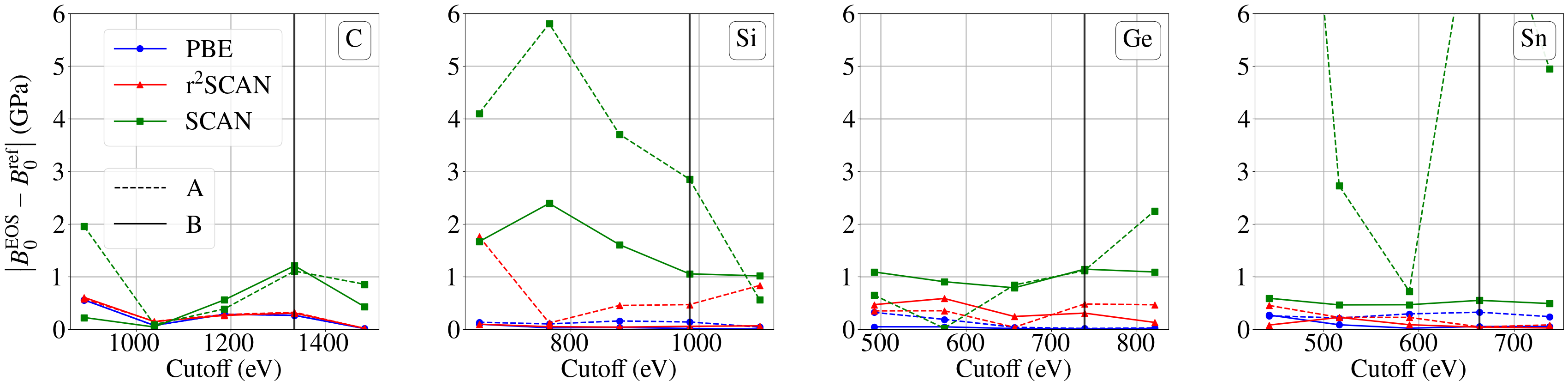}
        \begin{tikzpicture}[remember picture, overlay]
          \node[anchor=north west, inner sep=0pt] at (-0.935 \textwidth, 0.255\textwidth) {(a)};
        \end{tikzpicture}
        \label{fig:EOS_conv}} \\
      \subfloat{
        \hspace{-0.13cm}
        \includegraphics[width=0.97\textwidth]{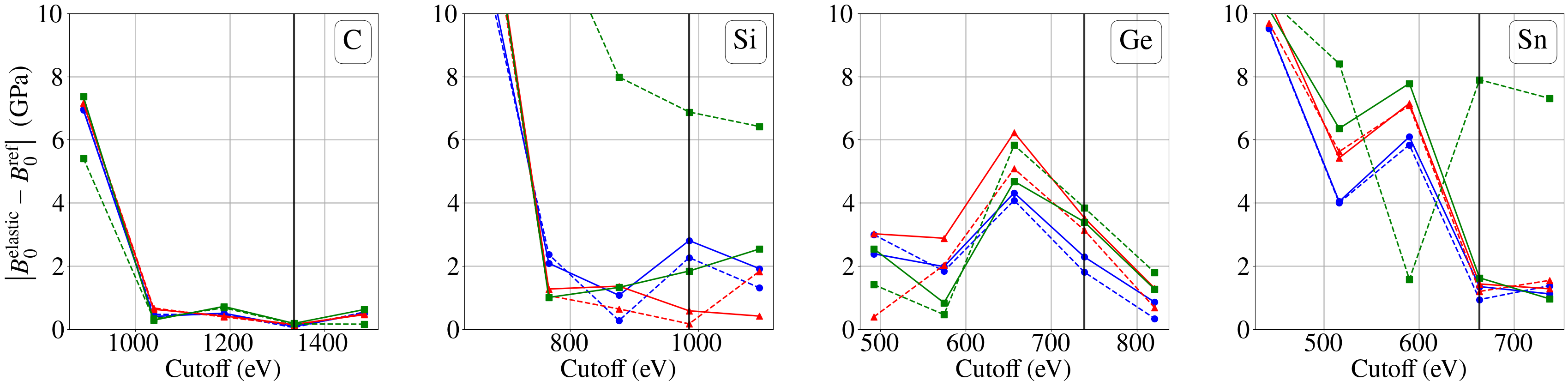}
        \begin{tikzpicture}[remember picture, overlay]
          \node[anchor=north west, inner sep=0pt] at (-0.935 \textwidth, 0.255\textwidth) {(b)};
        \end{tikzpicture}
        \label{fig:bulk_modulus_convergence}}\\
      \subfloat{
      \hspace{-0.4cm}
    \includegraphics[width=0.98\textwidth]{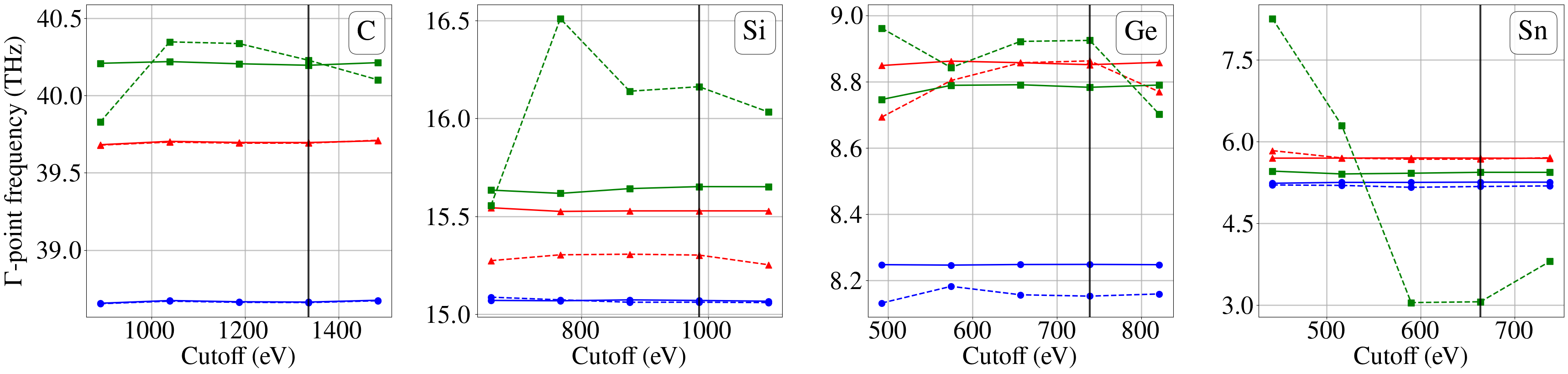}
        \begin{tikzpicture}[remember picture, overlay]
          \node[anchor=north west, inner sep=0pt] at (-0.935 \textwidth, 0.255\textwidth) {(c)};
        \end{tikzpicture}
        \label{fig:gamma_conv}}
    \caption{Convergence with respect to the plane-wave energy cutoff, varied from \(1.2\times \) ENMAX to \(2.0\times \)ENMAX, where ENMAX is the default cutoff of the respective pseudopotentials. All convergence curves are shown for the three functionals PBE, r$^2$SCAN, and SCAN, and the labels A and B in the legend refer to the numerical setups A and B listed in Table~\ref{tab:numerical setup}. Panel (a) shows the bulk modulus \(B_0^{\mathrm{EOS}}\) obtained from Birch-Murnaghan equation-of-state fits [Eq.\@~\eqref{eq:BM EOS}]. Panel (b) shows the bulk modulus \(B_0^{\mathrm{elastic}}\) evaluated from the elastic constants using Eq.~(\ref{eq:bulk-with-C}) of the main text. \(B_0^{\mathrm{ref}}\) in panels (a) and (b) denotes a reference bulk modulus obtained from an equation-of-state fit at a very high cutoff (4$\times$ENMAX). Panel (c) shows the convergence of the \(\Gamma\)-point phonon frequency. In all panels, the black vertical line marks the cutoff \(1.8\times\) ENMAX, which is the value adopted for the final elastic-constant calculations (Tables~\ref{tab:bulk moduli} and~\ref{tab:elastic_constants_EXPT} in the main text) and phonon-frequency calculations (Table~\ref{tab:vibrational_properties_combined} in the main text).}
    \label{fig:combined_convergence}
\end{figure*}

\begin{figure*}[thb]
  \begin{tikzpicture}
    \node[anchor=south west, inner sep=0] (img) at (0,0)
      {\includegraphics[width=0.97\textwidth]{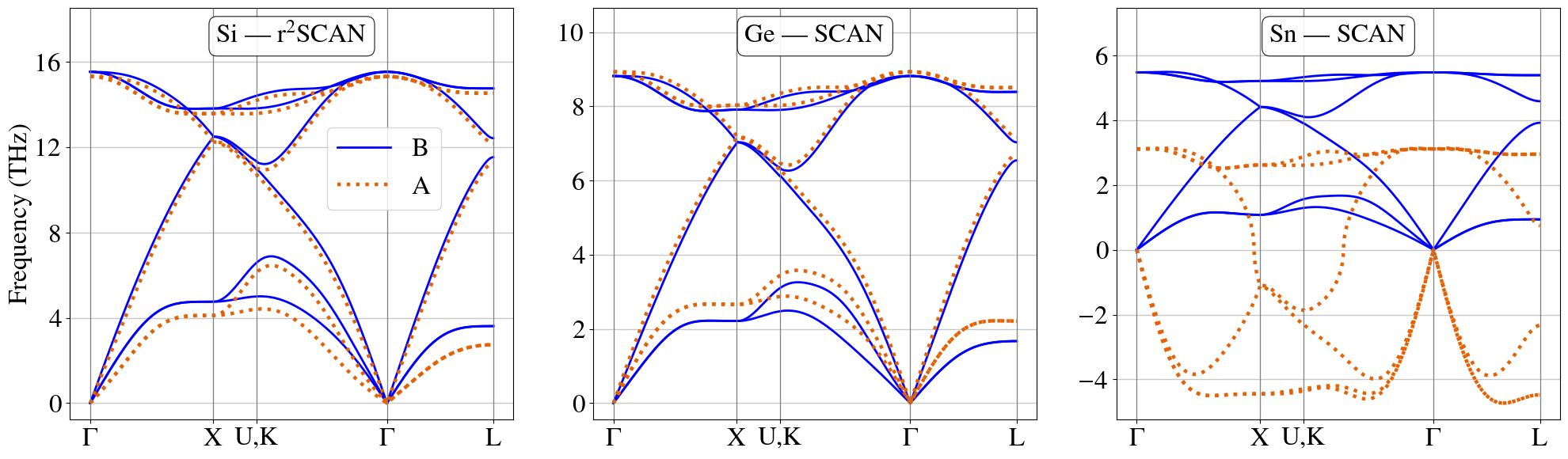}};
    \begin{scope}[x={(img.south east)}, y={(img.north west)}]
      \node[anchor=north west, inner sep=2pt]
        at (0.0388, 1.068) {(a)};
    \end{scope}
    \begin{scope}[x={(img.south east)}, y={(img.north west)}]
      \node[anchor=north west, inner sep=2pt]
        at (0.372, 1.068) {(b)}; 
    \end{scope}
    \begin{scope}[x={(img.south east)}, y={(img.north west)}]
      \node[anchor=north west, inner sep=2pt]
        at (0.706, 1.068) {(c)}; 
    \end{scope}
  \end{tikzpicture}
\caption{\label{fig:phonons_supercell_conv_P06} Demonstrations of numerical instabilities observed for phonon dispersions in the cases of (a) Si with r$^2$SCAN, (b) Ge with SCAN, and (c) Sn with SCAN. Dotted lines correspond to setup A in Table~\ref{tab:numerical setup}, solid lines correspond to setup B in Table~\ref{tab:numerical setup}. All phonon dispersions are calculated with 128-atom supercells and the plane-wave cutoffs indicated by the black vertical lines in Fig.\@~\ref{fig:combined_convergence}. The phonon dispersions obtained by setup B are equivalent to those used in Fig.\@~\ref{fig:phonons_vs_exp} of the main text.}
\end{figure*}
Next, we compare the plane-wave basis-set convergence for the two setups. We note that setup B is used throughout for the calculations in the main text.
Figure~\ref{fig:combined_convergence}(a)  shows the basis-set convergence of the bulk moduli $B_0^{\rm EOS}$ obtained by Birch-Murnaghan equation-of-state fits [Eq.\@~\eqref{eq:BM EOS}]. Using setup A, SCAN exhibits slow basis-set convergence, especially for Si and Sn (see green dashed lines). The convergence behavior is significantly improved with the double-grid technique (see full green lines). In contrast, both setups yield rapid convergence of $B_0^{\rm EOS}$ for PBE (blue lines) and r$^2$SCAN (red lines).
Figure~\ref{fig:combined_convergence}(b) shows the convergence of the bulk moduli $B_0^{\rm elastic}$ calculated via Eq.\@~\eqref{eq:bulk-with-C}. For large cutoffs, we observe that the identity $B_0^{\rm elastic}=B_0^{\rm EOS}$ holds very well, as it should [see Eq.\@~\eqref{eq:bulk-with-C}]. However, at lower cutoff energies, the basis-set incompleteness induces Pulay stress~\cite{Dacosta1986,G_P_Francis_1990}, which artificially alters the calculated pressure and leads to a systematic violation of the identity in Eq.\@~\eqref{eq:bulk-with-C} of the main text. Furthermore, numerical setup A shows again poor basis-set convergence for the SCAN elastic constants of Si and Sn, which does not occur with setup B. In contrast, setups A and B yield almost identical results for the PBE and r$^2$SCAN functionals.
Figure~\ref{fig:combined_convergence}(c) illustrates the basis-set convergence for the phonon frequencies at the $\Gamma$ point. Since these calculations do not involve volume changes, basis-set convergence can be more rapid than for $B_0$. The $\Gamma$-point frequencies obtained using PBE and r$^2$SCAN are indeed already converged at very low cutoffs. Furthermore, the full and dashed lines representing numerical setups A and B differ slightly even in the limit of large cutoffs. This is due to the different step sizes (0.015~\text{\AA } for setup A, and 0.060~\text{\AA } for setup B, see Table~\ref{tab:numerical setup}). In principle, the frequencies obtained with setup A are more accurate, since we want to simulate the linear response regime. Still, the differences between setups A and B are below 0.1~THz, within our desired numerical accuracy. We have also confirmed that PBE $\Gamma$-point frequencies obtained with setup A are in excellent agreement with density functional perturbation theory (not shown). The phonon frequencies obtained with SCAN are numerically stable only with setup B, while SCAN frequencies calculated with setup A exhibit poor basis-set convergence, particularly for Sn. These numerical instabilities are demonstrated in Fig.\@~\ref{fig:phonons_supercell_conv_P06}, where we use the same plane-wave cutoff as in the main text. We note that the phonon dispersion obtained with setup A is generally less stable. A drastic example is the SCAN phonon dispersion of Sn, where setup A produces large, unphysical imaginary frequencies [see panel (c) in Fig.\@~\ref{fig:phonons_supercell_conv_P06}]. Smaller instabilities for the acoustic modes are present for the SCAN phonon dispersions of Si and Ge, the r$^2$SCAN phonon dispersion of Si, and the PBE phonon dispersion of Sn. In contrast, numerical setup B manages to stabilize the phonon dispersions in all cases, see also Fig.\@~\ref{fig:phonons_supercell_conv_all}. 
In summation, we find that for the group-IV elemental solids, the r$^2$SCAN functional offers stable phonon frequencies, and favorable plane-wave basis-set convergence. In this regard, r$^2$SCAN performs comparably to the PBE functional, while SCAN exhibits poor numerical stability and slow basis-set convergence. This agrees with the findings of \citet{Ning2022}. These authors have correctly pointed out that it might be close to impossible to obtain converged SCAN results for numerically challenging materials. Given that r$^2$SCAN closely approximates the lattice dynamics of SCAN, but avoids the numerical problems, we agree with \citet{Ning2022} that r$^2$SCAN should be used instead of SCAN for high-throughput studies of lattice dynamics.

\section{Supercell convergence}\label{app: Supercell convergence}

\begin{figure*}[thb]
\includegraphics[width=0.97\textwidth]{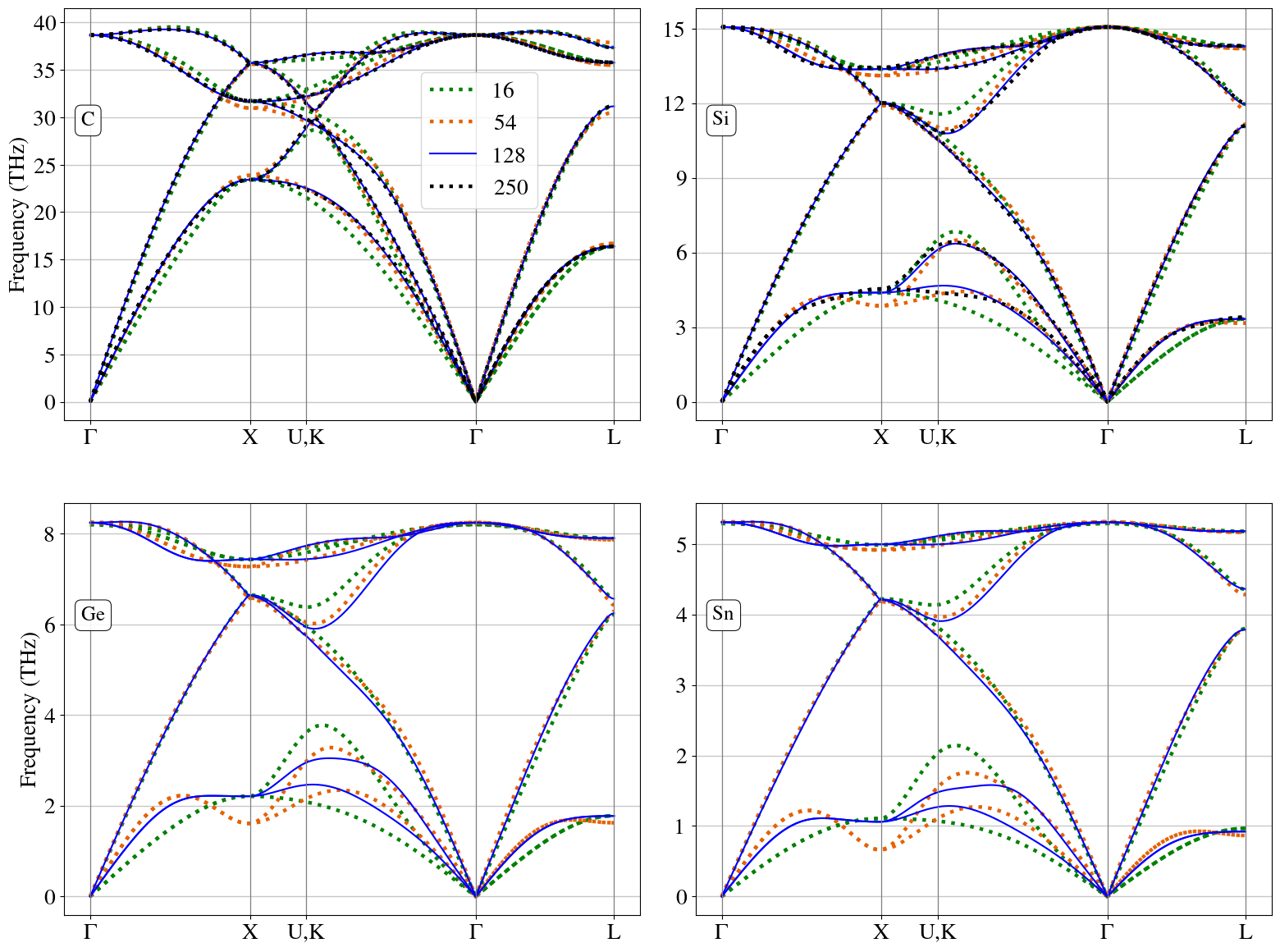}
        \begin{tikzpicture}[remember picture, overlay]
          \node[anchor=north west, inner sep=0pt] at (-0.929 \textwidth, 0.736\textwidth) {(a)};
        \end{tikzpicture}
        \begin{tikzpicture}[remember picture, overlay]
          \node[anchor=north west, inner sep=0pt] at (-0.935 \textwidth, 0.362\textwidth) {(c)};
        \end{tikzpicture}
        \begin{tikzpicture}[remember picture, overlay]
          \node[anchor=north west, inner sep=0pt] at (-0.462 \textwidth, 0.736\textwidth) {(b)};
        \end{tikzpicture}
        \begin{tikzpicture}[remember picture, overlay]
          \node[anchor=north west, inner sep=0pt] at (-0.468 \textwidth, 0.362\textwidth) {(d)};
        \end{tikzpicture}
\caption{\label{fig:phonons_supercell_conv_all} Supercell size convergence of phonon dispersions for group-IV elemental solids in the diamond phase with the PBE functional: (a) C, (b) Si, (c) Ge, and (d) Sn. The 128-atom supercell used in Fig.\@~\ref{fig:phonons_vs_exp} of the main text is indicated in bold.}
\end{figure*}

As discussed in Sec.\@~\ref{sec:Th:Phonons} of the main text, the phonon frequencies obtained by the direct method are explicitly calculated only at wave vectors $\textbf{q}$ that are commensurate with the supercell, whereas phonons at other wave vectors are interpolated. It is therefore important to check the convergence of the phonon dispersion with respect to the supercell size. 
Figure~\ref{fig:phonons_supercell_conv_all} demonstrates that the changes from the 54-atom supercell to the 128-atom supercell are only significant near the X point. Note that phonon frequencies at the X point are explicitly calculated for both the 16-atom supercell and the 128-atom supercell (negligible finite-size error, same for the $\Gamma$ and L points). Stringent agreement for the 16-atom and 128-atom supercells at these high-symmetry points is thus an important sanity check, demonstrating good numerical accuracy. In contrast, the 54-atom and 250-atom supercells are not commensurate with phonons at the X and L points, and phonon frequencies at these high-symmetry points are subject to larger finite-size errors.
We explicitly demonstrate that the phonon dispersions for C and Si obtained with 250-atom supercells  barely differ from the phonon dispersions obtained with 128-atom supercells. We note, however, that very large supercells ($\sim$1000 atoms)  are required to completely converge the phonon dispersions~\cite{Aouissi2006}. For instance, 1024 atoms are needed to obtain a supercell that is commensurate with the K point. As an example, we compute the elastic constants by fitting to the acoustic phonon modes [see Eqs.~\eqref{eq:v_100} and~\eqref{eq:v_110}]. 

\begin{figure*}[thb]
\includegraphics[width=0.97\textwidth]
{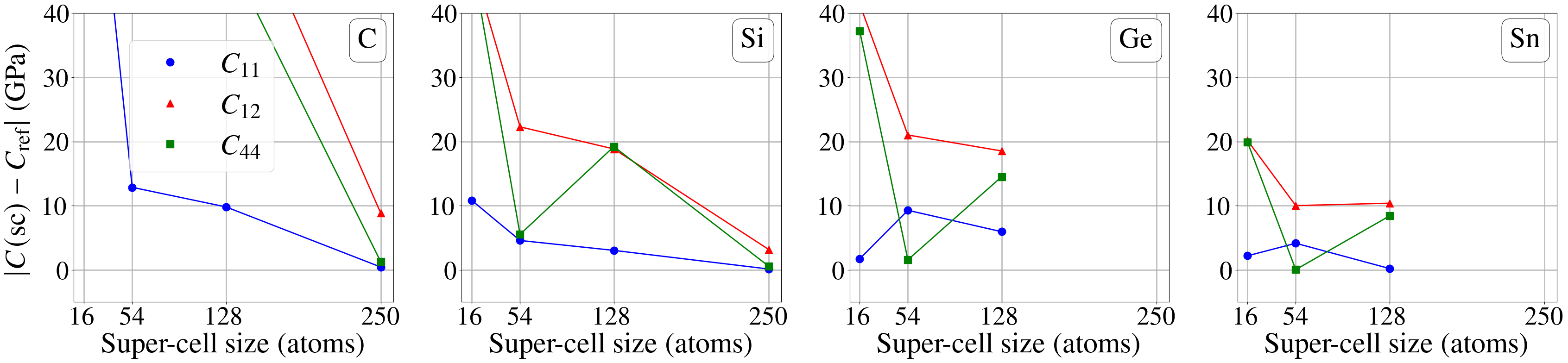}
\caption{Elastic constants obtained from linear fits to the acoustic modes of the phonon dispersions [see Eqs.\@~\eqref{eq:v_100} and~\eqref{eq:v_110}]. All calculations are done with the PBE functional, full phonon dispersions are shown in Fig.\@~\ref{fig:phonons_supercell_conv_all}. Reference elastic constants obtained by the finite difference strain-stress method are listed in Table~\ref{tab:elastic_constants_EXPT} of the main text.}\label{fig:elastic_constant_convergence}
\end{figure*}

Figure~\ref{fig:elastic_constant_convergence} demonstrates the slow supercell convergence of the $C_{ij}$. We remark that this approach has, in principle, some advantage over the finite difference method. Namely, it avoids slow basis-set convergence due to Pulay stress (see Fig.\@~\ref{fig:combined_convergence}). However, this advantage is, in practice, outweighed by the slow supercell convergence. To conclude, we choose the 128-atom supercells as a good compromise between accuracy and computational cost. This supercell size is also commensurate with phonons at the high-symmetry points $\Gamma$, X, and L. Similar supercell sizes were also used in previous studies (64-atom supercells for group-IV elemental solids in Ref.\@~\cite{Hummer2009}, 128-atom supercell for C in Ref.\@~\cite{Riemelmoser2025}, and 128-atom supercell for Sn in Ref.\@~\cite{Mehl2021}).

\section{Data visualization}

\begin{figure*}[thb]
  \centering
  \begin{minipage}[t]{0.49\textwidth}
    \centering
    \begin{tikzpicture}
      \node[anchor=south west, inner sep=0] (img) at (0,0)
        {\includegraphics[width=0.995\textwidth]{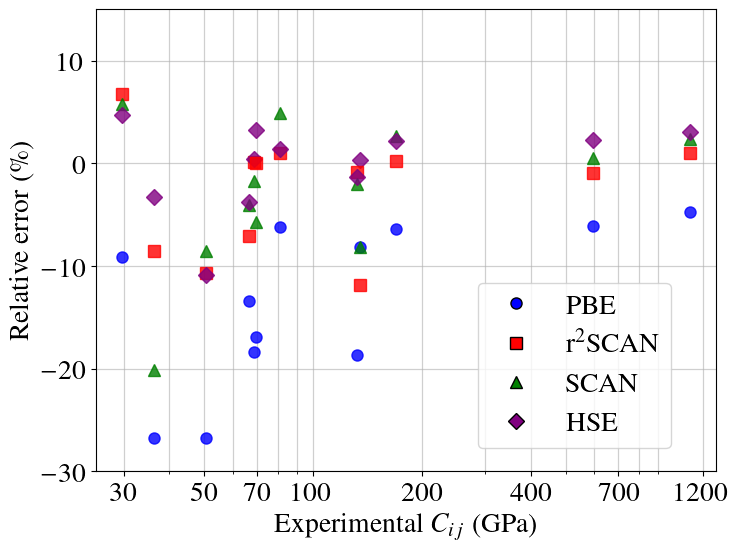}};
      \begin{scope}[x={(img.south east)}, y={(img.north west)}]
        \node[anchor=north west, inner sep=2pt]
          at (0.118, 1.05) {(a)}; 
      \end{scope}
    \end{tikzpicture}
  \end{minipage}
  \hfill
  \begin{minipage}[t]{0.49\textwidth}
    \centering
    \begin{tikzpicture}
      \node[anchor=south west, inner sep=0] (img) at (0,0)
        {\includegraphics[width=\textwidth]{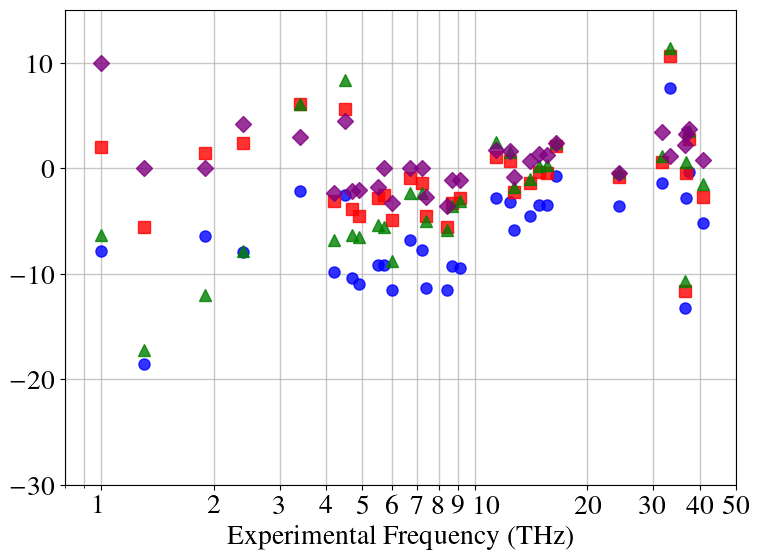}};
      \begin{scope}[x={(img.south east)}, y={(img.north west)}]
        \node[anchor=north west, inner sep=2pt]
          at (0.074, 1.05) {(b)};
      \end{scope}
    \end{tikzpicture}
  \end{minipage}
  \caption{\label{fig:combined_properties}
    Relative error of calculated properties for group-IV elemental solids in the diamond phase compared to ZPE-corrected experimental references.
    (a) Elastic constants ($C_{11}$, $C_{12}$, and $C_{44}$).
    (b) Phonon frequencies at high-symmetry points ($\Gamma$, X, and L).
    Results are shown for the PBE, r$^2$SCAN, SCAN, and HSE functionals. Note that a logarithmic scale is used for the abscissae.
    Numerical data are provided in Tables~\ref{tab:elastic_constants_EXPT} and~\ref{tab:vibrational_properties_combined} of the main text.}
\end{figure*}

Figure~\ref{fig:combined_properties}(a) shows the relative errors of the calculated elastic constants with respect to the ZPE-corrected experimental values (as in Table~\ref{tab:elastic_constants_EXPT} of the main text). Furthermore, Fig.\@~\ref{fig:combined_properties}(b) shows the relative errors of the calculated phonon frequencies at high-symmetry points in the Brillouin zone (as in Table~\ref{tab:vibrational_properties_combined} of the main text). In both cases, the PBE underbinding error and the improvements with meta-GGA and hybrid functionals are clearly apparent.

\end{document}